# Proposal to Build a 30 Kiloton Off-Axis Detector to Study $\nu_\mu \to \nu_e$ Oscillations in the NuMI Beamline

# NOνA

# N$_{uMI}$ O$_{ff\text{-}Axis}$ ν$_e$ A$_{ppearance}$ Experiment

March 21, 2005

## The NOνA Collaboration

Argonne, Athens, Caltech, UCLA, Fermilab, College de France,
Harvard, Indiana, ITEP, Lebedev, Michigan State,
Minnesota/Duluth, Minnesota/Minneapolis, Munich,
Stony Brook, Northern Illinois, Ohio, Ohio State, Oxford,
Rio de Janeiro, Rutherford, South Carolina, Stanford, Texas A&M,
Texas/Austin, Tufts, Virginia, Washington, William & Mary

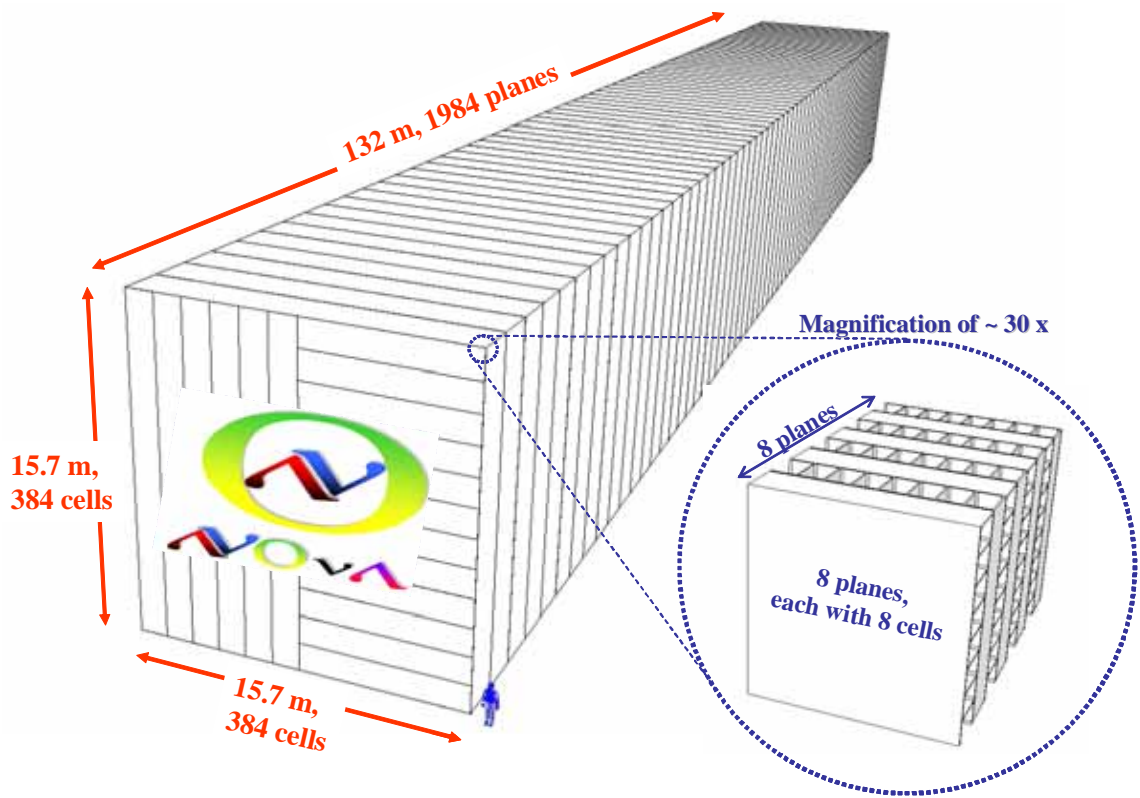

D. S. Ayres, J. W. Dawson, G. Drake, M. C. Goodman, J. J. Grudzinski,
V. J. Guarino, T. Joffe-Minor, D. E. Reyna, R. L. Talaga, J. L. Thron, R. G. Wagner
*Argonne National Laboratory, Argonne, IL*

D. Drakoulakos, N. Giokaris, P. Stamoulis, S. Stiliaris, G. Tzanakos, M. Zois
*University of Athens, Athens, Greece*

J. Hanson, C. Howcroft, D. G. Michael, H. B. Newman, C. W. Peck, J. Trevor, H. Zheng
*California Institute of Technology, Pasadena, CA*

D. B. Cline, K. Lee
*University of California, Los Angeles, CA*

R. Bernstein, G. Bock, S. Brice, L. Camilleri*, S. Childress, B. Choudhary, J. Cooper**,
D. Harris, R. Hatcher, J. Hylen, H. Jostlein, J. Kilmer, D. Koolbeck, P. Lucas,
V. Makeev, O. Mena, S. Mishra, S. Parke, R. Plunkett, R. Rameika, R. Ray, B.J. Rebel, R.
Schmitt, P. Shanahan, P. Spentzouris, R. Wands, R. Yarema
*Fermi National Accelerator Laboratory, Batavia, IL*

T. Patzak
*College de France, Paris, France*

J. Boehm, G. J. Feldman**, N. Felt, A. Lebedev, J. Oliver, M. Sanchez, S.-M. Seun
*Harvard University, Cambridge, MA*

C. Bower, M. D. Messier, S. Mufson, J. Musser, J. Urheim, A. Waldron
*Indiana University, Bloomington, IN*

I. Trostin
*Institute For Theoretical And Experimental Physics, Moscow, Russia*

V. Ryabov
*P. N. Lebedev Physical Institute, Moscow, Russia*

C. Bromberg, J. Huston, R. Miller, R. Richards
*Michigan State University, East Lansing, MI*

A. Habig
*University of Minnesota, Duluth, MN*

T.R. Chase, D. Cronin-Hennessy, K. Heller, P.J. Litchfield, M.L. Marshak,
W.H. Miller, L. Mualem, E.A. Peterson, D.A. Petyt, K. Ruddick, R. Rusack
*University of Minnesota, Minneapolis, MN*

M. Lindner
*Technische Universität München, Munich, Germany*

R. Shrock
*State University of New York, Stony Brook, NY*

C. Albright
*Northern Illinois University, DeKalb, IL*

J. Beacom
*Ohio State University, Columbus, OH*

C. R. Brune, S. M. Grimes, A. K. Opper
*Ohio University, Athens, OH*

G. Barr, J. H. Cobb, K. Grzelak, N. Tagg
*University of Oxford, Oxford, United Kingdom*

H. Nunokawa
*Pontifícia Universidade Católica do Rio de Janeiro, Rio de Janeiro, Brazil*

A. Belias, T. Durkin, T. Nicholls, G. F. Pearce, A. Weber
*Rutherford Appleton Laboratory, Chilton, Didcot, United Kingdom*

T. Bergfeld, A. Godley, J. Ling, S. R. Mishra, C. Rosenfeld
*University of South Carolina, Columbia, SC*

S. Avvakumov, G. Irwin, S. Murgia, S. Wojcicki, T. Yang
*Stanford University, Stanford, CA*

E. Tetteh-Lartey, R. Webb
*Texas A&M University, College Station, TX*

S. Kopp, K. Lang
*University of Texas, Austin, TX*

H.R. Gallagher, T. Kafka, W.A. Mann, J. Schneps, A. Sousa
*Tufts University, Medford, MA*

C. Dukes, K. Nelson
*University of Virginia, Charlottesville, VA*

J. Rothberg, T. Zhao
*University of Washington, Seattle, WA*

J. K. Nelson, F.X. Yumiceva
*The College of William and Mary, Williamsburg, VA*

\* On leave from CERN
\*\* Co-Spokespersons

# Table of Contents

# 1. Executive Summary

## 1.1. The Physics of NOvA

The past two decades have seen great advances in our understanding of neutrinos. Underground experiments detecting neutrinos produced in the sun and in the earth's atmosphere have shown that neutrinos have mass and that they oscillate from one species to another as they travel. These oscillations arise because the neutrino species produced in particle decays (electron, muon, and τ-type neutrinos) do not have specific masses but are combinations of neutrino species (simply called 1, 2, and 3-type neutrinos) that do have specific masses. The average distance a neutrino travels before it oscillates is proportional to its energy and inversely proportional to the difference of the squares of masses of the underlying species of neutrinos. The probability that an oscillation will occur is related to a parameter known as a mixing angle.

The neutrinos that come from the sun are electron-type neutrinos that oscillate to muon and τ-type neutrinos, characterized by the mixing angle $\theta_{12}$ and an oscillation length (normalized to an energy of 2 GeV) of approximately 35,000 km. Muon-type neutrinos produced by cosmic rays in the earth's atmosphere oscillate to τ–type neutrinos, characterized by the mixing angle $\theta_{23}$, and an oscillation length (again normalized to an energy of 2 GeV) of approximately 1,000 km. A third type of neutrino oscillation is possible: the oscillation of muon-type neutrinos to electron-type neutrinos at the atmospheric oscillation length. These neutrino oscillations, which so far have not been observed, would be characterized by the mixing angle $\theta_{13}$. The study of this last category of neutrino oscillations is the main goal of NOvA (**N**uMI **O**ff-Axis $\nu_e$ **A**ppearance Experiment)[1].

The significance of the search for these oscillations is that if they exist, i.e., if $\theta_{13}$ is not zero, then we will ultimately be able to determine the ordering of the neutrino masses and measure CP violation in neutrino oscillations. There is widespread belief that the very small neutrino masses are related to physics at an extremely high-energy scale, one that cannot be studied directly with accelerator beams. There is also theoretical speculation that CP violation by neutrinos could be one aspect of understanding why the universe is composed solely of matter, rather than equal amounts of matter and antimatter.

MINOS is one of the first generation of long baseline accelerator-based neutrino oscillation experiments.[2] This Fermilab experiment, which has a 735 km baseline, will start taking data next month, April 2005. The MINOS Far Detector is located in the lowest level of the Soudan mine in northern Minnesota and it sits directly on the center of the Fermilab NuMI neutrino beam line. The physics goals of the MINOS experiment are to verify the atmospheric neutrino oscillations, to improve the measurement of their parameters, and to perform a low-sensitivity measurement of $\theta_{13}$.

We are proposing NOvA to utilize Fermilab's investment in the NuMI beamline by building a second-generation detector, which will have the primary physics goal of measuring $\nu_\mu \to \nu_e$ with approximately a factor of 10 more sensitivity than MINOS. To accomplish this we make three major improvements on the MINOS detector design to optimize it for the detection of electron neutrinos:

---

[1] It is also possible that in addition to the three types of neutrinos produced in particle decays and interactions, there could exist additional types of neutrinos that are not produced in these decays and interactions. There is unconfirmed evidence for the existence of this type of neutrino, called a sterile neutrino, from an experiment at Los Alamos National Laboratory. This result is currently being checked by a Fermilab experiment, MiniBooNE. If the existence of sterile neutrinos is confirmed, it will greatly enrich the already rich physics of neutrino oscillations. Searching for evidence of sterile neutrinos will be part of the NOvA physics program.

[2] The other two first-generation experiments are K2K, an experiment in Japan over a 250 km baseline, now completed, and CNGS, an experiment in Europe over a 730 km baseline, that will start in 2006.

(1) We increase the mass of the far detector by a factor of 5.5, from 5.4 kT for MINOS to 30 kT for NOνA. At the same time, we decrease the cost per kT by about a factor of two.

(2) We design a detector that is optimized for the identification of electron-type neutrino events. Specifically, we increase the longitudinal sampling by an order of magnitude from once every 1.5 radiation lengths[3] in MINOS to once every 0.15 radiation lengths in NOνA. Further, 80% of the NOνA detector mass will be active detector, compared to about 5% for MINOS.

(3) We position the detector not directly on the NuMI beam, as MINOS is located, but 12 km off the axis of the beam. This provides more neutrino events in the energy range in which the oscillation takes place, and fewer background events.

Once a signal for electron-type neutrino appearance is seen, NOνA can run an antineutrino NuMI beam to attempt to measure the ordering of the neutrino masses. Whether this will be successful will depend on the parameters that nature has chosen. However, the sensitivity of NOνA can be markedly increased by a four-fold increase in the NuMI beam intensity created by the construction of the Fermilab Proton Driver. In the absence of a Proton Driver, smaller, but still quite significant, increases in NOνA sensitivity can be provided by less expensive investments in the Fermilab accelerator complex, for example, by reducing the Main Injector cycle time to give more protons per year on the NuMI beamline target.

Since there are three unknown parameters to be measured — $\theta_{13}$, the ordering of the mass states, and the parameter that measures CP violation — a third measurement may eventually be required in addition to neutrino and antineutrino measurements in NOνA to determine all three parameters. The third measurement could be done by building an additional detector on the NuMI beamline but further off-axis to measure the second oscillation maximum, or by combining NOνA measurements with those taken elsewhere on different length baselines. Such experiments are being contemplated in Europe and Japan.[4]

We view NOνA as a second step in a step-by-step Fermilab program to measure all of the unknown parameters of neutrino oscillations. Each step will provide guidance on the optimum direction for the succeeding step.

## 1.2. The NOνA Detectors

Like MINOS, NOνA will be a two-detector experiment. A small Near Detector, as identical in structure to the far detector as possible, will be constructed on the Fermilab site. Its function is to predict the expected rate of event types and their energy spectra in the Far Detector in the absence of oscillations. Differences seen between the events in the two detectors can then be attributed to oscillations.

The MINOS detectors are sandwich detectors with alternating layers of iron absorber and active detector made from solid scintillator strips. By contrast, the NOνA detectors will be of a "totally active" design.

The NOνA Far Detector will be composed solely of liquid scintillator encased in 15.7 m long 32-cell titanium dioxide-loaded PVC extrusions. The 3.9-cm wide, 6-cm deep liquid scintillator cells are read out by U-shaped wavelength-shifting fibers into avalanche photodiodes (APDs). This configuration gives better performance at lower cost than that of MINOS. The liquid scintillator is less expensive than solid scintillator and less costly to assemble.

---

[3] A radiation length is the average distance in which an electron loses 63% of its energy.

[4] T2K, a second-generation experiment being built in Japan, will send an off-axis beam from JPARC to the 50 kT SuperKamiokande detector over a 295 km baseline. It plans to begin operation in 2008. A possible future third-generation experiment on this baseline involves increasing the JPARC intensity by a factor of five and building a new detector with 20 times the mass of SuperKamiokande. There is discussion in Europe on building a third-generation experiment using a proposed CERN proton driver called the SPL. It would provide both a conventional neutrino beam and a beam based on the decay of accelerated ions (called a beta beam) over a 130 km baseline to a new, very massive detector to be built in the Frejus tunnel. It should be noted that neither of these proposed third-generation experiments would have a sufficiently long baseline to resolve the ordering of the neutrino mass states without NOνA data.

The APDs provide much higher quantum efficiency than photomultipliers and are cheaper. The high quantum efficiency of the APDs allows longer scintillator cells than those in MINOS.

However, this design is not without challenges. The APDs have low gain requiring low noise electronics. They must also be cooled to –15 C to reduce the dark noise to an acceptable level. Recently, we have verified that we can obtain adequate signals from prototype liquid scintillator-filled extrusions read out into APDs. Our ongoing R&D program will verify the performance of the full liquid scintillator system.

We have selected a site for the NOvA Far Detector near Ash River, Minnesota, about 810 km from the NuMI target. This site is the furthest site from Fermilab along the NuMI beam line in the United States.

Unlike MINOS, the NOνA Far Detector will sit on the earth's surface. Our calculations indicate that backgrounds from cosmic radiation will be acceptably low, largely due to the very short beam pulses from Fermilab, one 10 μs pulse every 1.5 seconds. Part of our R&D program is to verify these calculations with an experimental measurement in a prototype detector.

We have constructed a detailed cost estimate for the full experiment, including a generous contingency for items that have not yet been fully designed. The fully burdened cost in FY2004 dollars is 165 M$, of which 55 M$ is assigned to contingency.

Assuming a project start in October 2006, our technically driven schedule calls for 5 kT of the Far Detector to be constructed by February 2010 and the full detector by July 2011. Since the NuMI beam will be available throughout this entire period and the Far Detector is modular, we will be able to begin taking useful data in February 2010.

To meet this schedule NOνA needs a prompt approval and approximately $ 2–3M in R&D funds before the project start in October 2006. Our R&D request is detailed in the final chapter of this proposal.

# 2. Introduction

In recent years, underground experiments have provided convincing evidence of oscillations of both solar and atmospheric neutrinos. With these measurements, we have an emerging framework with a rich structure in the lepton sector, which we can compare with a structure in the quark sector that has been studied for more than 25 years. An intriguing possibility is that CP violation exists in the lepton sector and that this asymmetry is somehow related to the fundamental matter-antimatter asymmetry of our universe.

The flavor-changing transitions observed in atmospheric and solar neutrinos are most naturally described by a simple extension to the Standard Model, in which three types of neutrinos have masses and mix with each other. The three well-known flavor eigenstates, the electron, muon and tau neutrinos, are related to these mass eigenstates by the (3 × 3) unitary MNS matrix. The model explains the observed flavor-changing transitions as neutrino oscillations, described by mass differences $\Delta m_{ij}^2$ and mixing angles $\theta_{ij}$ (which are parameters of the MNS matrix). The model also provides for CP violation in a natural way through a phase ($\delta$) in the MNS matrix.

While measurements of atmospheric and solar neutrino oscillations have provided some information about the mass differences and two of the three mixing angles, we have (*e.g.*, from the CHOOZ reactor experiment) only an upper limit on the third mixing angle, $\theta_{13}$. Measuring this parameter is key to obtaining a complete picture of the structure of the lepton sector. In particular, a non-zero value for $\theta_{13}$ is a prerequisite to both the ability to probe CP violation in the leptonic sector and to resolve the ordering of neutrino mass states. The latter can only be determined by matter effects, which occur when electron-type neutrinos propagate long distances through the earth. These measurements are the goal of the NuMI Off-Axis $\nu_e$ Appearance experiment (NOvA) described in this proposal.

Chapter 1 of the proposal provides an Executive Summary. Chapter 2 is this introduction. The body of the proposal begins with Chapter 3, which is a concise discussion of the physics motivation. This chapter provides a framework for understanding how the results of this proposed experiment relate to the results of other lepton sector experiments.

An overview of the proposed experiment is provided in Chapter 4. Essentially, we intend to measure electron neutrino appearance in a 30,000 metric ton Far Detector that will be located about 810 km from Fermilab and 12 km off the central axis of the NuMI beam. This off-axis location provides a lower energy, more monoenergetic neutrino beam, which is better suited for this measurement than the on-axis beam.

A Near Detector will measure the electron neutrino content of the beam at Fermilab, characterize the detector response to neutrino events and perform crucial background studies. The NOvA detectors will be optimized to separate charged current electron-neutrino events from neutrino events producing neutral pions. The proposed detectors are "totally active" planar tracking calorimeters with 0.15 radiation length longitudinal segmentation.

Chapter 5 describes the Far Detector structure and its fabrication and assembly. A five-story structure constructed from plastic is not conventional. The detailed engineering studies we have done to assure ourselves of its stability are described in this chapter.

The process of collecting light from the scintillator is discussed in Chapter 6. Chapter 7 follows with a description of the photodetector, the electronics for its readout, and the data acquisition system.

The preferred site for the Far Detector is Ash River, Minnesota, close to the northernmost road in the United States near the NuMI beamline. Chapter 8 describes this site and the design of the building required to house the detector, as well as ES&H considerations.

The Near Detector design and the test beam program are covered in Chapter 9. The purpose of the Near Detector is to measure the process that will be the backgrounds to the signal in the Far Detector. These backgrounds and systematic uncertainties will be the subject of Chapter 10.

The termination of Tevatron Collider operations prior to the start of NOνA running allows higher NuMI beam intensity and repetition rate than had been earlier anticipated. Chapter 11 details how this improvement can be obtained and also discusses the improvement that would be possible with the construction of a Proton Driver to replace the Booster. Although the Proton Driver is not required for the first phase of the experiment, it provides a natural upgrade path for a Fermilab world-leading program in understanding the physics of the lepton sector.

Chapter 12 describes simulations of the NOνA detector performance and their results. Chapter 13 uses these results to assess the physics potential of the NOνA experiment. It outlines a step-by-step program through which NOνA can contribute to the determination of the mass ordering and the measurement of CP violation, if $\theta_{13}$ is in the range accessible to conventional neutrino beams. Chapter 13 also discusses the use of NOνA's high-resolution to make highly precise measurements of $\Delta m^2_{32}$ and $\sin^2(2\theta_{23})$. Finally, it discusses measurements that can be made with the near detector and the detection of galactic supernovae in the far detector.

Chapter 14 presents the cost and schedule of NOνA, and Chapter 15 presents our R&D request. Prompt approval and a FY 2007 construction start will allow NOνA to start data taking in February 2010, with the completion of the full Far Detector by July 2011. Substantial R&D funds will be needed prior to the start of construction and into the first year of construction. A prototype Near Detector will focus our efforts to address many detailed design issues.

In essence, we lay out in this proposal a major step in a program of experiments to study couplings in the lepton sector, with an eventual goal of measuring leptonic CP violation. NOνA is a natural next step after MINOS. Once NOνA determines the $\theta_{13}$-coupling, it will be possible at Fermilab, likely with the Proton Driver, to go on to the next phase of mass hierarchy and leptonic CP measurements.

The recent commissioning of the NuMI beamline represents a very significant step forward for particle physics. At a length of more than 800 km, the NOνA baseline will be nearly three times as long as the baseline in T2K and somewhat longer than the baseline from CERN to Gran Sasso. Thus, with the NuMI beam, Fermilab has a unique capability to answer some of the most important questions that can be asked in elementary particle physics, both today and in the foreseeable future.

# 3. Physics Motivation

## 3.1. Introduction

Recently the SuperKamiokande [1], K2K [2], and Soudan 2 [3] experiments have provided very strong evidence that the muon neutrino undergoes flavor changing transitions. These transitions are seen for neutrinos whose path length divided by energy (*L/E*) is of order ~500 km/GeV. SuperKamiokande also has some supporting evidence that muon neutrinos are transformed primarily into tau neutrinos. Although the SuperKamiokande detector has some sensitivity to flavor transitions of electron neutrinos, their data provides no evidence that electron neutrinos are involved in these transitions. In fact, the CHOOZ [4] reactor experiment provides a tighter constraint on the upper limit on the probability of electron neutrino flavor transitions of order 5-10% at the values of *L/E* for which SuperKamiokande sees muon neutrino flavor transitions. This leaves open the interesting and important question: What is the role of the electron neutrino in flavor transitions at these values of *L/E*? A measurement or stringent limit on the probability of $\nu_\mu \rightarrow \nu_e$ for values of *L/E* of order 500 km/GeV is an important step in understand these neutrino flavor transitions in atmospheric neutrinos. As the NuMI beam is primarily a $\nu_\mu$ beam, the observation of $\nu_e$ appearance would address this question directly. This measurement is the primary goal of the experiment described by this proposal.

The SNO [5] experiment has recently reported large transitions of solar electron neutrinos to muon and/or tau neutrinos both with and without salt added to the heavy water. SuperKamiokande [6], studying solar neutrinos, and KamLAND [7], studying reactor neutrinos, also see large electron neutrino flavor transitions. From a combined analysis, the *L/E* for these flavor transitions is a factor of ~30 times larger than the *L/E* for flavor transitions in atmospheric muon neutrinos. These transitions occur for an *L/E* such that the transition probability $\nu_\mu \rightarrow \nu_e$ measured by an experiment in the NuMI beam will also have some sensitivity to the flavor transitions associated with solar neutrinos through interference effects.

The LSND [8] experiment has reported small muon antineutrino to electron antineutrino transitions for values of *L/E* that are more than two orders of magnitude smaller than the transitions seen in atmospheric neutrinos. However this transition probability is very small, on the order of 0.3% of the one observed for atmospheric and solar neutrinos. If this result is confirmed by the MiniBooNE [9] experiment, this transition could be an important background for a measurement of $\nu_\mu \rightarrow \nu_e$ transitions at the larger values of *L/E* associated with atmospheric neutrinos.

## 3.2. Neutrino Mixing

Extensions to the Standard Model are required to explain the phenomena described here. The simplest and most widely accepted extension is to allow the neutrinos to have masses and mixings such that these phenomena are explained by neutrino oscillations. The masses and mixing of the neutrinos in these extensions would be the low energy remnant of some yet to be determined high energy physics. Thus, neutrino masses and mixing provide a unique window on physics that is inaccessible to current or near future collider experiments. One popular theory is the so called "seesaw" scenario, where the active left handed neutrinos seesaw off their heavier right handed (sterile) partners, leaving three very light Majorana neutrinos. It is already clear that the masses and mixings in the neutrino sector are very different from the masses and mixings in the quark sector and that a detailed understanding of the neutrino masses and mixings will be important in differentiating fermion mass theories. Also, they may provide a key to advancing our theoretical understanding of this fundamental question.

If the neutrinos have masses and mixings then the neutrino mass eigenstates, $\nu_i = (\nu_1, \nu_2, \nu_3, ...)$ with masses $m_i = (m_1, m_2, m_3, ...)$ are related to the flavor eigenstates $\nu_\alpha = (\nu_e, \nu_\mu, \nu_\tau, ...)$ by the equation

$$|\nu_\alpha\rangle = \sum_i U_{\alpha i} |\nu_i\rangle \qquad (1)$$

The charged weak current for the neutrino flavor states is given by $J_\lambda = \bar{\nu}_L \gamma_\lambda \ell_L$, where $\ell = (e, \mu, \tau)$ is the vector of charged lepton eigenstates. In the absence of light sterile neutrinos, the 3 × 3 lepton mixing matrix *U* is unitary. Lepton flavor mixing was first discussed (for the 2 × 2 case) by Maki, Nakagawa, and Sakata.

If we restrict the light neutrino sector to the three known active flavors and set aside the LSND results[1], then the unitary MNS lepton mixing matrix, $U$, can be written as

$$U = \begin{pmatrix} c_{13}c_{12} & c_{13}s_{12} & s_{13}e^{-i\delta} \\ -c_{23}s_{12} - s_{13}s_{23}c_{12}e^{i\delta} & c_{23}c_{12} - s_{13}s_{23}s_{12}e^{i\delta} & c_{13}s_{23} \\ s_{23}s_{12} - s_{13}c_{23}c_{12}e^{i\delta} & -s_{23}c_{12} - s_{13}c_{23}s_{12}e^{i\delta} & c_{13}c_{23} \end{pmatrix} \quad (2),$$

where $c_{jk} \equiv \cos\theta_{jk}$ and $s_{jk} \equiv \sin\theta_{jk}$.

With this labeling, the atmospheric neutrino oscillations are primarily determined by the $\theta_{23}$ and $\Delta m_{32}^2$ parameters, whereas the solar neutrino oscillations depend on $\theta_{12}$ and $\Delta m_{12}^2$, where $\Delta m_{ij}^2 = m_i^2 - m_j^2$. From SuperKamiokande[1] we already have some knowledge of $|\Delta m_{32}^2| = (1.5 - 3.4) \times 10^{-3}$ eV$^2$ and $\sin^2 2\theta_{23} > 0.92$ at the 90% confidence level. A SuperKamiokande analysis which concentrates on events with high resolution in $L/E$ yields $|\Delta m_{32}^2| = (1.9 - 3.0) \times 10^{-3}$ eV$^2$ and $\sin^2 2\theta_{23} > 0.90$ at the 90% confidence limit. The K2K experiment[2] results give $|\Delta m_{32}^2| = (1.9 - 3.6) \times 10^{-3}$ eV$^2$ for $\sin^2(2\theta_{23}) = 1$, at the 90% confidence level. The K2K lower limit on $\sin^2(2\theta_{23})$ is considerably less constraining than those from SuperKamiokande.

Note the substantial uncertainty in these atmospheric measurements. In contrast, the combined analysis of the SNO, SuperKamiokande and KamLAND experiments gives $\Delta m_{21}^2 = +7.9 \pm 0.6 \times 10^{-5}$ eV$^2$ and $\sin^2 2\theta_{12} = 0.82 \pm 0.07$. For the purposes of this experiment our knowledge of the solar parameters is already in good shape and is expected to improve with time.

CHOOZ (and SuperKamiokande) provide us with a limit on $\sin^2 2\theta_{13} < 0.18$. The CHOOZ limit is dependent on the input value used for $|\Delta m_{32}^2|$; for the current central value $2.5 \times 10^{-3}$ eV$^2$, this limit is $\sin^2 2\theta_{13} < 0.14$, while for $|\Delta m_{32}^2| = 2.0 \times 10^{-3}$ eV$^2$, it is $\sin^2 2\theta_{13} < 0.18$ [4]. Thus, the proposed long-baseline neutrino oscillation experiment to search for $\nu_\mu \rightarrow \nu_e$ should be sensitive to a substantial range below this upper bound.

The MINOS experiment [10] will provide a 10% measurement of the atmospheric $|\Delta m_{32}^2|$ but probably will not improve our knowledge of $\theta_{23}$. This experiment has sensitivity to $\sin^2 2\theta_{13}$ only about a factor of two below the CHOOZ bound. Any future reactor experiment to measure $\sin^2 2\theta_{13}$ could improve our knowledge of this important parameter but such an experiment has no sensitivity to $\theta_{23}$, the sign of $\Delta m_{32}^2$ or the CP violating phase $\delta$. Therefore, such a reactor experiment is complementary to long-baseline experiments to observe $\nu_\mu \rightarrow \nu_e$.

The appearance probability of $\nu_e$ in a $\nu_\mu$ beam in vacuum is given, to leading order, by

$$P_{vac}(\nu_\mu \rightarrow \nu_e) = \sin^2\theta_{23} \sin^2 2\theta_{13} \sin^2 \Delta_{atm} \quad (3),$$

where $\Delta_{atm} \approx 1.27\left(\dfrac{\Delta m_{32}^2 L}{E}\right)$, where $\Delta m_{32}^2$ is measured in eV$^2$, $L$ is measured in km, and $E$ is measured in GeV. If the experiment is performed at one of the peaks of this probability, that is, when $\Delta_{atm} = \dfrac{\pi}{2} + n\pi$, and $\sin^2(\theta_{23}) = \dfrac{1}{2}$ then

$$P_{vac}(\nu_\mu \rightarrow \nu_e) = \dfrac{1}{2}\sin^2 2\theta_{13} = 2.5\% \left(\dfrac{\sin^2 2\theta_{13}}{0.05}\right) \quad (4)$$

The first peak occurs at neutrino energy,

$$E = 1.64 \text{ GeV} \left(\dfrac{\Delta m_{32}^2}{2.5 \times 10^{-3} \text{eV}^2}\right)\left(\dfrac{L}{810 \text{ km}}\right) \quad (5)$$

The constraint on $\sin^2(2\theta_{13})$ from the CHOOZ experiment varies from 0.14 to 0.18 depending on the atmospheric $\Delta m_{32}^2$, therefore the maximum appearance probability ranges from ~7 to 9%. To be effective any $\nu_e$ appearance experiment has to aim to exclude or convincingly see a signal at least an order of magnitude below this 7% limit.

### 3.3. Matter Effects

The neutrinos in the NuMI beam propagate through the Earth and matter induced contributions to the propagation amplitude are non-negligible. These matter effects have opposite sign for neutrinos and antineutrinos and for the normal versus inverted neutrino mass hierarchies. The matter effects can be thus used to distinguish the two possible three-neutrino mass hierarchies, see Fig. 3.1. If the experiment is performed at the first peak in the oscillation, as above, the matter effects are primarily a function of the energy of the neutrino beam and the transition probability in matter can be approximated by

$$P_{mat}(\nu_\mu \to \nu_e) \approx \left(1 \pm 2\frac{E}{E_R}\right) P_{vac}(\nu_\mu \to \nu_e) \quad (6),$$

where $E_R$ is the matter resonance energy associated with the atmospheric $\Delta m^2$, that is

$$E_R = \frac{\Delta m^2_{32}}{2\sqrt{2} G_F N_e} = 12 \text{ GeV} \left(\frac{\Delta m^2_{32}}{2.5 \times 10^{-3} \text{ eV}^2}\right)\left(\frac{1.4 \text{ g cm}^{-3}}{Y_e \rho}\right) \quad (7),$$

where $N_e$ is the electron number density in the earth, $\rho$ is the matter density (2.8 g.cm$^{-3}$) and $Y_e$ is the average $Z/A$.

---

[1] In the 3+1 neutrino mass hierarchy the LSND result can be accommodated as a perturbation on the pure active 3 neutrino hierarchy. The 2+2 mass hierarchy would require major modifications.

For the normal hierarchy, matter effects enhance (suppress) the transition probability for neutrinos (antineutrinos) and vice versa for the inverted hierarchy. For a 2 GeV neutrino energy, matter effects give a 30% enhancement or suppression in the transition probability.

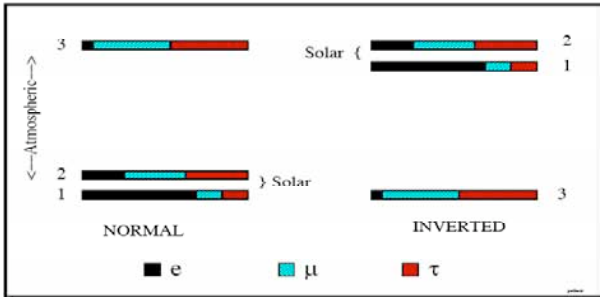

Fig. 3.1: The two allowed three-neutrino mass squared spectra that account for the oscillations of solar and atmospheric neutrinos. The normal spectrum has $\Delta m^2_{32} > 0$ and the inverted has $\Delta m^2_{32} < 0$. The $\nu_e$ fraction of each mass eigenstate is indicated by the black solid region, whereas the $\nu_\mu$ ($\nu_\tau$) fraction is indicated by the blue-green right-leaning (red left-leaning) hatching. The $\nu_e$ fraction in the mass eigenstate labeled, 3, has been enhanced for clarity.

### 3.4. CP Violation

The "Large Mixing Angle" (LMA) solution for solar neutrino oscillations, now the only viable solution, has the property that the $\nu_\mu \to \nu_e$ transition probability is sensitive to sub-leading effects and in particular to the CP violating phase $\delta$.

In vacuum, the shift in the transition probability associated with the CP violating phase is given by

$$\Delta P_\delta(\nu_\mu \to \nu_e) \approx J_r \sin \Delta_{sol} \sin \Delta_{atm} (\cos\delta \cos\Delta_{atm} \mp \sin\delta \sin\Delta_{atm}) \quad (8),$$

where the minus (plus) sign is for neutrinos (antineutrinos),

$$J_r = \sin 2\theta_{12} \sin 2\theta_{23} \sin 2\theta_{13} \cos\theta_{13}$$
$$J_r \approx 0.9 \sin 2\theta_{13} \quad (9)$$

$$\Delta_{sol} = 1.27 \frac{\Delta m^2_{21} L}{E} = \frac{\Delta m^2_{21}}{\Delta m^2_{32}} \Delta_{atm} \approx \frac{1}{36} \Delta_{atm}. \quad (10)$$

At the first oscillation maximum of the atmospheric $\Delta m^2$ scale, the shift in the transition probability dependent on $\delta$ is of order

$$|\Delta P_\delta(\nu_\mu \to \nu_e)| \sim 0.6\% \sqrt{\frac{\sin^2 2\theta_{13}}{0.05}} \quad (11)$$

Note that the shift is proportional to $\sqrt{\sin^2 2\theta_{13}}$, while the leading term is proportional to $\sin^2 2\theta_{13}$. Thus, the relative importance of the sub-leading terms grows as $\sin^2 2\theta_{13}$ gets smaller.

The full transition probability, in vacuum, is given by

$$P(\nu_\mu \to \nu_e) = \left|\sum_{j=1}^3 U^*_{\mu j} U_{ej} e^{-i(m_j^2 L/2E)}\right|^2$$
$$= \left|2 U^*_{\mu 3} U_{e3} e^{-i\Delta_{32}} \sin\Delta_{31} + 2 U^*_{\mu 2} U_{e2} \sin\Delta_{21}\right|^2 \quad (12)$$

The second form of this probability is especially illuminating as the first term is the amplitude for $\nu_\mu \to \nu_e$ associated with the atmospheric $\Delta m^2$ and the second term the amplitude associated with the solar $\Delta m^2$. The interference between these two amplitudes differs for neutrinos and antineutrinos because for antineutrinos the $U$ matrix is replaced with $U^*$. This difference in the interference term leads to the difference in the transition probability $\nu_\mu \to \nu_e$ between neutrino and antineutrinos. Such an effect is an example of CP violation.

Using the MNS mixing matrix given in Eq. 2,
$$2U_{\mu 3}^{*}U_{e3} = e^{-i\delta}\sin 2\theta_{13}\sin\theta_{23}$$
$$2U_{\mu 2}^{*}U_{e2} = \sin 2\theta_{12}\cos\theta_{23}\cos\theta_{13} + O(\sin\theta_{13}) \quad (13)$$

Since the $O(\sin\theta_{13})$ term is multiplied by $\sin(\Delta_{21})$ in the amplitude, it is quadratic in the small quantities $\sin\theta_{13}$ and the solar $\Delta m^2$ and therefore can be neglected.

$$P(\nu_\mu \to \nu_e) = $$
$$| e^{-i(\Delta_{32}+\delta)}\sin 2\theta_{13}\sin\theta_{23}\sin\Delta_{31} \quad (14)$$
$$+ \sin 2\theta_{12}\cos\theta_{23}\cos\theta_{13}\sin\Delta_{21} |^2$$

$$P(\nu_\mu \to \nu_e) = $$
$$\sin^2\theta_{23}\sin^2 2\theta_{13}\sin^2\Delta_{31}$$
$$+ \cos^2\theta_{13}\cos^2\theta_{23}\sin^2 2\theta_{12}\sin^2\Delta_{21} \quad (15)$$
$$+ J_r \sin\Delta_{21}\sin\Delta_{31}$$
$$(\cos\Delta_{32}\cos\delta - \sin\Delta_{32}\sin\delta)$$

The first and second terms are the probability of $\nu_\mu \to \nu_e$ associated with the atmospheric and solar $\Delta m^2$ 's respectively, whereas the third term is the interference between these two corresponding amplitudes. The term proportional to $\sin\delta$ is responsible for CP violation since it changes sign when going from neutrinos to antineutrinos[2].

To show the growing importance of the CP violating term as $\sin^2 2\theta_{13}$ gets smaller we have plotted the neutrino antineutrino asymmetry, $|P_\nu - P_{\bar\nu}|/(P_\nu + P_{\bar\nu})$ versus $\sin^2 2\theta_{13}$ in Fig. 3.2 at the first oscillation maximum assuming maximum CP violation, i.e. $\Delta_{31} = \pi/2$ and $\delta = \pi/2$. The asymmetry grows as $\sin^2 2\theta_{13}$ gets smaller until the amplitude for $\nu_\mu \to \nu_e$ from the atmospheric $\Delta m^2$ is equal in magnitude to the amplitude from the solar $\Delta m^2$. At this value of $\sin^2 2\theta_{13}$ there is maximum destructive (constructive) interference for neutrinos (antineutrinos) and therefore a maximum asymmetry of unity. The value of $\sin^2 2\theta_{13}$ at this peak asymmetry is given by

$$\sin^2 2\theta_{13}|_{peak} \approx \frac{\sin^2 2\theta_{12}}{\tan^2\theta_{23}}\left(\frac{\pi}{2}\frac{\Delta m_{21}^2}{\Delta m_{31}^2}\right)^2 \sim 0.002 \quad (16)$$

Even at the CHOOZ bound for $\sin^2 2\theta_{13}$ the asymmetry is greater than 20%. This asymmetry scales as $\sin\delta$ for values of $\delta$ away from $\pi/2$.

### 3.5. Ambiguity Resolution

The effects of matter can easily be included in our expression for $P(\nu_\mu \to \nu_e)$ by replacing $\sin^n\Delta_{21}\sin^n\Delta_{21}$ and $\sin^n\Delta_{31}$ for all n in all three terms using

$$\sin\Delta_{ij} \to \frac{\Delta_{ij}}{(\Delta_{ij} \mp aL)}\sin(\Delta_{ij} \mp aL) \quad (17)$$

where

$$a = \frac{G_F N_e}{\sqrt{2}} \approx (3700 \text{ km})^{-1}\left(\frac{\rho}{2.8 \text{ g cm}^{-3}}\right) \quad (18)$$

The minus (plus) sign is for neutrinos

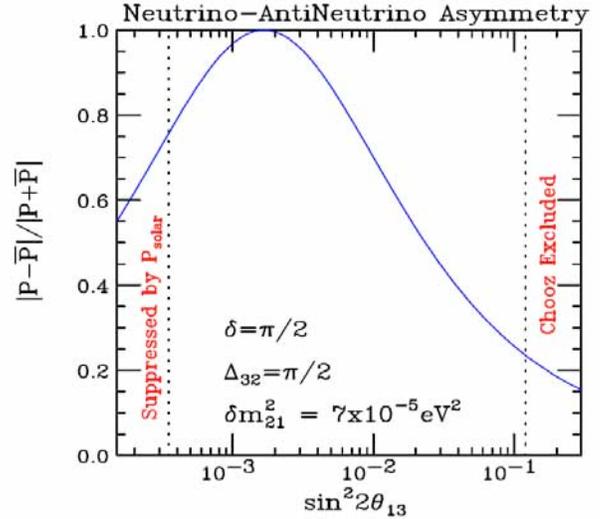

Fig. 3.2: The vacuum asymmetry
$|P(\nu_\mu \to \nu_e) - P(\bar\nu_\mu \to \bar\nu_e)|/|P(\nu_\mu \to \nu_e) + P(\bar\nu_\mu \to \bar\nu_e)|$
versus $\sin^2 2\theta_{13}$ at oscillation maximum, $\Delta_{32}$ assuming that the CP violation is maximal, $\delta = \pi/2$. At the peak of this asymmetry the amplitudes for $\nu_\mu \to \nu_e$ from the atmospheric and solar $\Delta m^2$'s are equal in magnitude. Above (below) the peak the atmospheric (solar) amplitude dominates.

---

[2] The inclusion of the $O(\sin\theta_{13})$ terms in $U_{\mu 2}^{*}U_{e2}$ gives the full expression for $P(\nu_\mu \to \nu_e)$ by multiplying the first term by $(1 - 2\sin^2\theta_{12}\sin\Delta_{12}\cos\Delta_{32}/\sin\Delta_{31})$ and the second term by $|1 - e^{-i\delta}\sin\theta_{13}\tan\theta_{12}\tan\theta_{23}|^2$, while the third term is unchanged. Both of these factors are very close to unity for any reasonable NuMI experimental setup. Equivalent expressions for $P(\nu_\mu \to \nu_e)$ can be found in [11].

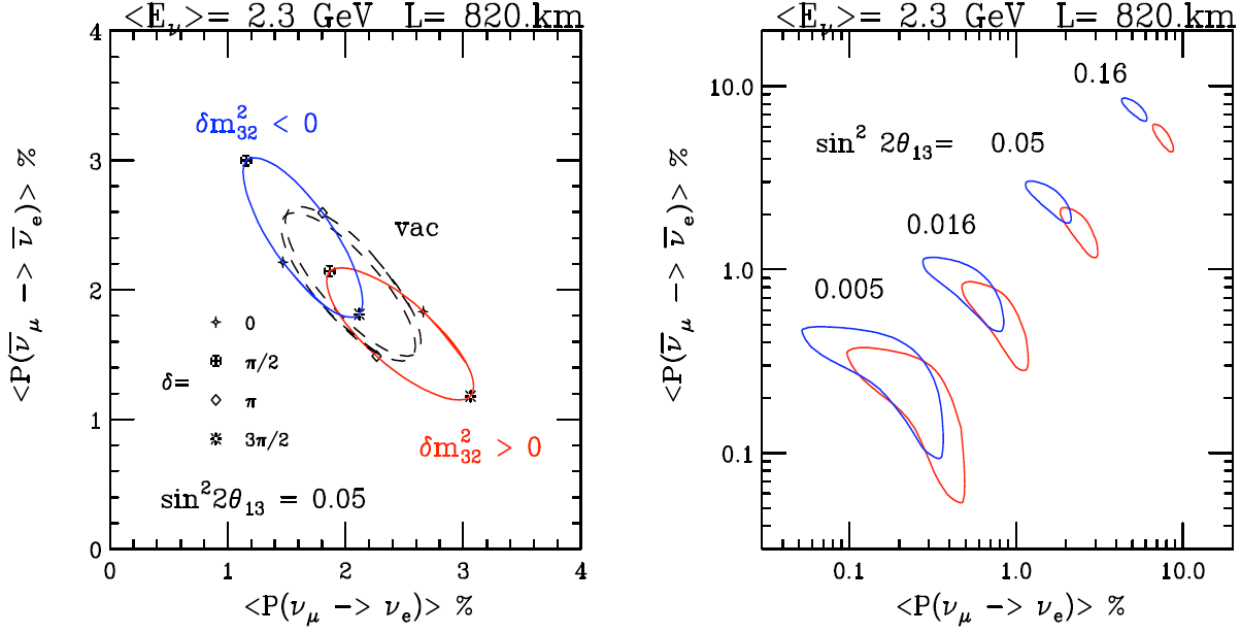

Fig. 3.3: The bi-probability plots $P(\nu_\mu \to \nu_e)$ versus $P(\bar{\nu}_\mu \to \bar{\nu}_e)$, assuming a constant matter density of $\rho = 2.8$ g. cm$^{-3}$ at a distance of 820 km and an average energy of 2.3 GeV with a 20% Gaussian spread. The mixing parameters are fixed to be $|\Delta m_{31}^2| = 2.5 \times 10^{-3}$ eV$^2$, $\sin^2 2\theta_{23} = 1.0$, $\Delta m_{21}^2 = +7 \times 10^{-5}$ eV$^2$, $\sin^2 2\theta_{12} = 0.8$ with the labeled values of $\sin^2 2\theta_{13}$ and $\delta$.

(antineutrinos). The factors $\sin \Delta_{32}$ and $\cos \Delta_{32}$ remain unchanged by matter effects. This algorithm comes from the invariance of the product $\Delta m_{ij}^2 \sin 2\theta_{ij}$ evaluated in matter and in vacuum.

A useful and instructive way to present the combined effects of matter and sub-leading terms is in the bi-probability plots of $P(\nu_\mu \to \nu_e)$ versus $P(\bar{\nu}_\mu \to \bar{\nu}_e)$, invented by Minakata and Nunokawa [13]. Fig. 3.3 shows an example of such a plot for a NuMI case.

At the larger values of $\sin^2 2\theta_{13}$, the ellipses associated with the two possible mass hierarchies separate in matter, whereas they are approximately degenerate in vacuum. There is also a significant sensitivity to the CP violating phase, $\delta$. It is the sensitivity to the sign of $\Delta m_{32}^2$ and the CP violating phase in these plots which allows for the determination of these parameters in a sufficiently accurate experiment. For a single experiment there can be a degeneracy in the determined parameters but this degeneracy can be broken by further experimentation.

In particular the normal and inverted hierarchies may also be able to be distinguished by a comparison of the probability of $\nu_\mu \to \nu_e$ between two different experiments at different baselines, e.g. NuMI and JPARC [12]. If both experiments operated at the first oscillation maximum and both run neutrinos then

$$P_{mat}^N(\nu_\mu \to \nu_e) \cong \left(1 \pm 2\frac{E^N - E^J}{E_R}\right) P_{mat}^J(\nu_\mu \to \nu_e)$$
(19)

where $(P^N, E^N)$ and $(P^J, E^J)$ are the neutrino transition probabilities and energies for NuMI and JPARC respectively. $E_R$ is the matter resonance energy associated with the atmospheric $\Delta m^2$, about 12 GeV, given by Eq. 7. The plus sign is for the normal hierarchy and the minus sign for the inverted hierarchy. For antineutrinos these signs are reversed. If either experiment is significantly away from oscillation maximum, the relationship between the two probabilities is more complicated, see [14].

### 3.6. Other NOνA Measurements

A high precision measurement of $\nu_\mu \to \nu_\mu$ can be used to determine the atmospheric $\Delta m^2$ to the $10^{-4}$ eV$^2$ level. Also $\sin^2 2\theta_{23}$ will be determined from 1 to 2%. Such a measurement can determine how much $\theta_{23}$ differs from maximal mixing, *i.e.*, $\pi/4$. This difference is a measure of the breaking of a $\nu_\mu \to \nu_\tau$ symmetry at some high-energy scale. Since matter effects are suppressed in the channel $\nu_\mu \to \nu_\mu$ compared to $\nu_\mu \to \nu_e$, a comparison of $\nu_\mu \to \nu_\mu$ to $\bar{\nu}_\mu \to \bar{\nu}_\mu$ is a sensitive test of CPT in the neutrino sector.

### 3.7. Neutrino Oscillations in 2010

While we have discussed the current status of neutrino oscillations, NOνA will not likely acquire data for a number of years. Thus, although speculative, it is likely worthwhile to attempt to predict the state of knowledge in 5 to 7 years time. There is considerable ongoing activity with respect to solar neutrino oscillations. Thus, by 2010, it is reasonable to expect that the solar $\Delta m^2$ and $\sin^2 2\theta_{12}$ will be known well enough that they will not be a major source of uncertainty in the interpretation of NOνA results. We also presume that MINOS will have made a 10% measurement of $\Delta m^2_{32}$. The T2K experiment has been delayed to 2008, so it may have only preliminary results by 2010. There has been considerable recent discussion of new reactor-based neutrino oscillation experiments, but in the absence of an approved experiment, it is difficult to predict a time scale for the results of such an experiment.

### 3.8. Summary

The important measurements that could be made by NOνA are
- Observation of $\nu_\mu \to \nu_e$ at an *L/E* in the range of $10^2$ to $10^3$ km/GeV, which would determine the $\nu_e$ role in atmospheric neutrino flavor transitions. In the neutrino oscillation scenario this is a measure of $\sin^2 2\theta_{13}$.
- Matter effects can be used to distinguish the two mass hierarchies and therefore determine the sign of $\Delta m^2_{32}$.
- For the Large Mixing Angle solution to the solar neutrino puzzle there is sensitivity to the CP violating phase in the channel $\nu_\mu \to \nu_e$.
- Precision measurements in the $\nu_\mu \to \nu_\mu$ channel can measure how close $\theta_{23}$ is to $\pi/4$, that is maximal mixing. A comparison of $\nu_\mu \to \nu_\mu$ to $\bar{\nu}_\mu \to \bar{\nu}_\mu$ is a sensitive test of CPT violation since matter effects are suppressed in this channel.

Thus, there is a very rich neutrino physics program to be explored in a $\nu_e$ appearance experiment using the NuMI beam. Details of experimental and beam possibilities will be explored in subsequent chapters.

# 4. Experiment Overview

## 4.1. NuMI Beam

As of this writing, the NuMI neutrino beam [1] is currently being commissioned. The beamline begins with 120 GeV protons extracted from the Main Injector accelerator, which are transported downward at a 158 mrad angle to the NuMI Target Hall. Before striking the production target the beam is bent upward to a 58 mrad downward angle, so that it is aimed at the MINOS far detector in Minnesota. Two parabolic magnetic horns, each about 3 m long and pulsed at 200 kA, focus secondary pions and kaons emitted from the target. The secondary beam subsequently travels with the same downward 58 mrad angle through an evacuated decay pipe, which is 675 m in length and 2 m in diameter. The decay pipe ends in the Hadron Absorber Hall where residual protons and non-decayed secondary mesons are absorbed in the Al-Fe water-cooled beam stop. The muons resulting from pion and kaon decays are absorbed in 240 m of earth shielding, which separates the Absorber Hall from the Near Detector Hall. Three muon alcoves, located within this shielding downstream of the Absorber Hall, contain muon detectors to monitor the beam intensity and shape on a pulse-to-pulse basis. Fig 4.1 shows the plan and elevation views of the NuMI beamline.

A unique feature of the NuMI neutrino beam is the ability to change the focusing optics configuration and hence the neutrino energy band accepted. Specifically, one can change the relative positions of the target and the first horn and the separation between two horns. These configurations are illustrated in Fig. 4.2, together with the spectra for three possible beam element arrangements, referred to as low, medium, or high energy beam tunes. While the movement of the second horn is logistically complex and requires several weeks downtime, the target position can be varied remotely. Accordingly, one also has a method of readily changing the energy spectrum in a continuous fashion by moving just the target at a small sacrifice of the neutrino flux as compared to a fully optimized configuration [2].

Full optimization for a given energy also involves adjusting the target length. The initial beam for the MINOS experiment is the low energy tune, with the front end of the target located 0.34 m upstream of the first horn and a horn separation of 7 m. The target is 0.95 m long and is composed of 47 graphite sections, each 20 mm in length, with 0.3 mm air gaps between sections.

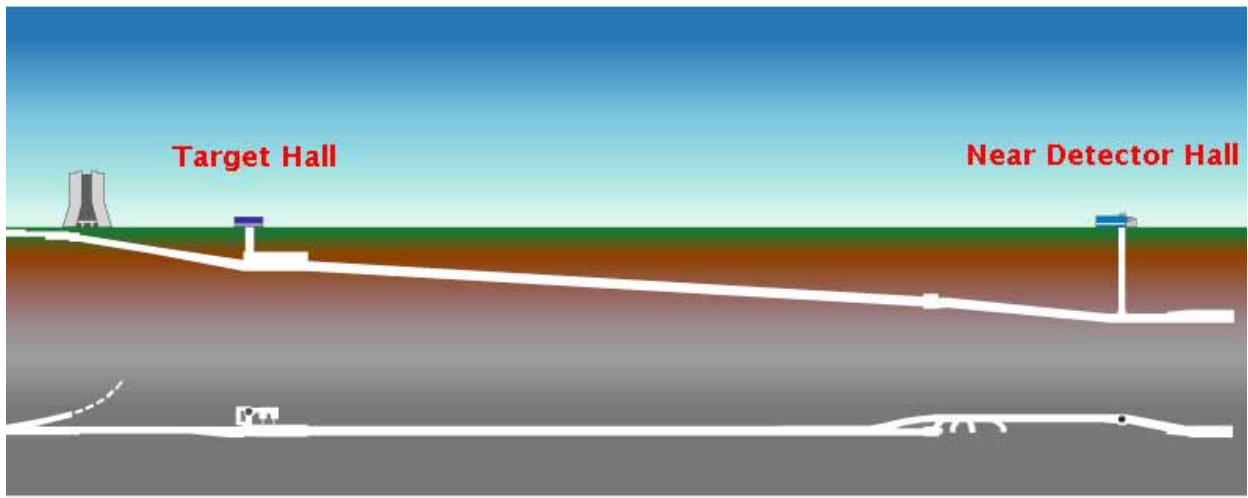

Fig. 4.1: Plan (bottom) and elevation (top) views of the NuMI beam line.

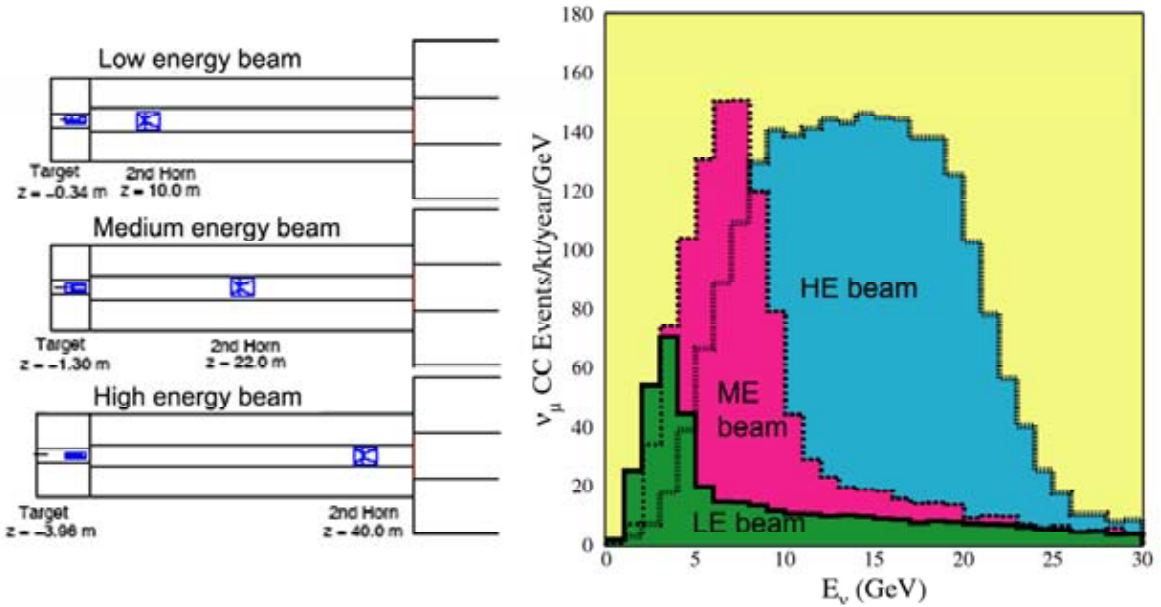

Fig. 4.2: Left: The locations of the target and second horn for the three NuMI beam configurations. Right: The expected neutrino interaction rates at the MINOS far detector site for each of the three beam tunes assuming $2.5 \times 10^{13}$ protons on target per year.

Under the assumption that it would run compatibly with Tevatron Collider, the NuMI beam was designed for a proton intensity of 4 x $10^{13}$ protons per pulse every 1.9 sec, roughly 0.4 MW of beam power. Since Tevatron Collider operations will cease prior to the start of NOvA, more protons will be available and the Recycler can be used to hide the filling time from the Booster. As is explained in Chapter 11, this should allow $6 \times 10^{13}$ protons per pulse every 1.467 sec, or 0.8 MW.

This intensity will stress the present target and beam components. However, Section 11.5 discusses these issues in detail and concludes that with some additional cooling, NuMI will be able to handle this power level.

### 4.2. Off-Axis Concept

Pions and kaons decay isotropically in their centers of mass resulting in a relatively broad neutrino beam energy spectrum. For small angles, the flux and energy of neutrinos produced from the decay $\pi \to \mu + \nu$ in flight and intercepted by a detector of area A and located at distance $z$ are given in the lab frame by:

$$F = \left(\frac{2\gamma}{1+\gamma^2\theta^2}\right)^2 \frac{A}{4\pi z^2} \quad (1)$$

$$E_\nu = \frac{0.43 E_\pi}{1+\gamma^2\theta^2}, \quad (2)$$

where θ is the angle between the pion direction and the neutrino direction, $E_\pi$ the energy of the parent pion, $m_\pi$ the mass of the pion and $\gamma = E_\pi/m_\pi$. The expressions for the neutrinos from the corresponding charged K decays are identical except that 0.43 is replaced by 0.96 resulting in a more energetic and broader distribution for identical meson energies. The neutrino flux peaks in the forward direction for all meson energies, which is the reason that, in general, neutrino detectors are placed on axis. Furthermore, in the forward direction there is a linear relationship between neutrino and meson energies. As the neutrino direction deviates from the meson direction, however, the relationship between the pion energy and neutrino energy flattens. At some angles, a wide energy band of pions contributes to roughly the same energy neutrinos. Fig. 4.3 illustrates both features.

The angle-energy relationship illustrated in Fig. 4.3 can be utilized to construct a nearly mono-energetic neutrino beam by viewing the NuMI beam with a detector at a location off the beam axis. This concept was first proposed for the experiment E-889 at the Brookhaven National Laboratory [3]. Fig. 4.4 shows the implementation of this scheme at locations of 5, 10, and 20 km (corresponding to the angles of 7, 14, and 21 mrad) off the NuMI beam axis at a distance of 800 km from the target.

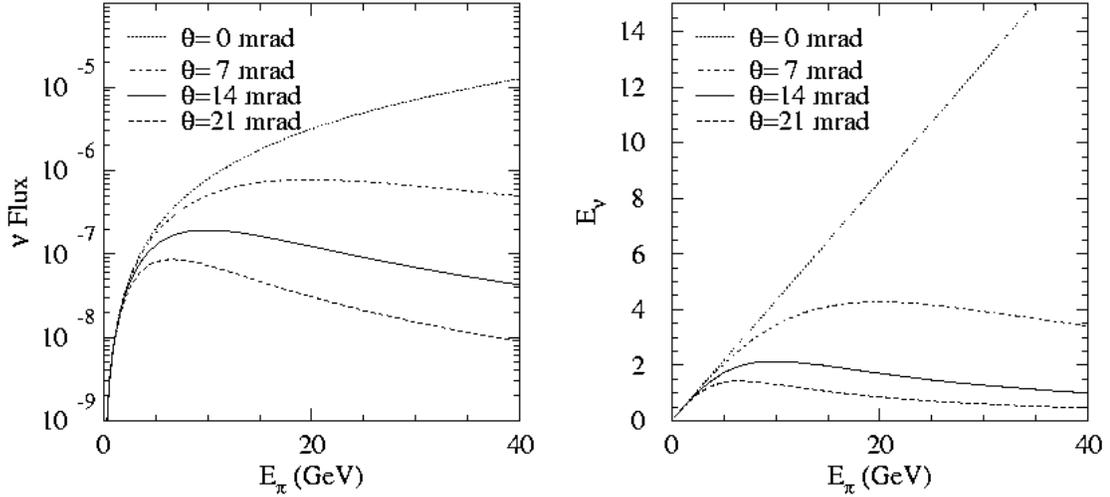

Fig. 4.3: Left: The neutrino flux from a pion of energy $E_\pi$ as viewed from a site located at an angle $\theta$ from the beam axis. The flux has been normalized to a distance of 800 km. Right: The energy of the neutrinos produced at an angle $\theta$ relative to the pion direction as a function of the pion energy.

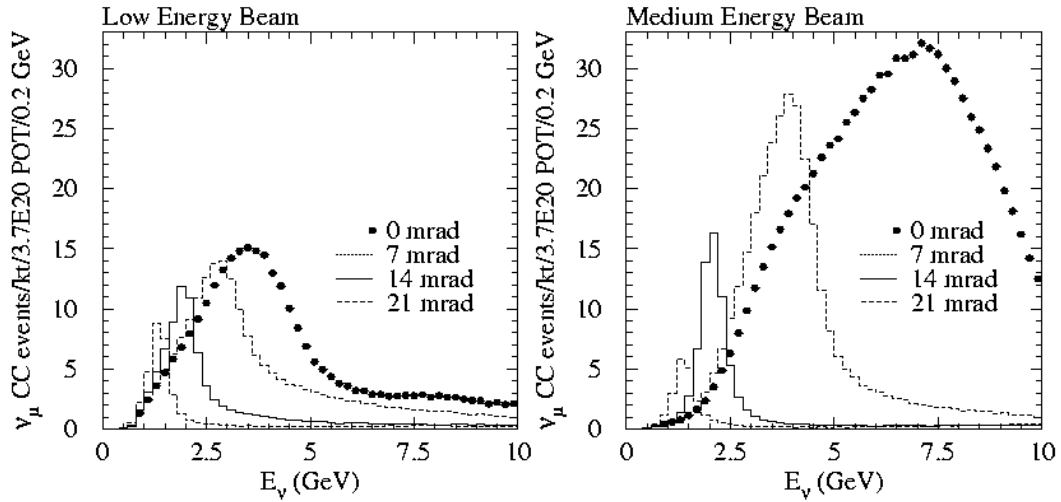

Fig. 4.4: CC $\nu_\mu$ event rates expected under a no-oscillation hypothesis at a distance of 800 km from Fermilab and at various transverse locations for the NuMI low-energy beam configuration (left) and medium-energy beam configuration (right).

The off-axis configuration has several important advantages for a $\nu_\mu \to \nu_e$ oscillation experiment. Among the most important ones are:
• The central energy of the beam can be tuned to the desired energy by selecting an appropriate angle with respect to the beam axis for the location of the detector.

- The spectrum in the peak is quite narrow which helps to reduce the backgrounds, which tend to have much broader energy distributions.
- The high energy tail is considerably reduced with respect to the on-axis beam, which reduces the neutral current and τ backgrounds

These features are quite apparent from Figs. 4.3 and 4.4.

Finally, we would like to make several additional observations about the properties of the off-axis configuration:
- The energy of the beam is determined primarily by the transverse location of the detector. The dependence on the focusing optics is relatively mild.
- The focusing optics configuration affects primarily the intensity of the beam.
- The main peak is composed almost exclusively of the neutrinos from pion decay; K decays give neutrinos at significantly wider angles. Thus, prediction of the spectrum is very insensitive to knowledge of the K/π production ratio.

For the current range of $\Delta m^2_{32}$ values and the three nominal NuMI beam configurations, the medium energy one gives the optimum spectrum for the $\nu_\mu \to \nu_e$ oscillation experiment. Additional fine tuning of the optics as well as the target geometry around the medium energy configuration should yield some additional optimization.

## 4.3. Detector Design Considerations

*4.3.1. General Goals:* The challenge for next generation neutrino experiments is to observe $\nu_\mu \to \nu_e$ oscillations in the atmospheric neutrino mass squared range down to the level of few parts per thousand. The CHOOZ experiment gives a limit on $\nu_e$ disappearance probability in that experiment of about 0.1 – 0.2 [4], the exact limit depending on the value of $\Delta m^2_{32}$. That translates into a limit on $\nu_e$ appearance probability of 0.05 – 0.1. MINOS is expected to improve this by a factor of two to three. There are no clear reliable theoretical guidelines as to the most likely value of this parameter.

Charged current $\nu_e$ interactions can be identified by the presence of an electron in the final state. The experimental backgrounds to the $\nu_\mu \to \nu_e$ oscillation signals arise from two general sources. There are genuine events with electrons resulting from the intrinsic $\nu_e$ component in the beam and from τ decays produced in the charged current $\nu_\tau$ interactions from $\nu_\mu \to \nu_\tau$ oscillations. The latter background is very small for NOνA since most of the $\nu_\mu$ flux is below τ production threshold. In addition there are potentially misidentified neutral-current events or high $y$ $\nu_\mu$ charged-current events where one or more $\pi^0$'s in the final state masquerades as an electron or, less likely, that a hadron is misidentified as an electron.

The intrinsic $\nu_e$'s in the beam come from μ decays and $K_{e3}$ decays (charged and neutral). They are of the order of 0.5-1.0% of $\nu_\mu$'s, but can be reduced further by an appropriate energy cut. $K_{e3}$ contamination is typically of the order of 20% of the μ decay background in NOνA.

The experimental challenge has two parts:
- reducing these two backgrounds as much as possible (discussed below)
- measuring these backgrounds well enough that the principal ultimate uncertainty comes from the statistical fluctuations in the event sample of interest (discussed in Chapter 10).

*4.3.2. Design Optimization Issues:* Background and signal $\nu_e$ events are identical except for their energy spectrum. The background events have a broader energy spectrum than that of the potential signal events, whose width is determined by the spectrum of $\nu_\mu$'s convoluted with the oscillation probability (see Fig. 4.5). Thus, the background from $\nu_e$'s can only be reduced by good energy resolution.

The $\nu_\mu$ neutral and charged-current backgrounds can be reduced by a well-designed detector. The challenge is to suppress them to levels comparable or lower than the intrinsic $\nu_e$ background level with minimum impact on the signal detection efficiency.

The need to separate out the electromagnetic component in a hadronic jet from the remaining hadrons is common to many high-energy experiments. In the calorimetric method, this is generally achieved by having a high Z electromagnetic calorimeter in front of the hadron section. Clearly that technique is not suitable for electron/$\pi^0$ separation. The latter has been traditionally done in open geometry experiments by using a Cherenkov counter. In the recent neutrino experiments: IMB, Kamiokande and SuperKamiokande, this general method was implemented by water Cherenkov detectors. The other technology of choice in those experiments (*e.g.* CHARM II and the BNL oscillation experiment) has been use of low Z calorimeters, which facilitate identification of the electron by tracking.

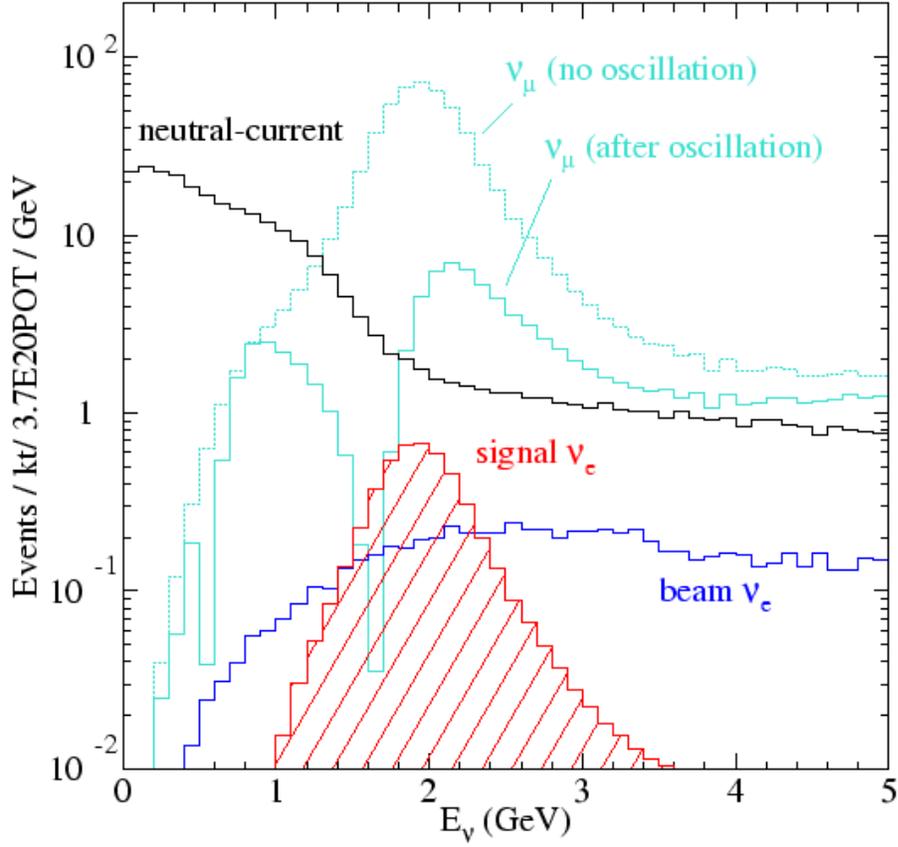

Fig. 4.5: Simulated energy distributions for the $\nu_e$ oscillation signal, intrinsic beam $\nu_e$ events, neutral-current events and $\nu_\mu$ charged-current events with and without oscillations. The simulation used $\Delta m_{32}^2 = 2.5 \times 10^{-3}$ eV$^2$, $\sin^2(2\theta_{23}) = 1.0$, and $\sin^2(2\theta_{13}) = 0.04$. An off-axis distance of 12 km at 810 km was assumed.

*4.3.3. Tracking calorimeter design issues:* In principle, at least, a highly segmented detector can separate electrons from $\pi^0$'s by utilizing several experimental characteristics:
• finite separation between the vertex and conversion points of the $\gamma$'s from the $\pi^0$,
• two electromagnetic showers (for $\pi^0$) *vs* one (for electrons),
• double pulse height right after a $\gamma$ conversion.

Success of the separation based on these characteristics requires fine segmentation: longitudinally, a fraction of a radiation length, and transversely, finer than the typical spatial separation of the two gammas from the $\pi^0$ decay. The transverse segmentation also has to be such that individual tracks in the final state can be separated from each other.

Besides the need to distinguish electrons from $\pi^0$'s, one must also distinguish electrons from hadrons and muons. This is harder in a low Z material and relies on absence of hadronic interactions for electrons and a generally broader pattern of hits along the track for electrons due to the electron shower.

The other important characteristic of a good $\nu_e$ detector for $\nu_\mu \to \nu_e$ oscillations is its energy resolution. One can reduce the intrinsic beam $\nu_e$ background utilizing the fact that the events from $\nu_\mu \to \nu_e$ oscillations will have a sharp energy spectrum at a predictable energy in contrast to the backgrounds that will exhibit a much broader spectrum. This is an important feature of an off-axis experiment, where the detector sees neutrinos in a narrow energy band. Electron-type neutrinos from μ decays will be in roughly the same energy range as the oscillated $\nu_e$'s but have a much broader distribution. $K_{e3}$ decays will give higher energy neutrinos covering a broad energy range whereas the τ decay electrons will peak towards low energies. The shape and the level of backgrounds as well of a possible signal are shown in Fig. 4.5.

## 4.4. Evolution of the Detector Design

The first NOνA proposal [5], submitted in March 2004, called for a 50 kT sandwich far detector. The detector in that proposal had alternating planes of absorber and active elements. The absorber consisted of eight inches of either particleboard or oriented strand board. In the baseline design, the active element was a plane of 30-cell PVC extrusions containing liquid scintillator. The cell size was 3.96 cm wide, 2.56 cm deep, and 14.6 m long. An appendix to the proposal described an alternative design in which the active elements were resistive plate chambers.

At around the time that the proposal was submitted, Stan Wojcicki suggested an alternative design in which the passive absorber was removed and the liquid scintillator cell dimensions were reoptimized [6]. The rationale for this suggestion was that in this "totally active" design, the higher resolution of the detector would lead to a higher efficiency for detection of $\nu_\mu \to \nu_e$ oscillation events along with a greater rejection of background events. This, in turn, would allow a detector with a smaller total mass to have as good or better performance for the same cost.

The reoptimization of the liquid scintillator cells consisted of making the cells deeper. The deeper cells produced more light per track, allowing longer cells with corresponding longer attenuation factors. However, the cell length was limited to 15.7 m, since 53 feet is the longest length that can be transported in the United States without substantial extra cost.

Preliminary investigations of this design showed that it was promising and Appendix B [5] to the proposal was submitted to the PAC for its June 2004 meeting.[7] At that meeting the PAC was told that more detailed simulations and engineering studies were needed before the collaboration could decide on whether to substitute the totally active design for the baseline design.

The decision to propose the totally active design was made at the January 2005 NOνA collaboration meeting. The reasons for this decision were

(1) The sensitivity for measuring $\nu_\mu \to \nu_e$ oscillations was the same for a 30 kT totally active detector and a 50 kt detector sandwich detector and the costs were comparable.
(2) It was anticipated that the ability to see almost all of the energy deposition, the finer longitudinal segmentation, the higher resolution, and the increased signal to background ratio[8] would yield a number of advantages, including
    (a) eventual improvement in the figure of merit as more sophisticated analyses use more aspects of the finer segmentation and higher resolution,
    (b) better understanding of the backgrounds and confidence in their subtraction,
    (c) increased precision in measuring $\sin^2(2\theta_{23})$, measuring neutrino cross sections in the Near Detector, and galactic supernovae in the Far Detector (see Sections 13.6, 13.8, and 13.9, respectively), and
    (d) reduction in backgrounds due to cosmic rays to a negligible level (see Section 10.7).
(3) Faster detector assembly time at the Far Site.

---

[7] The detector described in Appendix B was a 25 kT detector with 17.5 m long and 4.5 cm wide cells. The shipping limitation was ignored at the time of writing Appendix B. In this proposal, the mass has been increased to 30 kT and the cells deepened to 6 cm as a result of the simulations described in Chapter 12.

[8] For $\sin^2(2\theta_{13}) = 0.1$, the typical signal to background increases from 4.8 to 7.3 in going from the sandwich design to the totally active design.

## 4.5. Far Detector

The NOνA Far Detector will be located in a new surface laboratory approximately 810 km from Fermilab and displaced approximately 12 km from the central axis of the NuMI beam. The detector will be a low density, low Z, 30,000 metric ton, tracking calorimeter, comprised of approximately 24,000 metric tons of mineral-oil based liquid scintillator as an active detector and 6,000 tons of rigid polyvinyl chloride (PVC) extrusions, loaded with 15% titanium dioxide, to contain the liquid scintillator.

The liquid-scintillator filled extrusions will be arranged in 1984 planes, oriented normal to the axis pointing towards Fermilab. Each plane will be 15.7 m wide by 15.7 m high by 6.6 cm thick. The planes alternate horizontal and vertical alignments. Thirty-two planes are glued together into a block with a one-cm gap between blocks for structural reasons. The total length of the detector is 132 m.

The liquid will be contained in the PVC extrusions, which will be 1.3 m by 6.6 cm by 15.7m long. Each extrusion will be divided into 32 cells, each cell having an inner cross-section of 3.87 cm by 6.00 cm, with a total length of 15.7 m. The scintillation light in each cell will be collected by a looped 0.8 mm diameter wavelength-shifting plastic fiber. Light from both ends of the fiber will be directed to a single pixel on an avalanche photodiode (APD).

APDs are low cost photodetectors providing high quantum efficiency. Their main difficulties are low amplification and electronic noise. High gain preamplifiers, such as those developed for the LHC CMS detector, can provide the necessary signal output levels. Noise will be reduced to a feasible level by use of Peltier-effect coolers to reduce the operating temperature of the APDs to -15 C.

## 4.6. Near Detector

The purpose of the NOvA Near Detector is to increase the sensitivity of our search for $\nu_\mu \rightarrow \nu_e$ appearance by improving our knowledge of backgrounds, detector response and the off-axis neutrino beam energy spectrum. The Near Detector would be located about 12 m off the NuMI beam axis, in the access tunnel upstream of the MINOS Near Detector Hall. This site provides a neutrino-beam energy spectrum that is quite similar to that at the far-detector.

Although a primary design requirement is that the near detector be as similar as possible to the far detector, of necessity it will have smaller transverse and longitudinal dimensions. As described in Chapter 9, the active area of the detector will be 3.25 m wide and 4.57 m high. The first eight meters of the detector will be composed of the exact same extrusion cells as in the Far Detector. It will be logically be divided into three sections: the first 0.53 m will be a veto region, the second 2.64 m will be the target region, and the final 4.75 m will be a shower containment region. The fiducial volume of the target region will be the central 2.5 m horizontally and 3.25 m vertically. The final section of the detector will be a muon catcher with 10 10-cm plates of iron interspersed with additional planes of liquid scintillator cells. The total mass of the Near Detector will be 262 tons, of which 145 tons are in the totally active region. The fiducial region will have a mass of 20.4 tons.

As discussed in Chapter 10, the Near Detector will be modular, so that it can be placed in the MINOS Surface Building prior to NOvA running, where it will be illuminated by the 75-mrad off-axis NuMI beam. There it will see a $\nu_\mu$ beam peaked at 2.8 GeV and a $\nu_e$ beam peaked at 1.8 GeV, both from kaon decay. Running the Near Detector on the surface will also allow us to verify our calculation of the level of cosmic ray backgrounds.

## Chapter 4 References

# 5. The NOνA Far Detector

## 5.1. Overview

The NOνA Far Detector is optimized for detecting low-energy (~2 GeV) electron showers while rejecting background events. High signal efficiency and good background rejection require frequent sampling in low-Z materials. The NOνA detector has 80% active material and fine segmentation, providing good discrimination between signal and background.

The detector is a 30 kton tracking calorimeter, 15.7 m by 15.7 m by 132 m long, with alternating horizontal and vertical rectangular cells of liquid scintillator contained in rigid polyvinyl chloride (RPVC) extrusion modules. One plane of the detector is constructed from 12 extrusion modules as shown in Figure 5.1. Each extrusion module contains 32 cells and is sealed with a closure block on one end and a readout manifold on the other end. Individual cells in each extrusion have an interior cross section of 3.87 cm by 6.0 cm along the beam direction as shown in Figure 5.2. Each cell is 15.7 m long.

Extrusion modules will be made in three factories, operated by NOνA collaborators, and trucked to the Far Detector site. The factories assemble a rigid PVC module complete with a looped 0.8 mm diameter wavelength shifting (WLS) fiber in each cell to collect scintillation light and route it to a single photodetector pixel. The details of the light collection by the WLS fibers are discussed in Chapter 6.

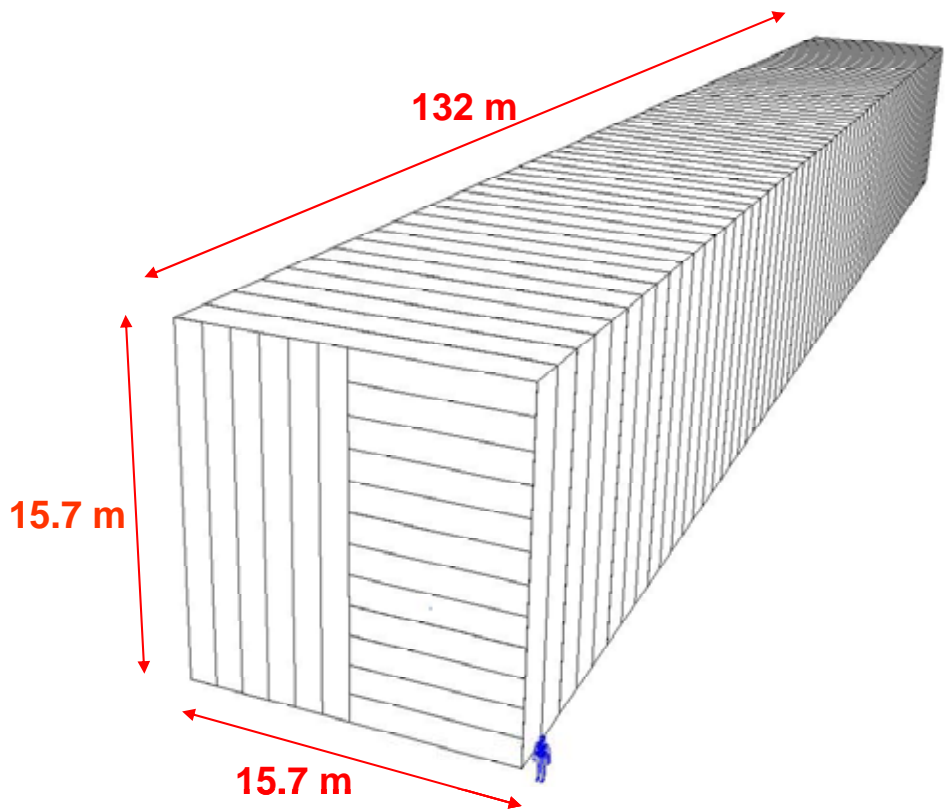

Fig. 5.1: Overview of the NOνA detector structure, showing the 132 meter length of the detector separated into blocks of 32 planes. The cut-away view of the front plane shows the alternating layers of horizontal and vertical extrusion modules.

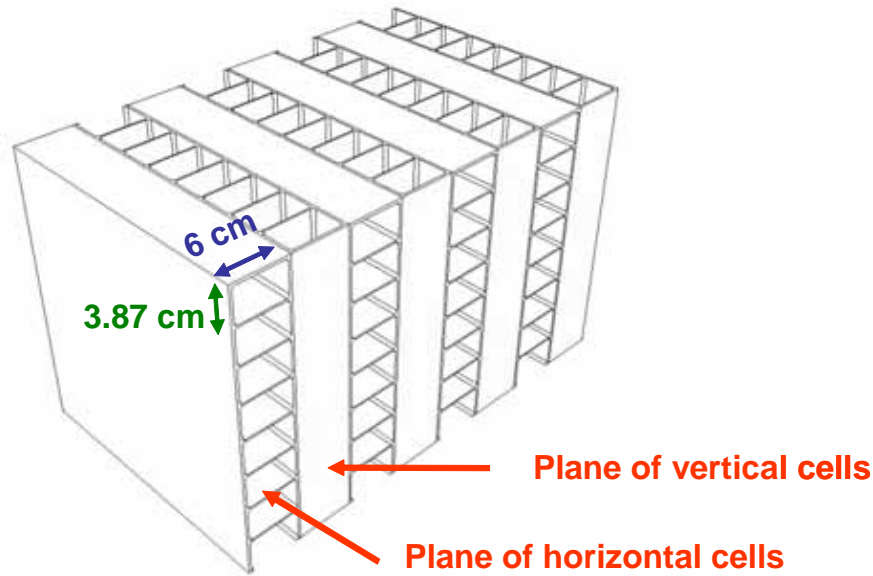

Fig. 5.2: Close-up of the detector structure, showing the cell structure of alternating horizontal and vertical extrusion modules.

The empty extrusion modules are assembled into planes and the planes are assembled into larger blocks at the far site. The photodetectors and readout electronics are mounted on each extrusion module, on the top and on one side of the detector. Details of the readout electronics are discussed in Chapter 7.

The extrusion module assembly process takes 44 months. The Far Detector installation can be completed in 26 months. Chapter 14 discusses how these schedules are interleaved, while the details of the assembly times are discussed here. The Far Detector gets filled with liquid scintillator as the last step of the assembly process, with liquid filling following the plane construction by about one month.

The Far Detector parameters are summarized in Table 5.1.

| | |
|---|---|
| Total mass | 30,090 tons |
| Mass of RPVC extrusions | 5,970 tons |
| Mass of liquid scintillator | 23,885 tons |
| Liquid scintillator | Bicron BC517L (or equivalent) |
| Active mass fraction | 80% |
| Active height × width | 15.7 m × 15.7 m |
| Active length | 132 m |
| Number of layers | 1984 |
| Radiation length per layer | 0.15 |
| Mass of epoxy between layers | 222 tons |
| Extrusions per layer | 12 |
| Extrusion outer wall thickness | 3 mm |
| Extrusion inner web thickness | 2 mm |
| Extrusion width | 1.3 m |
| Extrusion length | 15.7 m |
| Maximum pressure in vertical cells | 19.2 psi |
| Cells per extrusion | 32 |
| Cell width × depth | 3.87 cm × 6.00 cm |
| Total number of cells | 761,856 |
| Total number of extrusions | 23,808 |
| Wavelength-shifting fiber | Kuraray, Y-11 fluor, S-type (or equivalent) |
| WLS fiber diameter | 0.8 mm |
| Total WLS fiber length | 25,629 km |
| Total WLS fiber mass | 13.5 tons |

Table 5.1: Summary of Far Detector parameters.

## 5.2. Extrusion Module Fabrication

A 32-cell extrusion forms the body of each module, which has a WLS fiber manifold and photodetector assembly at one end and is sealed with a closure plate on the other end. The closure plate is a grooved RPVC block that is glued across the extrusion end, as shown in Figure 5.3. Before installing the closure plate a series of small circular holes are made in the interior webs so that the extrusion consists of a single liquid volume.

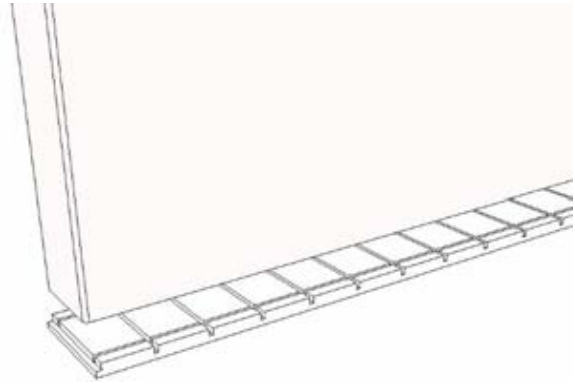

Fig. 5.3: Closure plate (bottom piece) for the "closed" end of an RPVC extrusion module.

Identical WLS manifolds, shown in Figures 5.4 and 5.5, are used on both the vertical and horizontal modules. The vertical extrusion manifolds have room for thermal expansion of the liquid scintillator. The horizontal extrusions have external overflow canisters for that purpose. Clips are used to position the fibers at the top of each extrusion and routing grooves align the fibers on the connector, control the fiber bend radii and facilitate assembly. The manifolds provide filling and venting ports, seal the extrusions, and guide the fibers to the photodetector connector [1].

Extrusions for modules arrive at three assembly factories, cut to length, from the commercial extruder. The factories perform the following tasks:
1. Inspect the incoming extrusions,
2. Install the looped WLS fiber in all cells,
3. Install end closures and manifolds,
4. Pot fibers in connectors and fly-cut faces,
5. Check fiber loops for continuity,
6. Leak test modules,
7. Pack modules and ship to detector site.

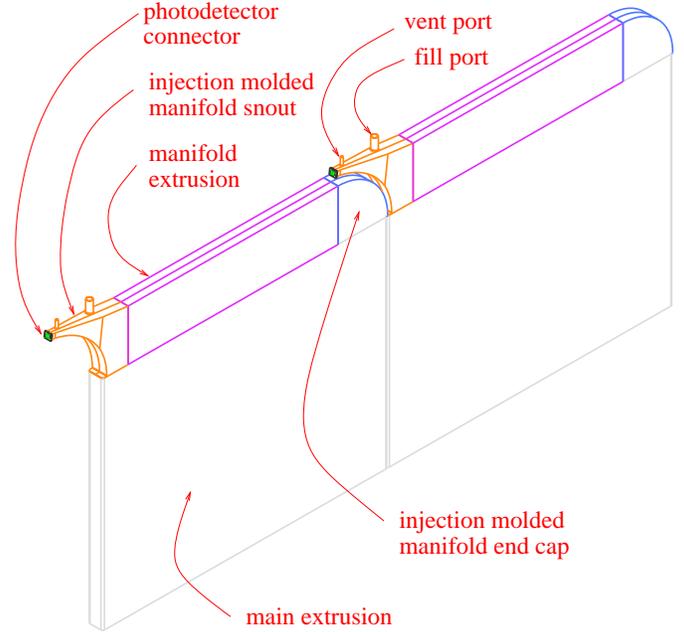

Fig. 5.4: WLS fiber manifolds mounted on adjacent RPVC extrusion modules.

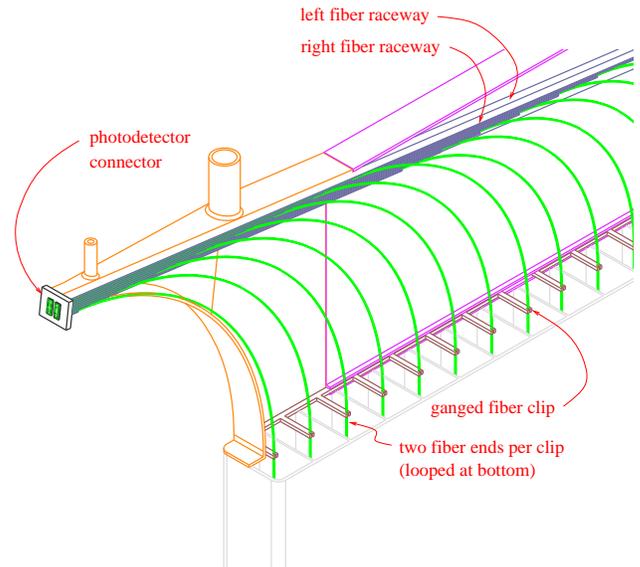

Fig. 5.5: WLS fiber routing within a manifold.

To complete the 23,808 modules during 4 years with 200 working days per year of construction, each of the three factories will complete 10 modules per day with a crew of 4. This schedule is derived from a time and motion study [2] based on the assembly of MINOS modules.

## 5.3. Detector installation at the Far Site

Detector installation starts with a steel frame bookend. Eight planes of 12 (empty) modules each are glued together on a horizontal assembly tables to form a sub-block. Each 26-ton sub-block is raised and glued to the previously installed sub-block. Four sub-blocks are glued together to form a 32-plane block. Small (~1 cm) expansion gaps are left between the 32-plane blocks to accommodate expansion of the RPVC when the modules are filled with liquid scintillator. PVC spacers (1 cm thick by 30 cm wide by 15.7 m long) are glued in place to maintain the gaps. Figure 5.6 shows the geometry of blocks and sub-blocks. Figure 5.7 shows the assembly tables and the "Block Raiser" lifting fixture.

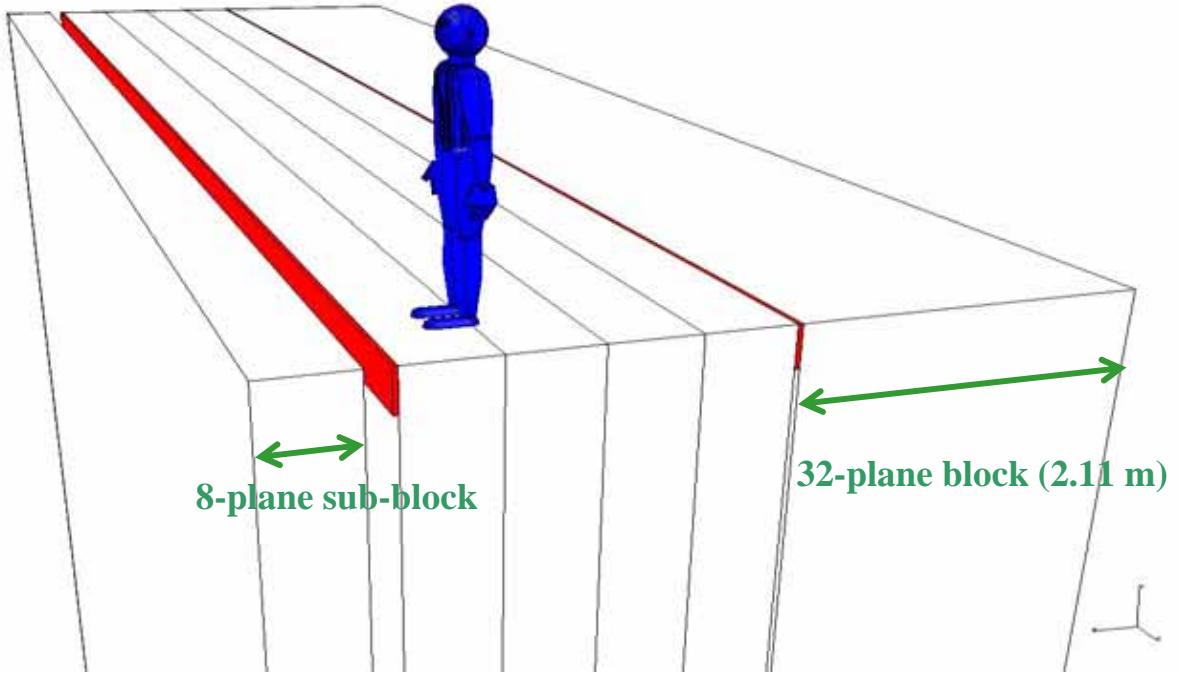

Fig. 5.6: Geometry of the NOvA detector structure. The 32-plane-blocks are made up of four 8-plane sub-blocks and are separated by 1 cm by 30 cm PVC spacers (shown in red) to create expansion gaps.

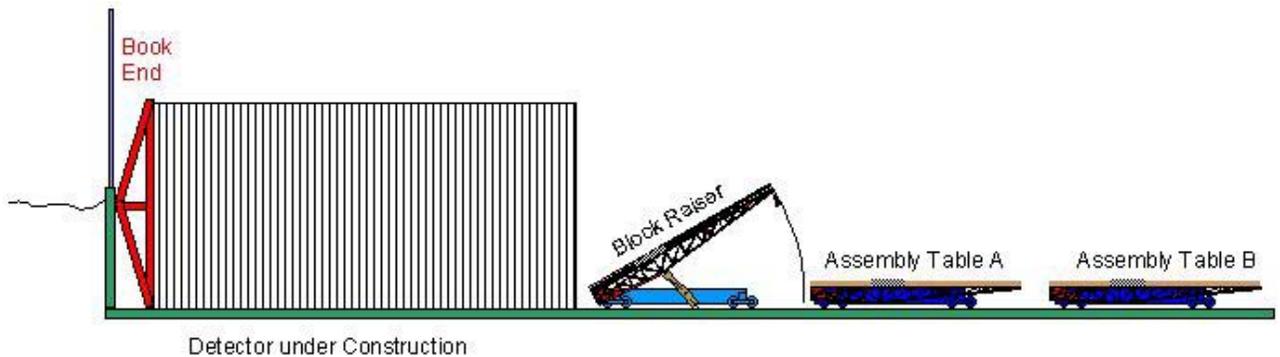

Fig. 5.7: Side elevation view of the detector during assembly, showing the bookend, the Block Raiser and two assembly tables. The Block Raiser and the two assembly tables operate on rails installed on the floor of the detector building.

In Figure 5.7 both assembly tables and the Block Raiser ride on rails installed in the floor along the length of the detector. The assembly tables are stationed near the receiving dock, while the Block Raiser shuttles completed blocks to the end of the detector, raises them up, aligns them, and holds them while the glue sets. After a startup period, crews of three technicians will be able to install one 8-plane sub-block in two shifts.

To illustrate the details of the assembly procedure, let us start with the first four planes of an 8-plane sub-block already completed on Table A (nearest to the detector) and an empty Table B (nearest to the receiving dock). Two operations now occur in parallel: The crew brings the next extrusion module to Table A using the crane and a vacuum lifting fixture, and a semi-automatic machine dispenses glue dots over the area where this module will be placed. The glue dispensing takes 4 minutes and moving the module into place takes 8 minutes. Stops on the table are used to align the module. The lifting fixture (one of 13) will be left on top of the module as a weight until the glue is sufficiently cured. Built-in arms push the module against its neighbor to straighten out the extrusion if necessary. This operation is repeated for the remaining 11 modules. To complete a plane takes a little under 2 hours.

A crew of outfitting technicians works on the newly completed plane on Table A while the glue cures. They install a signal concentrator box on the plane, install an electronics box on each module and plug in the power and readout cables. A test fixture is used to check all modules and cables for correct performance. The crew also installs the liquid scintillator fill panels and routes the liquid fill lines and air return lines to them (This is discussed in more detail in section 5.4).

Meanwhile, the assembly crew begins installing modules on Table B. The first layer gets set down without glue. The second layer is installed and glued in place as described above. Vacuum fixtures are retrieved from Table A as needed. Each fixture will have served as a glue-curing weight for about 2 hours before it is removed.

After completion of the glued layer on Table B, the crew resumes work on Table A, and so on. Eight layers will be stacked in a two-shift, 16-hour day.

By the next morning the epoxy has cured to sufficient strength for safe handling. A separate block installation crew comes in 4 hours before the day shift starts and retrieves the Block Raiser from the end of the detector where it has been holding the last completed block while the glue cured. The crew parks the block raiser next to Table A and transfers the 8-layer block from Table A onto the Block Raiser table. This is done on an air cushion, by pressurizing a plastic pipe system in the table to 0.1 psi to "float" the block. A set of wheels, inset into both tables, pushes and controls the block during the transfer.

Next the 4-plane half-block from the previous day's work is transferred from Table B to Table A and the block assembly cycle starts over when the day crew arrives.

The block installation crew uses a laser system to check alignment of the block that was installed the day before. Based on the results, they select spacers of appropriate thickness to keep the next block exactly vertical. Then they drive the Block Raiser with the completed block on its rails to the end of the detector. They glue the selected spacers to the block using a fast setting epoxy. They mix and spread grout on the floor and dispense epoxy on the spacer board. Now they are ready to raise the next block and align it, using the laser system and the control cylinders embedded in the Block Raiser. The Block Raiser holds the block in position overnight, pressed firmly against the existing detector stack, until the epoxy has cured.

**5.4. Filling the Detector with Liquid Scintillator**

The detector holds 23,885 tons (about 7.5 million gallons) of liquid scintillator. To match the overall assembly time at the far site, the detector will be filled in 20 months (333 8-hour shifts), requiring a fill rate of 46 gallons (174 liters) per minute. Time must be allowed for the liquid level to equalize between module cells through the 0.5-inch diameter holes in the internal webs. This requires the fill rate to be 3 liters/minute or less for a single module, so 48 modules must be filled simultaneously.

We will use an automated filling machine to fill 12 modules at once, metering the liquid mass output and fill rate in each module. The system will shut off the flow when the desired liquid level is reached or if any unusual situation occurs. To simplify the procedure, the fill and air return lines from each module will be routed to common fill panels located on the 8-plane sub-blocks near the building catwalks (see Chapter 8). The filling machine receives liquid scintillator from a pipeline installed along each catwalk.

Each filling machines takes 6.5 hours to fill 12 modules and 5 machines will be required to fill the entire detector in 333 shifts. One machine needs to be moved every 75 minutes, so one worker can handle the whole filling job.

Vendors will deliver pre-mixed liquid scintillator to the detector site in standard tanker trucks. We will use 88,000 gallons of scintillator mix a week, requiring 3.3 trucks per workday. In-line quality assurance will be used at both the mixing plant and the receiving site. Some intermediate storage, possibly in the form of leased tanker trailers, will be used as buffers while we verify product quality before injecting the scintillator into the distribution system. These also provide a steady supply of liquid for distribution.

## 5.5. Structural Considerations

The following sections describe our analyses of the composite detector structure. The structure is designed to be mechanically stable for the lifetime of the experiment and allows the completed planes of the detector to be filled with liquid scintillator and operated while the remaining planes are being installed. The design process includes testing sample portions of the structure to validate the engineering calculations.

*5.5.1. Rigid Polyvinyl Chloride (RPVC):* RPVC is an inexpensive, high-strength, readily available material. It has a high glass transition temperature (making it less prone to creep) and industrial extruders find it easy to work with. NOvA will use 5970 tons of RPVC, which represents less than one day of U.S. production capacity.

The ASTM D1784 "Standard Specification for Rigid PVC Compounds" defines six grades, with allowable design stresses ranging from 1000 psi to 2000 psi [3]. The Plastic Pipe Institute defines the design stress as the hoop stress in a pipe that, when applied continuously, will cause failure of the pipe at 100,000 hours (11.43 years) due to long-term creep. We have chosen a grade of RPVC with a design stress of 2000 psi for NOvA. Figure 5.8 shows a yield stress measurement of approximately 6000 psi in the kind of RPVC that we expect to use.

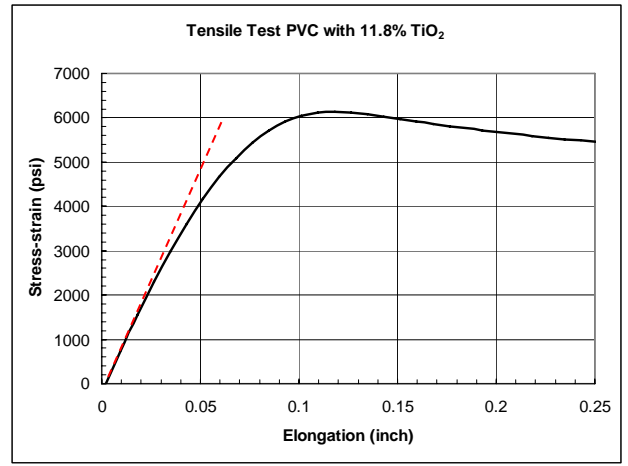

Fig. 5.8: Measured tensile test of RPVC with 11.8% $TiO_2$.

In RPVC, the elastic range extends to approximately 2000 psi, beyond which the material creeps [4]. RPVC is a tough material with low brittleness that limits crack propagation. For this reason, the choice of an allowable stress depends on the stress pattern. For NOvA, our Finite Element Analyses (FEAs) show that the high stress areas are limited to small, isolated regions [5]. If creep occurs, it will re-distribute the forces over a larger area, relaxing the stresses without a failure occurring. We use a design stress of 1500 psi, without any additional safety factors. Preliminary results from our long-term creep tests [4] indicate the choice of 1500 psi to be acceptable.

*5.5.2. Extrusion cell parameters:* The cell dimensions used for the NOvA RPVC extrusions have been optimized for signal efficiency and background rejection using the simulation studies described in Chapter 12. We used FEA calculations to determine the extrusion wall thicknesses that would provide mechanical stability of the far-detector structure at all stages of the detector construction and during filling with liquid scintillator. The effects of long-term creep in the RPVC material were also taken into account. The 6-cm extrusion cell depth we have chosen is the maximum allowed for the 3-mm extrusion wall thickness and 2-mm web thickness used [6], [7].

*5.5.3. Hydraulic forces:* The weight of the liquid scintillator in the vertical extrusion modules is transferred to the floor by the hydraulic pressure on their base plates.

Within one extrusion module, all 32 cells are hydraulically connected to allow the flow of liquid scintillator and displaced air during filling. Adjacent extrusion modules are not hydraulically connected to one another. The 15.7-m high vertical extrusions have a hydrostatic pressure of 19.2 psi at the bottom. The horizontal extrusions are only 1.3 m high and have maximum pressures of only 1.6 psi. The greatest forces are exerted by the hydraulic pressure on the outer cell walls of the vertical extrusions.

Each vertical extrusion will swell during filling by 2 to 5 mil near the bottom, where the hydrostatic pressure is highest, due to bowing of the outer walls and stretching of the webs. Our FEA calculations show that friction will prevent the bottom plates of the vertical extrusions from sliding on the floor [8], so stresses will build up during filling. Fig. 5.9 shows how this affects a stack of planes.

Our FEA analysis has determined that local stresses in the RPVC will exceed our design stress if more than 80 planes are assembled in one block [9]. We therefore plan to use 32-plane blocks separated by expansion gaps to limit the buildup of hydraulic stress during filling.

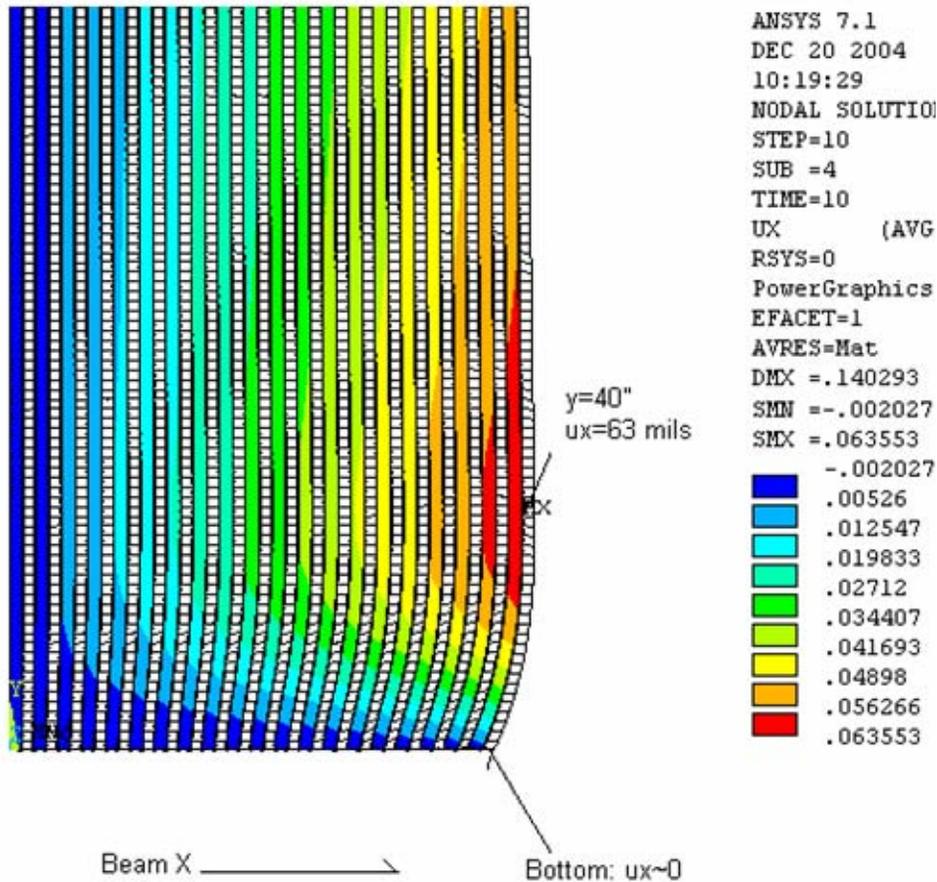

Fig. 5.9: The displacement along the beam direction after 40 planes are filled (80 planes total due to symmetry condition). Only the bottom 3 m out of the 15.7 m height is shown, and that the deformation is highly exaggerated. The deformation and stress is concentrated at the bottom of the detector.

*5.5.4. Vertical extrusions:* Each vertical extrusion is filled with liquid scintillator to a height of 15.7 m, subjecting the bottom of the extrusion to 19.2 psi. The pressure creates a downward force on the bottom closure plate that is transmitted to the floor. This force also bows out the outer walls and stretches the webs between adjacent cells.

For an outer extrusion wall thickness of 3 mm and a web thickness of 2 mm, the FEA gives a maximum stress of 1400 psi for the interior cell, as shown in Fig 5.10.

*5.5.5. Horizontal extrusions:* Although the maximum hydrostatic pressure in each horizontal cell is only 1.6 psi, the lower extrusions cannot support the load of the filled extrusions above them. For this reason the horizontal extrusions are glued to the adjacent vertical extrusions, which support their weight and prevent it from being transferred to the horizontal extrusions below [7], [10]. Table 5.2 shows that gluing horizontal and vertical extrusions together further reduces the stresses from hydrostatic pressure.

|  | Fully glued | No glue |
|---|---|---|
| Deflection (mils) | 1.5 | 5.7 |
| Maximum Stress (psi) | 560 | 1,400 |
| Maximum shear stress in the mid plane (psi) | 70 | Not applicable |

Table 5.2: Deflections and forces calculated for planes of vertical and horizontal extrusions, with and without glue. The deflections at the outer edge of the horizontal extrusion are given. The pressures shown are conservative in that they do not take this deflection into account.

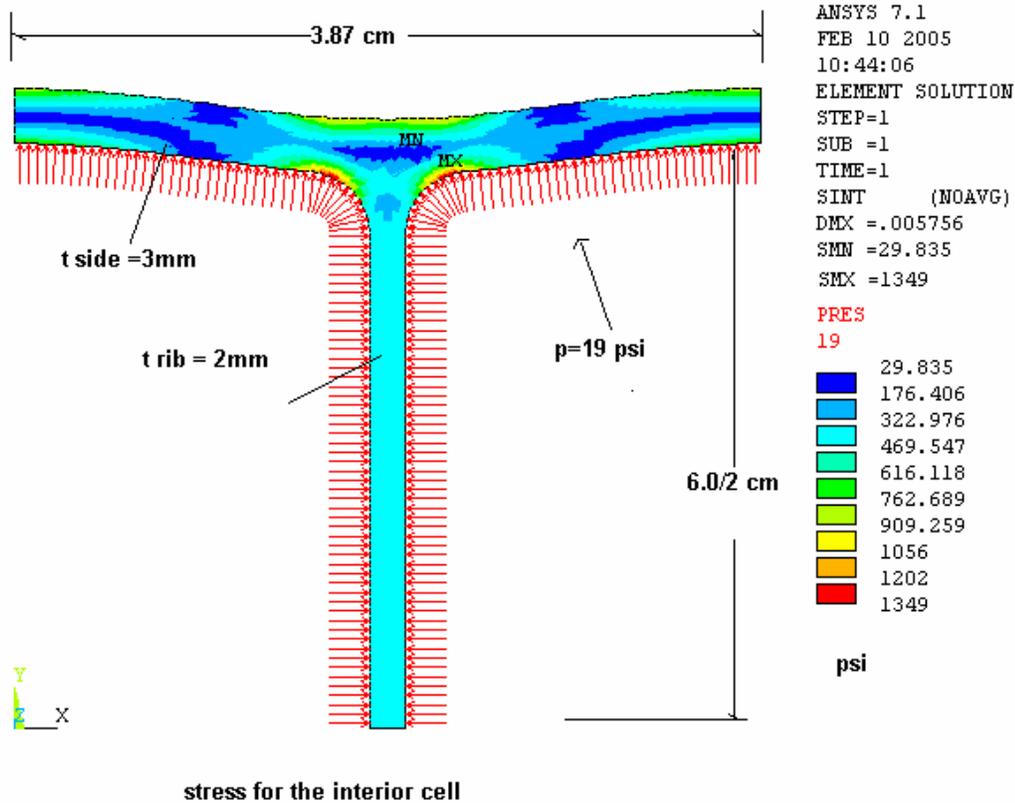

Fig. 5.10: FEA results for interior cell stresses.

*5.5.6. Buckling stability:* FEA calculations show that the 32-plane blocks do not buckle under their own weight [10]. Figure 5.11 shows that a 32-plane block, when free standing, has a safety factor of 2.3 against buckling. This safety factor applies to each block and demonstrates that a collective failure, where all blocks would buckle together, is extremely unlikely.

For additional stability, successive 32-plane blocks will be connected along their top edges using the PVC spacer blocks described earlier. This increases the buckling safety factor for each block to 2.9, as shown in Fig. 5.12 [9].

*5.5.7. Thermal expansion:* RPVC has a thermal expansion coefficient of 67 ppm per $^0$C. If we take a design temperature range of $20 \pm 10\ ^0$C, then the complete detector RPVC stack will try to expand or contract by 9 cm. The bottom plates of each extrusion module are held in place by friction and will not move. The top spacer plates are glued to the adjacent blocks and hence the whole top moves as a unit as the temperature changes. Thermal expansion will tilt the planes slightly, starting with an exactly vertical plane at the first bookend, and ending with a 9 cm tilt at top of the far end of the complete detector (for a 10 C change). The tilt creates a force parallel to the detector axis proportional to the angle. Summing over all planes, the force is ±15 tons at the extremes of the design temperature range. The bookend will be designed to resist that force. The top spacer board is 30 cm by 15.7 m in area, and will see a stress of just ± 4.3 psi, which is small compared to typical epoxy yield strength of 2000 psi.

## 5.6. Summary

Our initial detector engineering studies have led to a design that meets conservative structural safety standards while providing the excellent performance of a highly segmented, totally active liquid scintillator detector. Chapter 15 outlines our plan and funding request for extending this work to develop a complete conceptual design of the NOνA detector.

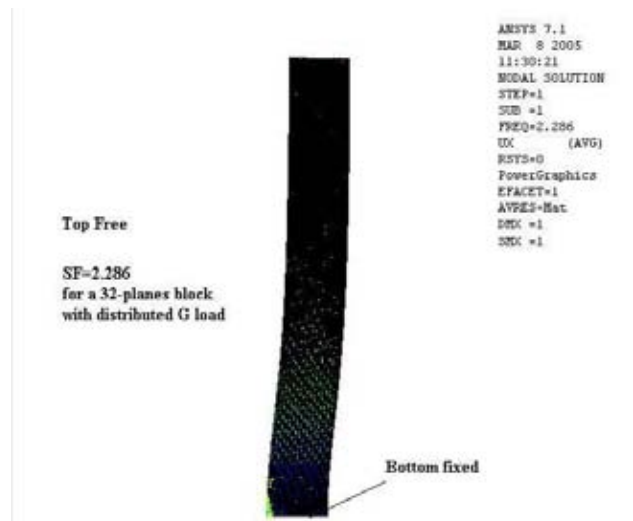

Fig. 5.11: Buckling calculation for a 32-plane free-standing block.

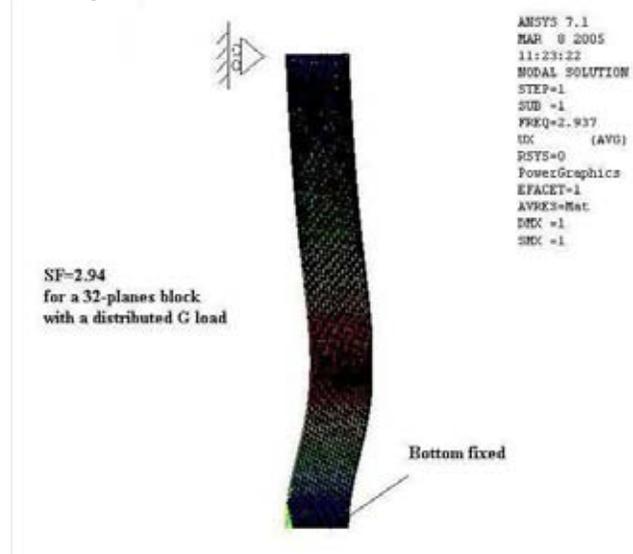

Fig. 5.12. Buckling calculation for a 32-plane block that is constrained at the top. Beginning with the first block, the tops of all blocks are connected together and also tied to the bookend, to increase the safety factor against collective buckling.

# 6. Light Collection

## 6.1. Introduction

The signal resulting from the passage of a charged particle through a cell of the detector depends on the type of scintillator used, the wavelength shifting fiber, the size of the cell, the cell wall reflectivity, and the photodetector response.

## 6.2. Liquid Scintillator

The scintillator we propose to use is a mixture equivalent to Bicron BC517L [1] (also sold as Eljen EJ321L), essentially pseudocumene in a mineral oil base. BC517L has a moderate light output, 39% of anthracene, when fresh, and 27% of anthracene, when fully oxygenated. The advantages of this mixture include stability, low cost, availability in large quantities, low toxicity, high flashpoint and low potential as an environmental hazard. Previous work has shown that this scintillator attacks neither wavelength shifting fiber nor PVC over lifetimes exceeding this experiment [2]. Formulations with significantly higher light output are less stable and do interact with the WLS fiber.

Oxygen will pass through the PVC cell walls so that the scintillator will become oxygenated. The oxygenation of BC517L generally proceeds to a stable light output within a few months. Fig. 6.1 shows the results of measurements we have made of pulse-height spectra from solid scintillator, fresh liquid scintillator and oxygenated liquid scintillator. Our light yield calculations and measurements are with fully oxygenated scintillator.

We have tested BC517L to determine how its light yield changes with temperature. Over temperature ranges between 15 $^0$F and 110 $^0$F degrees F (-10 $^0$C to 48 $^0$C), there was no measurable change in light yield (less than 1%). At the low end of the temperature range, mineral oils experience a rapid increase in viscosity as the temperature is reduced below their pour point, which is typically around 10 $^0$F (-12 $^0$C). This is due to gelling of the oil as crystalline wax is precipitated. Below 15 $^0$F (-9 $^0$C) our measurements showed the scintillator begins to get cloudy. At -22 $^0$F (-30 $^0$C) wax balls precipitated out of the scintillator. These wax balls do not dissolve when the liquid is warmed to room temperature. On the high end of the temperature range, the light yield of the scintillator decreased when the pseudocumene began to vaporize, above its flashpoint of 118 $^0$F (48 $^0$C). Although the performance of liquid scintillator is very robust to temperature changes, extremes need to be avoided both in its transportation and storage. The scintillator is likely to be permanently damaged at temperatures below about -20 $^0$F and above 110 $^0$F. Both extremes are possible in a building without climate control in Northern Minnesota. We discuss the building for this detector in Chapter 8.

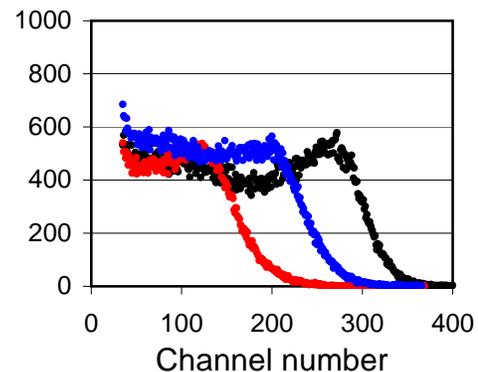

Fig. 6.1: Pulse-height spectra for scintillators showing Compton edge of $^{137}$Cs gammas. Black (right): MINOS scintillator; blue (middle): fresh BC517L; red (left): 5-year old BC517L.

## 6.3. Wavelength-Shifting (WLS) Fiber

WLS fiber provides an efficient method for collecting light from the long narrow cells used in this detector. The WLS shifts light from shorter wavelengths to green (~525 nm) and traps it within the fiber. The MINOS Far Detector provides considerable experience on the construction and operation of this light collection design. Suitable multiclad WLS fiber is available from Kuraray, the same type of fiber and the same vendor used for MINOS.

One adjustable design parameter of the fiber is its diameter. Diameters greater than ~1.5 mm are difficult to spool and ship. For fiber diameters around 1 mm, the light collection efficiency depends approximately linearly on the radius of the fiber as shown in Figure 6.2, while the cost of the fiber depends on its volume ($r^2$). Thus, in terms of photons per dollar, two thinner fibers are more efficient than one thicker fiber.

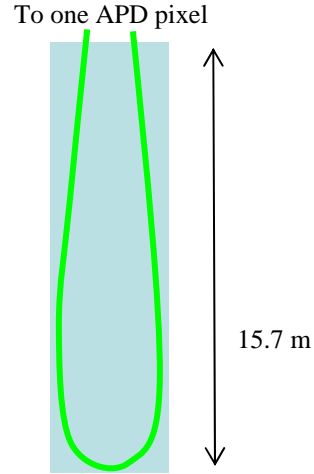

Fig. 6.3: A single liquid scintillator filled PVC cell with a 0.8 mm diameter looped WLS fiber shown in green.

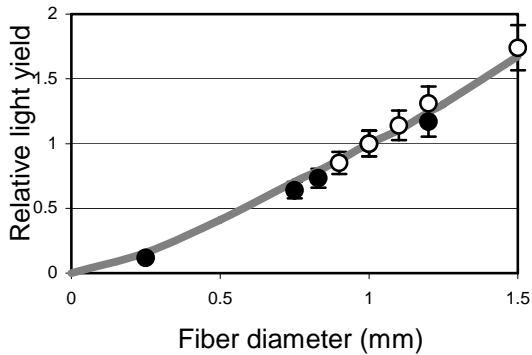

Fig. 6.2: Relative light yield as a function of WLS fiber diameter. Open circles - from measurements made for MINOS Detector; closed circles - recent measurements; solid line - Monte Carlo simulation. (Data are normalized to unity at 1 mm diameter.)

The looped fiber design shown in Figure 6.3 effectively provides two fibers with a no cost, perfect mirror at one end. This gives a factor of two more light from the far end of each cell, where light output is most important, than from the far end of two individual fibers with nonreflecting far ends. The two ends of the looped fiber will be brought together in an optical connector and connected to one pixel of an avalanche photodiode (APD). Our current design uses 0.8 mm diameter looped fiber, which satisfies our requirements for cost, light yield, and handling. During the proto-typing phase we will examine these parameters to determine the optimal fiber diameter.

Figure 6.4 shows the attenuation of light in a 0.8 mm diameter Kuraray multiclad fiber. It demonstrates the light collection advantage of a looped fiber design. Note that the ratio of light output for a looped fiber to a single fiber is larg-est at the far end of the liquid scintillator cell where light collection is at a premium.

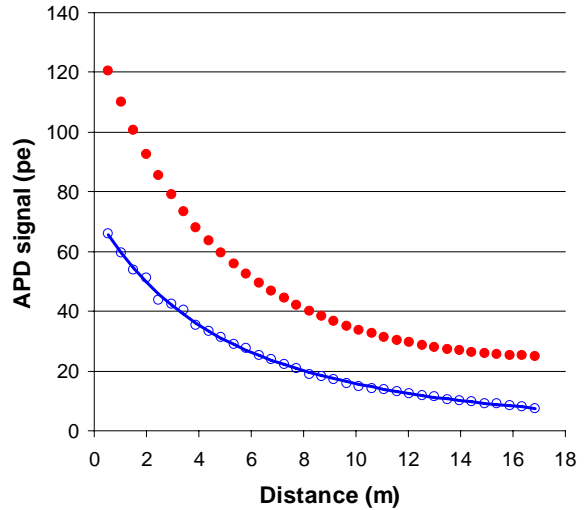

Fig. 6.4: APD signal for looped (red) and single (blue) 0.8 mm diameter fibers. The red curve is scaled from the measured blue curve using the measured shadowing factor of 1.8 for two fibers. At the end near the APD the ratio of the single fiber signal to that of the looped fiber is 1.8 while at the far end it is 3.3.

The attenuation of light in a fiber is a function of its wavelength since short wavelengths are attenuated more strongly than long wavelengths. Figure 6.5 shows the spectrum of light transported through a fiber as a function of the fiber length. Also shown is the quantum efficiency of an APD and, for comparison, a bialkali photocathode photomultiplier tube. The high quantum efficiency of the APD and its flat response over the wavelengths transmitted from the far end of the fiber, where the signal is smallest, make it the ideal for this detector.

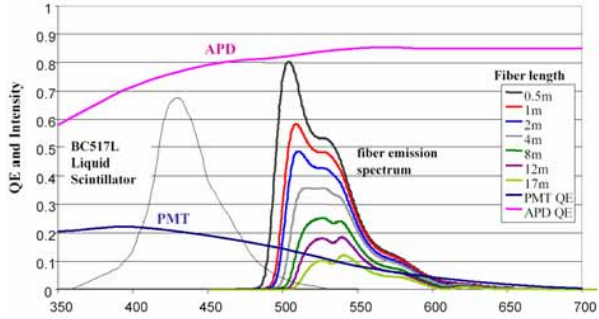

Fig. 6.5: WLS fiber emission spectra measured at lengths of 0.5, 1, 2, 4, 8, 16 m, respectively illustrating the shift of the average detected wavelength as attenuation (fiber length) increases. Also shown are the quantum efficiencies of APD and PMT (bialkali photocathode) as a function of wavelength. The emmission spectrum of the liquid scincillator is also shown..

## 6.4. Cell Structure – Reflectivity and Geometry

The PVC in the extrusions will be loaded with titanium dioxide for reflectivity. $TiO_2$ is the additive that gives commercial PVC its white color. We have tested prototypes of a multi-cell extrusion with a 12% content of $TiO_2$ and measured their reflectivity as a function of wavelength. Pictures of these prototypes are shown in Figures 6.6 and 6.7. The reflectivity measurements are shown in Figure 6.8. This figure also shows the reflectance is reasonably matched to the spectrum of light being reflected, the liquid scintillator emission spectrum. Since photons are, on the average, reflected over 10 times before hitting the fiber, good reflectance is an important component of light yield. For reference, we remeasured the reflecting layer of the MINOS plastic scintillator and the prototype PVC extrusion discussed in ref. [2], and these data are also shown in Figure 6.8.

Within a cell, the light captured by a fiber is fairly independent of its position in a cell but decreases significantly when it is actually touching a cell wall. Figure 6.9 shows a simulation of light capture as a function of the location of a single, unmirrored WLS fiber within a liquid scintillator cell. Our test setup does not constrain the fiber position, so over most of its length the fiber will be at or near an extrusion cell wall. As part of our R&D program, we intend to explore economical ways to control the fiber location.

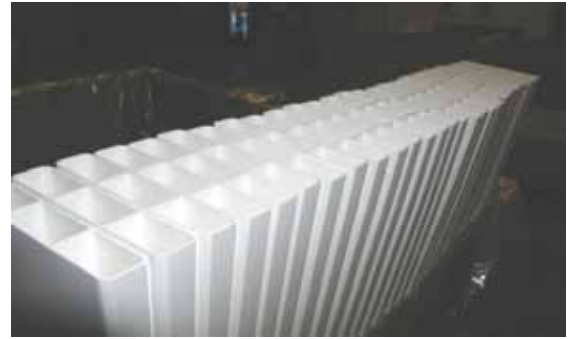

Fig. 6.6: Cell structure of the three cell prototype of extruded PVC with 12% $TiO_2$.

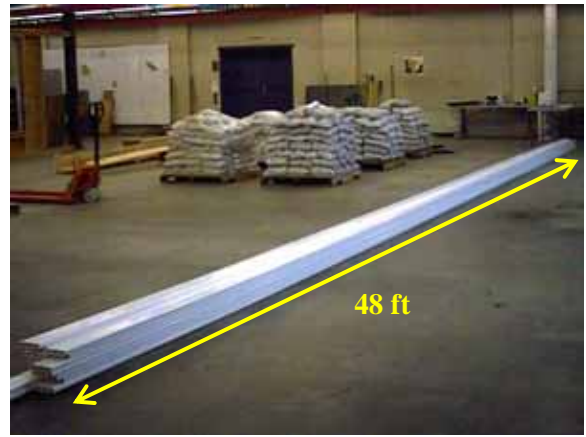

Fig. 6.7: 48 foot long three cell prototype of extruded PVC with 12% $TiO_2$ delivered for testing.

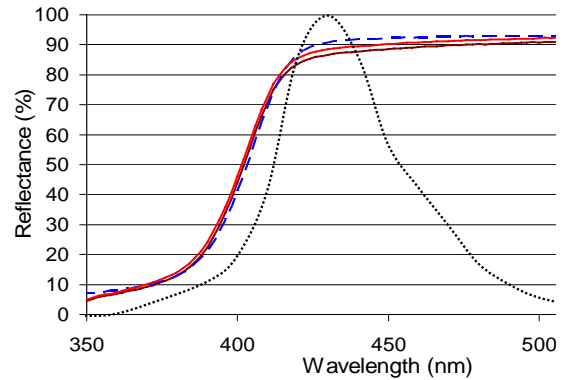

Figure 6.8: Reflectance of the prototype multicell PVC extrusion with 12% TiO2 (red) together with the liquid scintillator emission spectrum (black dots). Also shown is the reflectance of the MINOS plastic scintillator cap (blue dashes) and the MINOS liquid scintillator prototype extrusion (purple) from ref. [1].

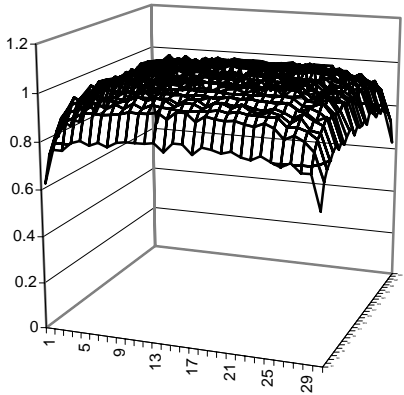

Fig. 6.9: Relative light yield for a single fiber as a function of location within a scintillator cell.

### 6.5. Light Yield

We have measured the light from minimum ionizing cosmic rays passing through 2.2 cm of BC517L liquid scintillator in a prototype multi-cell extrusion (12% $TiO_2$) shown in Figure 6.6. The cell size is 2.2 cm deep by 4.2 cm wide. The light was captured by a looped 0.8 mm WLS fiber and transported 16.4 m to an APD operated at a temperature of -15$^0$C. This prototype setup tests the basic NOvA detector cell using cosmic ray muons. The average signal for muons traversing the cell perpendicular to its walls was 13 photoelectrons (pe). Previous short sample measurements using a photomultiplier tube had predicted 15 pe. Based on the measured attenuation curve shown in Figure 6.4, the signal from the very end of the 16.7 m length would be 12.5 pe.

For the NOvA detectors proposed here, the cell size is modified from the prototype 2.2 cm (along the beam) by 4.2 cm (wide) to 6.0 cm (along the beam) by 3.9 cm (wide). The increase cell size will increase the energy deposited since the charged particle traverses almost three times more scintillator. In the larger cell, however, the fiber is typically farther from the reflecting walls, decreasing the light collected. Since we do not yet have an extrusion of this size, we have used our light collection simulation to determine a relative light yield from the two geometries. The simulation program includes the emission spectrum of the liquid scintillator, defuse reflection from the cell walls, and the absorption and reflection characteristics of the fiber. The simulation program predicts a ratio of 1.75 for the two cell geometries. Thus we expect that a minimum ionizing particle traveling in the beam direction and traversing a NOvA cell perpendicular to its walls 16.7 m from the photodetector will give a signal of 22 pe.

Event simulations show that 22 pe is sufficient for event discrimination. However we expect several simple improvements will increase the light yield by 20 to 30%. For example, increasing the reflectance by 1%, increases the light yield by about 10%. We expect at least this increase when the $TiO_2$ content of the PVC is increased from the 12% in our prototype extrusion to the design value of 15%. Figure 6.8 shows the reflectance measurement of the MINOS plastic scintillator cap, which has a $TiO_2$ content of 15% in polystyrene. This $TiO_2$ content is 3% higher than our current prototype extrusion in the relevant part of the spectrum but in a different plastic. Another light yield increase of greater than 10% can be achieved by controlling the position of the fiber in the cell as predicted by simulations shown in Figure 6.9 and verified by our measurements. With these changes, we expect at least 25 pe from the far end of the detector. Using the electronics described in Chapter 7 would then give a signal to noise ratio in excess of 10 to 1.

The prototype electronics used in our test setup have a noise level of about 350 electrons. Figure 6.10 shows the measured photoelectron distribution for a light yield of 25 pe with the prototype cell, fiber, APD, and electronics. The peak at 0 shows the pedestal caused by random coincidence triggers. The pedestal is separated from the single minimum ionizing signal.

A 10 to 1 signal to noise ratio is illustrated by Figure 6.11. In this case the cosmic ray muon signal was measured at 8.4 meters which gives a signal of about 35 pe and a noise of about 350 electrons. The ASIC proposed in Chapter 7, which is matched to the APD capacitance, has a noise of 250 electrons thus achieving the 10 to 1 signal to noise ratio with a signal of 25 pe.

**Chapter 6 References**

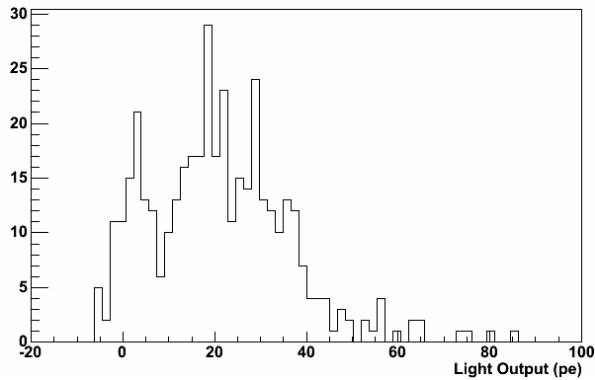

Fig. 6.10: Histogram of the cosmic ray muon signal from an 0.8 mm fiber with an average signal of 25 pe with a noise of about 350 electrons. The peak at 0 is the pedestal from random triggers.

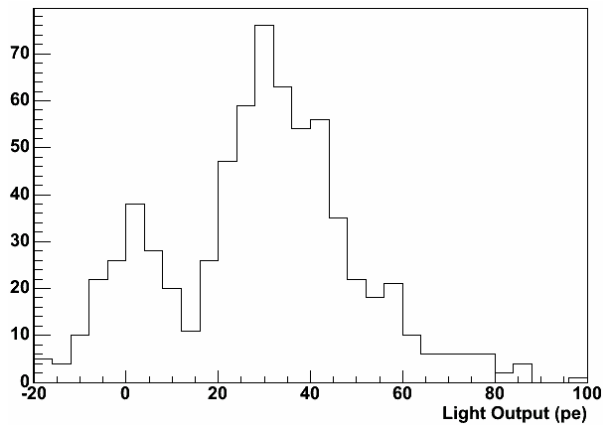

Fig. 6.11: Histogram of the cosmic ray muon signal from 8.4 m along a looped 0.8 mm fiber with the prototype cell, fiber, APD, and electronics. The average signal of 36 pe shows a 10 to 1 signal to noise with the existing electronics (350 electrons). The peak at 0 is the pedestal.

# 7. The Photodetector and Readout

## 7.1. Introduction

The function of the readout is to convert the optical signal from the wavelength shifting fibers into an electrical signal. The readout of the NOvA detector has two distinct tasks: (1) read out events caused by neutrinos from Fermilab and (2) operate between spills in a triggerless mode to collect cosmic ray events for calibration and supernova events if they occur. A trigger generated from the early stages of the Main Injector cycle can be used to form the gate for the in-spill events, while the other mode requires fast signal processing.

In our design the phototransducer is an avalanche photodiode (APD), one per detector channel, readout though a low-noise preamplifier. The APDs are in 32-channel arrays, with one array being coupled to all the fibers from a 32-channel detector module. The signal processing behind the preamplifier allows for two data acquisition modes to be operated alternatively: one on-spill and another between the spills.

## 7.2. Avalanche Photodiodes (APDs)

*7.2.1. Overview:* The proposed light detectors for the baseline design are avalanche photodiodes (APDs) [1] manufactured by Hamamatsu. They are similar to the 5 mm × 5 mm APDs developed for use in the Compact Muon Solenoid (CMS) detector at the CERN Large Hadron Collider [2]. Table 7.1 summarizes the key parameters of the NOvA APDs.

APDs have two substantial advantages over other photodetectors: high quantum efficiency and low cost. The high APD quantum efficiency enables the use of very long scintillator modules, thus significantly reducing the electronics channel count, while the per channel cost is about a factor of four less than that of a multi-channel photomultiplier tube (PMT). Figure 6.5 compares the quantum efficiency of a Hamamatsu APD to that of the PMT used in the MINOS Far Detector. In the wavelength region relevant to the output of the wavelength shifting (WLS) fibers described in Chapter 6, 500 to 550 nm, the APD quantum efficiency is 85% *vs.* 10% for the PMT. As shown in that figure, the quantum efficiency advantage of the APD increases with wavelength and thus the length of the fiber. This gives the APD an even greater advantage over a PMT for long fibers as shown in Figure 7.1.

| Manufacturer | Hamamatsu |
|---|---|
| Pixel Active Area | 1.8 mm × 1.05 mm |
| Pixel Pitch | 2.3 mm |
| Array Size | 32 pixels |
| Die Size | 15 × 15 mm$^2$ |
| Quantum Efficiency (>525 nm) | 85% |
| Pixel Capacitance | 10 pF |
| Bulk Dark Current ($I_B$) at 25 C | 10 pA |
| Bulk Dark Current ($I_B$) at -15 C | 0.15 pA |
| Peak Sensitivity | 600 nm |
| Operating Voltage | 400 ± 50 volts |
| Gain at Operating Voltage | 100 |
| Operating Temperature (with Thermo-Electric Cooler) | -15°C |
| Expected Signal-to-Noise Ratio (Muon at Far End of Cell) | 10:1 |
| APD channels per plane | 384 |
| APD arrays per plane | 12 |
| Total number of planes | 1,984 |
| APD pixels total | 761,856 |

Table 7.1 Avalanche Photodiode Parameters.

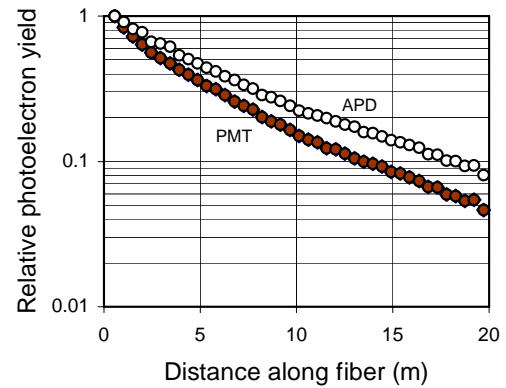

Fig. 7.1: Relative photoelectron yield from 1.2 mm diameter WLS fiber, for APD and a PMT with a bialkali photocathode. The data have been normalized at 0.5 m to illustrate the effect of the longer wavelength response of the APD.

The commercially-available Hamamatsu APD has a pixel size of 1.6 mm by 1.6 mm. A photograph of the 32 pixel APD package is shown in Figure 7.2. We will use a 32 pixel array of APD's in the bare die form and mount the chip, cooler, electronics and optical coupler on a printed circuit board.

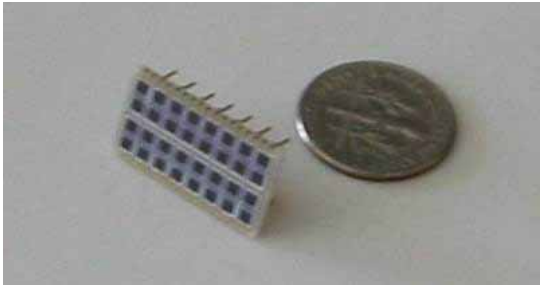

Fig. 7.2: A commercially available Hamamatsu APD package shown with a dime. Two 16 pixel arrays are packaged together.

*7.2.2. Photodetector Requirements:* Photodetectors for the NOvA Far Detector must be able to efficiently detect single minimum ionizing particles traversing the far ends of scintillator strips, ~16.7 m (of fiber length) away. Each photodetector pixel should be large enough to collect the light from both ends of a 0.8 mm diameter looped fiber.

Based on the measurements described in Section 6, we estimate that a single minimum ionizing particle, normally incident at the far end of a liquid scintillator tube, will produce ~30 photons at the face of the APD. The quantum efficiency for an APD in the region of the spectrum where the light is emitted is 85%, giving a signal for such a particle of ~25 photoelectrons. This signal must be distinguishable from the noise with high efficiency. One of the operational characteristic of APDs, and, in fact, all silicon devices, is the thermal generation of electron hole pairs which mimic the signal. The thermally generated electrons are amplified at the diode junction and appear at the input to the pre-amplifier and thus contribute directly to the noise. To reduce this generation rate to a manageable level we will lower the operating temperature to -15° C using thermo-electric (TE or Peltier-effect) coolers. These are very common commercially available devices.

*7.2.3. Fundamentals of APD operation:* The general structure of an APD is shown in Figure 7.3. Light is absorbed in the collection region, electron-hole pairs are generated and, under the influence of the applied electric field, electrons propagate to the p-n junction. At the junction, the electric field is sufficiently high that avalanche multiplication of the electrons occurs. The multiplication (M) of the current is determined by the electric field at the junction, and by the mean-free-path of electrons between ionizing collisions, which depends on both the accelerating field and on the temperature. This temperature dependence occurs because the probability of electron-phonon scattering increases with temperature.

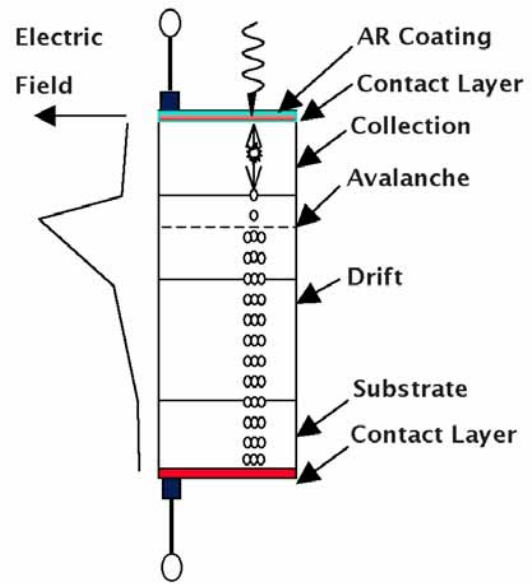

Fig. 7.3: The basic structure of a blue/green sensitive APD. Light crosses the anti-reflection coating at the surface and is absorbed in the collection region. Photoelectrons drift in the electric field to the junction where they undergo avalanche multiplication.

APDs, like PIN diodes, have an intrinsic noise that comes from the electron-hole pairs generated thermally in the depletion region of the diode. Since the current from the positive carriers is amplified about fifty times less than the negative carrier current at the junction, only the current from electrons generated in the photo-conversion region ($I_B$), or the bulk current, needs to be considered in the noise current estimation. As it is a thermally generated cuurent, it can be reduced by lowering the operating temperature of the APD. We will operate the APDs in the NOvA detector at

-15° C to keep the noise contribution from $I_B$ small in comparison to the front-end noise. This choice is based on measurements obtained with the prototype readouts.

Besides this source of noise, the amplification mechanism is itself subject to noise, characterized by the excess noise factor $F$, with such factors as device non-uniformities and the ratio of the positive to negative impact ionization coefficients contributing. This factor is well modelled and has been included in our signal to noise calculations.

One of the attractive features of APDs is that once they have been calibrated, the gain can be easily determined from the applied bias voltage and the operating temperature. In the NOνA detector, we will maintain the operating bias to a precision of 0.2 Volts and control the temperature to 0.5° C and thus hold the gain stability to about 3%.

*7.2.4. Experience with the CMS APDs:* The CMS experiment is using 124,000 Hamamatsu APDs, with 5 mm × 5 mm pixels, to read out the lead-tungstate calorimeter. The full order has been delivered to the experiment and tested. The quantum efficiency for these devices is consistently at 85% at 550 nm as can be seen in the Figure 7.4.

*7.2.5. APDs for the NOνA Detector:* We have purchased Hamamatsu's of-the-shelf APDs for our measurements. The measured dark current, pixel gain and pixel separation for one of the sample arrays are shown in Figures 7.5, 7.6 and 7.7. The dark current is consistent with expectations from CMS APD measurements, and the gain is uniform from pixel to pixel on the same chip and within an individual pixel. The fall-off on the pixel edges in Figure 7.8 mostly reflects the finite spot size used to illuminate the APD pixels.

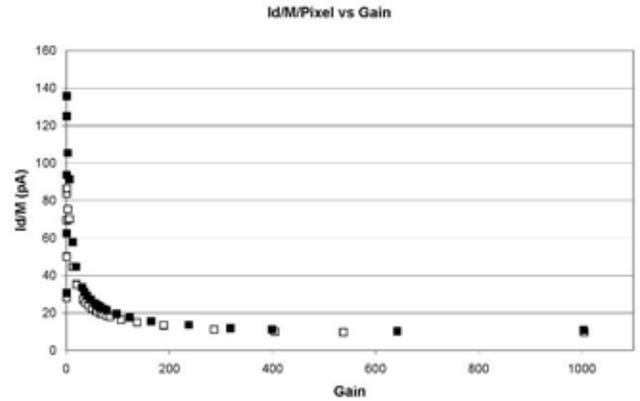

Fig. 7.5: Dark current $I_d$ divided by gain *vs.* gain in a commercial Hammatsu APD at 25 C. The asymptotic value of the current is $I_B$, which is 10 pA for this sample.

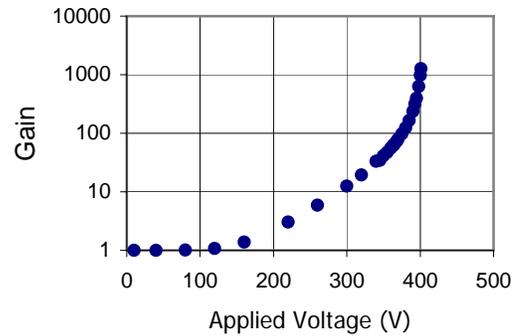

Fig. 7.6: Gain *vs.* applied voltage at 25 C.

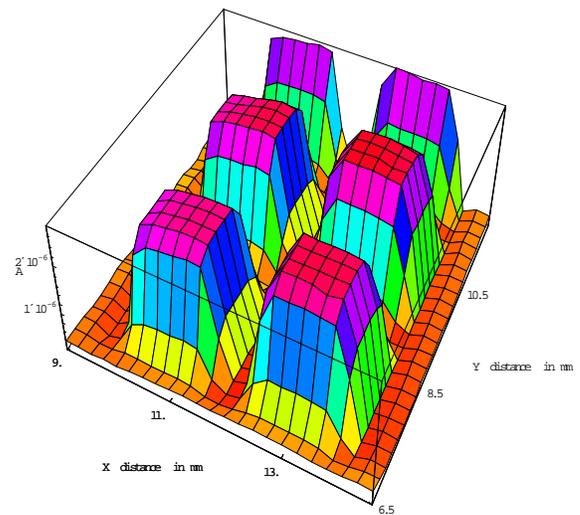

Fig. 7.7: Fine point scan across part of the APD array.

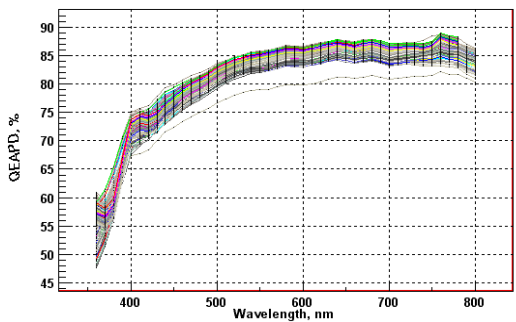

Fig. 7.4: Quantum efficiency of several hundred CMS APDs.

## 7.3. The Readout Electronics.

The readout of the APDs requires a preamplifier that can sample the signal throughout a 10 μs spill gate. The proposed architecture is based on the Fermilab MASDA (Multi-Element Amorphous Silicon Detector Array) chip [4,5,6] and the SVX4 (a multi-channel amplifying and digitizing chip developed for CDF and D0). An ASIC has been designed and simulated specifically for this readout. It has two operating modes which can be selected electronically, one for gated in-spill collection and another for the triggerless mode. In the design the dual correlated sampling (DCS) method or a multiple correlated sample method is used to remove the common mode noise.

*7.3.1. Signal-to-Noise:* We have investigated the performance of the APD coupled with the MASDA ASIC, which uses the dual correlated sampling (DCS) technique, to investigate the noise performance that can be achieved with a cooled APD. The MASDA is optimized for 70 pF input capacitance, rather then the 10 pF of the APD, so these measurements are upper limits. Figure 7.8 shows the measured noise of several APD's operating at a gain of 100. At about -10° C the noise plateaus at around 300 electrons, indicating that the contribution from the dark current has become negligible.

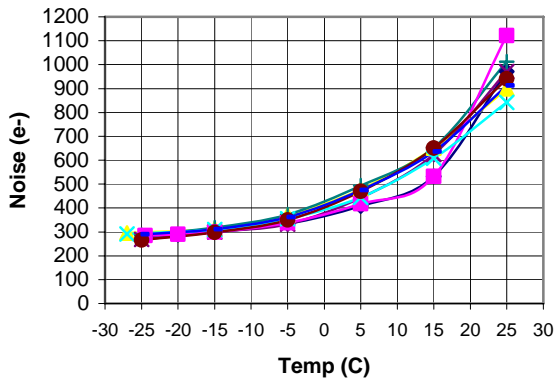

Fig. 7.8: Noise measurements of the readout of an APD operating at a gain of 100, as a function of temperature. The readout used in this test was the MASDA ASIC which was optimized for 70 pF input capacitance.

The computed noise level for the chip that we have designed specifically for this application operating with an APD at a gain of 100 and cooled to -15°C is 150 electrons. From this we estimate that the real noise levels for production devices will be ~ 200 electrons. This is much lower than normally associated with readout electronics because we are using the DCS method, which eliminates much of the correlated noise.

For our discussion we will always refer to the noise level at the photo-electron level, thus with the APD at a gain of 100, the 200 equivalent noise charge (ENC) reduces to 2.0 photoelectrons at the photodetector input. This is to be compared with the 25 photoelectron signal we expect from a muon at the far end of the channel.

We have measured the bulk dark current ($I_B$) for several APDs and the average value per pixel as 10 pA at 23°C. This is consistent with the bulk dark current of the CMS APD: 5 pA/mm$^2$, corresponding to 12 pA/pixel. A current of 10 pA corresponds to a current of 62 electrons every microsecond. At our operating temperature of -15 C, the APD background rate is then 2 thermally-generated electrons in our 1 μs sampling time with a rms noise of 1.4.

The requirement for the readout is then to detect a signal with an average value of 25 photoelectrons spread over a short time interval, with a background rate of 2 thermally-generated electrons per microsecond using an amplifier with an effective ENC of 2.0 electrons. Figure 7.9 shows the estimated signals from one and two minimum ionizing particles, considering all noise factors, including amplification noise, compared with the noise. In making this graph we have assumed that the ENC of the amplifier is 2.5. The graph shows good discrimination between zero, one and two normally incident muons crossing the far end of the scintillator strip. For comparison, Figure 7.10 shows the actual signal measured using the current prototype described in Chapter 6 using light injection to generate a signal corresponding to one and two minimum ionizing particles at the end of the proposed detector (25 and 50 pe).

*7.3.2. Digitizing and Readout Architecture:* We have examined several different readout architectures and have settled on a baseline design based on the SVX4 structure that makes use of the DCS method of the MASDA chip. The ASIC has several modes of operation. In one mode – the high precision mode - the data are stored in a 32-channel wide 64 deep switched capacitor array (SCA) during the spill and digitized afterwards. This minimizes any risk of noise from the conversion appearing at the signal inputs. The signal from each APD is amplified by a high gain integrating amplifier with a shaping time of ~350 ns and the output is stored in the SCA every 500 ns. After the beam spill, the SCA contains 64 samples taken 500 ns apart for the 32 APD channels. The difference in the stored signals, taken 1 µs apart for all 32 channels, are compared in parallel with a linear ramp and the crossover times stored, as in a Wilkinson ADC. When this digitization is complete, the difference between the next pair of stored signals is converted. At the end of the conversion every digitized difference is stored in an field programmable gate array (FPGA) for transmission to the data acquisition (DAQ) system. The chip's architecture is shown in Figure 7.11. In this design the gate width can be up to 30 µs; this is wide enough to accommodate uncertainties in the foreknowledge of the beam arrival time.

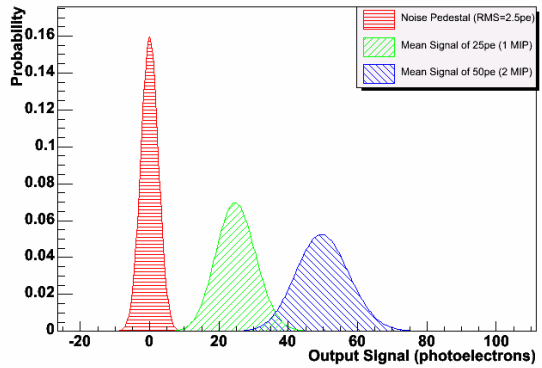

Fig. 7.9: Expected APD signals from noise, 1 and 2 minimum ionizing particles. The calculation uses a total noise of 250 electrons and signal levels of 25 and 50 photoelectrons.

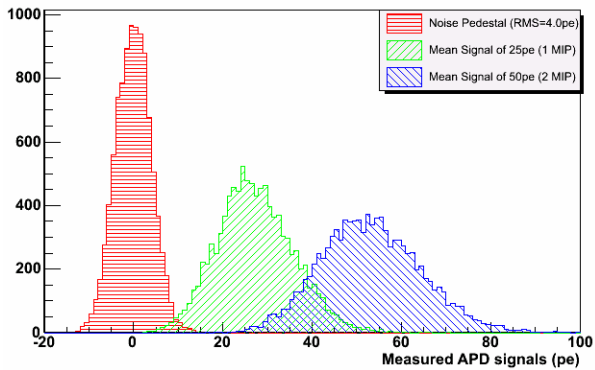

Fig. 7.10: Measured APD signals from noise, 1 and 2 minimum ionizing particles using the current prototype electronics. This measurement has a total noise of about 400 electrons and signal levels of 25 and 50 photoelectrons generated by light injection.

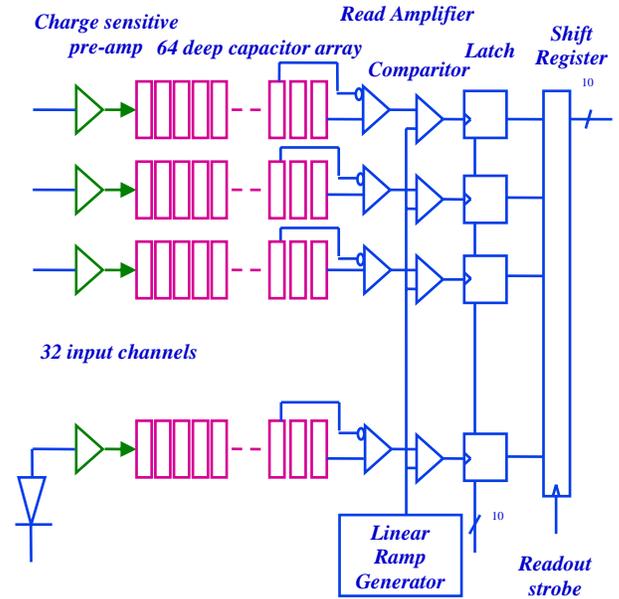

Fig. 7.11: Proposed architecture of the front-end electronics ASIC. Shown here is the configuration that would be used for data collection during the spill and digitization after the spill.

In addition to the in-spill data taking mode we have designed the possibility of taking data in a triggerless mode so that the detector can be used to search for supernova events and to collect cosmic ray data. A supernova signal would be a large number of small energy depositions spread over an interval of several seconds. To collect data in this mode the linear ramp is stopped earlier, reducing the dynamic range from 100 to 8 MIP's, or a pseudo-logarithmic ramp is used which achieves the full dynamic range at the expense of resolution of large signals. Here one MIP is defined as the signal of a minimum ionizing particle crossing a liquid scintillator cell at the far end of that cell, a 25 photoelectron signal.

*7.3.3.Further improvements:* We have been investigating a method to further reduce the noise levels by using multiple, rather than just two, correlated samples. Since the data rate in the NOνA detector is low, we can use considerably more signal information than would be available with DCS and by using many consecutive samples we can further reduce the noise. The degree to which the signal to noise ratio (SNR) can be improved will depend on the detailed noise spectrum of the front end integrator and APD combination. If the noise is dominated by APD leakage currents then any gains will be small, but if the noise is completely dominated by the integrator front end and the APD capacitance, then the signal to noise ration would improve by $\sqrt{N}$, where N is the number of pairs of data points used. As discussed above, both noise components will be present. Tests performed on a prototype system have achieved a 25% improvement in the signal to noise ratio over dual correlated sampling and we anticipate an improvement of this order could be achieved if we implement this method. This would entail converting the individual samples stored in a SCA rather than the differences. The most significant difference would be in the complexity of the FPGA firmware.

In addition to improving the noise performance, time resolution can also be improved. To study this we have conducted tests to establish the limits of the timing resolution that can be achieved. In our baseline with 500ns sampling of the integrator waveform, a timing resolution of no better than the sampling interval divided by $\sqrt{12}$, or about 145 ns can be achieved. We have examined various digital signal processing techniques for timing resolution improvement. The most promising of these is a combination of "matched filtering" and "interpolation" filters. The matched filter output is the cross-correlation between the incoming signal and an ideal version taking by averaging over many signals. This yields a fairly symmetric output, upon which a low-pass interpolation filter is applied. This filter supplies a 10:1 interpolation between the 500ns data points, providing computed points every 50ns. These calculations, while currently done off-line, can be easily done in firmware on the FPGA. As one might expect, the resulting timing resolution depends on pulse height. For very small signals, the timing resolution is about the same as DCS, while for larger pulse height, it is up to five times better.

In summary, in our baseline design there is a FPGA on each front end board to handle control and data transmission. Digital signal processing algorithms could be encoded in the FPGA firmware to improve both SNR and timing resolution. This would require that data collected during the beam spill are the stored as signals instead of differences. We will continue to evaluate the possibility of employing these advanced signal processing methods as needed.

### 7.4. Mechanics

Each module will have a single readout box mounted on it, with a single 32-channel amplifier reading out a 32 pixel APD which will be connected to the 32 channels in the detector module. The operating voltage ($400 \pm 50$ V) to bias the APD array will be supplied from an on-board high voltage generator designed for this purpose, an integrated circuit based on the Cockroft-Walton technique. The APD array will be cooled by a single-stage TE cooler. The thermal power generated in the APD array is ~25 μW, so the most significant thermal load will be from local conduction along the fibers and through the electrical interconnects. The TE cooler will produce less than 2W of heat for each 32 channel liquid scintillator module. Temperature monitoring and control, clock regeneration and I/O functions will be controlled with a low-power FPGA. The APD array will be mounted on the opposite side of the board from the other electronic components to minimize the thermal load. The mounting will be done with flip-chip technology, so the active area will be facing a hole cut out in the electronics board (PCB) where the fiber ends will be located. The flip-chip method provides an accurate way to align of the APD to the PCB, to which the fiber connector will also be aligned.

A box housing the APD and the associated electronics will be connected to the end of each scintillator manifold. The APD box has several functions: (a) align the fibers to the APD array, (b) provide a light tight connection to the scintillator module, (c) house the APD and the associated electronics, (d) remove heat from the electronics and the TE-cooler, (e) protect the cold surfaces from humid air to prevent condensation and (f) provide structural strength. The manifolds are designed such that a module can be connected into a single APD box. This modularity allows for testing of the complete system prior to installation.

The APD arrays, the PCB, the heat sink, and the electronics are housed in an aluminum sheet-metal box that serves as a Faraday cage. The box also contains connectors for the low voltage, clock signals and electronics readout. The APD box will also be light tight. A schematic of our concept of the APD housing showing is shown in Figure 7.12.

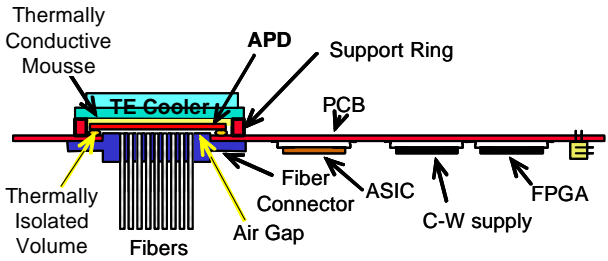

Fig. 7.12: Side view of the components in an APD electronics box. The box receives the signals from a scintillator module through an optical connector. A TE cooler is connected to the APD through an electrically insulating, thermally conducting material. The hot side of the TE cooler is in thermal contact with a heat sink to remove the heat from the box. The APD is mounted on the PCB using the flip-chip method.

## 7.5. Data Acquisition

*7.5.1 Data Acquisition Modes:* The primary task for the readout and data acquisition system is to concentrate the data from the large number of APD channels into a single stream, which can be analyzed and archived. The specifications of the (DAQ are given in Table 7.2. The complexity of the DAQ electronics comes from the requirement that system is both externally triggered, for in-spill events, and triggerless for cosmic ray and supernovae events. We will discuss both these modes of operation in turn.

| APD boxes per plane | 12 |
|---|---|
| APD channels per box | 32 |
| Digitization | 10 bits every 0.5 µsec |
| Digitization in triggerless mode | 7 bits every 0.5 µsec |
| Noise rate per channel | <$10^3$ Hz |
| Bytes per hit (channel ID, TDC, ADC, status) | ≤8 |

Table 7.2: Specifications for DAQ system.

The externally triggered system is "live" for only a short period of time, ~30 µs surrounding the neutrino beam spill. The actual beam spill will be 10 µs allowing a large margin for predicting the arrival time of the neutrino pulse. The upper limit is determined by the depth of the SCA memory. We plan to use the $23 and $A5 signals from the Main Injector which occur 1.4 s before the spill and predict the beam arrival time to within ± 5 µs.

For triggerless operation, the data will be input continuously to an FPGA where it will be sparsified and stored. We would use trigger processors to analyze the data stream looking for hit clusters that might indicate an interesting event.

The data rate per APD box is ~0.5 MB/s, so that an average of 10 bytes is produced per APD box per 20 µs readout, yielding approximately 100 kB per readout for the entire detector. If the readout is triggered randomly at ~100 Hz to measure cosmic ray background, the total data rate is ~10 MB/s. In comparison, the total data rate for the entire detector with a continuous readout mode is estimated to be 5 GB/s.

The DAQ threshold is set to satisfy two requirements: efficient detection of a minimum ionizing particles and a low noise rate so that the DAQ system is not overwhelmed by spurious hits. Since the system will digitize everything in a spill gate, the threshold can be adjusted to meet these goals. For example, assume an electronics noise level of 250 electrons, an APD gain of 100 and a mean signal from a minimum ionizing particle of 25 photoelectrons, or 2500 electrons after the APD. If we set a threshold of 1200 electrons, we expect greater than 99% efficiency for a minimum ionizing particle with a probability for a noise hit of less than $3 \times 10^{-6}$ in 1 microsecond.

*7.5.2. System Architecture:* The overall concept of the readout and DAQ system is similar to that of other experiments. Digitized signals from each ASIC are input into a FPGA. This applies zero suppression and timestamps, and then buffers the digitized values before serialization and transmission to the DAQ. The FPGA can also provide control and monitoring of the APD box. The APD-box FPGA provides an external interface using standard Ethernet protocols. The baseline design specifies less expensive electronic Ethernet interconnections using standard Cat5 cabling. Optical interconnections have the advantages of higher bandwidth and no ground loops at somewhat higher system cost. The final choice will require value engineering.

The overall organization of the DAQ system will be as a collection of local rings readout through Readout Concentrator Nodes (RCN) as shown in Figure 7.13. The advantage of the ring architecture is that the loss of any single ring member disables only that element and not the entire ring. For design simplicity and to reduce requirements for spares, each APD box will have a switchable capability to act as either a ring master or a ring slave. The baseline design is to connect 96 APD boxes from 8 successive planes into each local ring. This gives 1994/8 = 250 rings. Since the total detector data rate is 10 MB/s, the rate per ring is ~100kB/s.

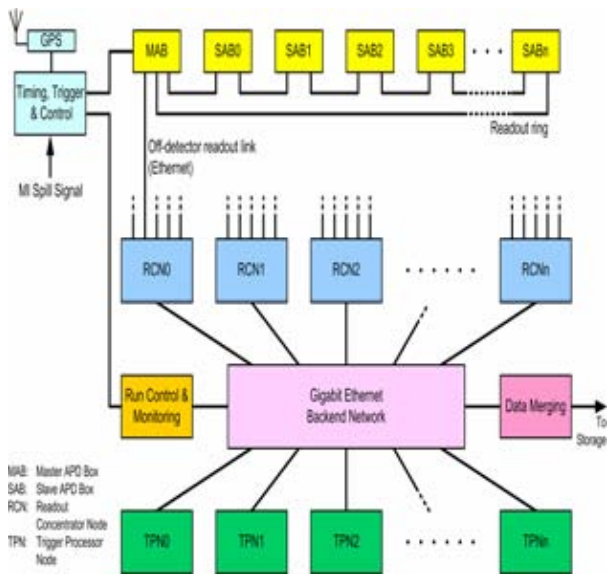

Fig. 7.13: Overview of the entire DAQ system: The data from a number of Slave APD Boxes (SAB) will be collected and transmitted by a Master APD Box (MAB) via Ethernet. Data from a number of MAB will be funneled via Ethernet into a Readout Concentrator Node (RCN). The RCNs will transmit this data via Ethernet to trigger processor nodes (TPNs). The TPNs will run trigger algorithms on this data to decide which data to write to the data storage. A timing system will distribute clock signals (locked to the GPS time) to all MAB. These signals would be redistributed by the MABs to the SABs. The timing system also receives the Main Injector spill signal for redistribution.

We expect to use ~25 Readout Concentrator Nodes (RCNs) to collect data from the APD box Ethernet rings. The RCNs will be PC's with multiple Ethernet cards. Each ringmaster APD box will be connected to a dedicated Ethernet interface card on a RCN. The RCNs will direct all data from a specific trigger to one of several Trigger Processor Nodes (TPNs). The TPN that receives all the data from one particular trigger will then determine whether and how the data from that trigger should be archived for later off-line analysis.

Control information will follow an inverse path via the same network. Detector Control System (DCS) computers will send data to the RCNs, which will then distribute control signals to the master APD boxes which will then pass control information around the readout rings.

*7.5.3. Timing System:* The synchronous readout of data from the detector in the system proposed here requires distribution to the APD boxes of (a) 2 MHz clock, (b) a 1 pulse per second (PPS) signal to reset the hit timestamp counter and (c) a readout trigger ("spill") signal.

These signals are easily modulated onto a 10 MHz carrier frequency, so only a single pair of cables is needed to distribute them. The timing signals are centrally generated and fanned out to the master APD boxes. These boxes distribute the timing signals to all other APD boxes in the ring.

The clock and PPS signals would be locked to a GPS receiver, providing a stable, high-quality absolute time reference for the detector. In order to trigger a readout in time with a beam spill, the spill signal generated at the Main Injection must arrive at the central timing unit around 1 ms before the neutrinos arrive at the detector. A well-defined route for this signal is therefore necessary; either via a reliable, low-latency network connection from FNAL, or possibly via a dedicated radio link.

**Chapter 7 References**

# 8. Far Detector Site, Building, and ES&H Issues

## 8.1. Detector Site Criteria

We have chosen a location near Ash River, Minnesota as the NOvA Far Detector site. Ash River is about 810 km from Fermilab. We examined more than a dozen possible sites for the NOvA Far Detector as well as multiple detector locations within several particular sites. Possible sites begin ~710 km from Fermilab, near the city of Aurora MN, and continue to the north-northwest until a point in Ontario that is about 900 km from Fermilab. Sites more distant than ~900 km are too far off-axis to have desirable beam characteristics because of the beam's upward inclination of 3.3° and the curvature of the Earth. The sites we examined were all near the half-dozen or so east-west all-weather roads that cross the NuMI beamline.

Our principal site selection criteria were:
• Availability of land approximately 10-14 km (12 – 17 mrad) off-axis from the NuMI beam.
• As far as practical from Fermilab. A longer baseline is more sensitive to resolution of the mass hierarchy.
• A site with year-round road access at the maximum trunk highway weight limit, adequate electrical power and T-3 capable communications access. Other geographic criteria included access to workers, road transportation and airports and proximity to support services such as hotels, restaurants, gasoline and other retail outlets.
• A site with at least 20 and more likely 40 acres of usable land (not wetlands) which would permit a layout of a ~200 m by ~40 m footprint for a detector building oriented with its long axis pointing towards Fermilab.
• A site which would likely enjoy strong local support. The selection should not result in land use controversies or litigation. The characteristics of the site must also facilitate a straightforward environmental permitting process.

## 8.2. The Ash River Site Characteristics

The Ash River site is on the Ash River Trail (St. Louis County Highway 129) near the entrance to Voyageur's National Park. The site is west of the NuMI beam centerline and has the unique property of being the furthest site from Fermilab in the United States. See Figure 8.1.

Ash River is located about 15 km east of U.S. Highway 53, about 40 km east south east of International Falls MN. By car, it is about an hour drive from International Falls which is served by a Northwest Airlines affiliate from Minneapolis with four flights per day. By car, the site is about a 2 hour drive from the airport at Duluth and about a 4 hour drive from the Minneapolis airport. Driving time from Soudan to Ash River is about 1.5 hours.

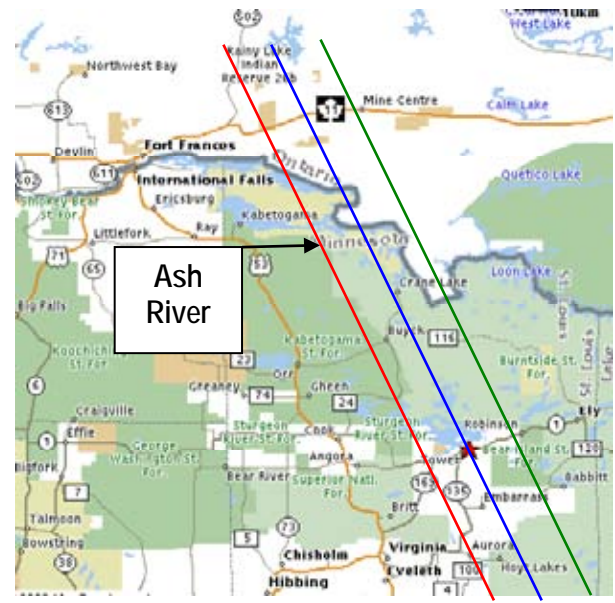

Fig. 8.1: Map showing our preferred site. The red star indicates the site of the Soudan Underground Laboratory. The beam centerline passes through Soudan; the left line is ~13 km (~ 16 mrad) west of the centerline, while the right line is ~13 km east of the centerline.

The actual detector laboratory locations at the Ash River Trail site are in Sections 12, 13 and 14 of Township 68 North, Range 15 West, St. Louis County MN. These locations are shown in Fig. 8.2 on the 1:24000 USGS topographic map. All locations would require upgrading of the access road, mostly with an improved gravel base and culverts for drainage (or a new road in the case of Site F).

The sites are located near Voyageur's National Park, but GIS studies by the National Park Service

suggest that the Detector Laboratory would be essentially invisible from the Park because of intervening high terrain (except for Site F). These sites are all ~810 km from Fermilab. The detailed parameters of all six locations are listed in Table 8.1.

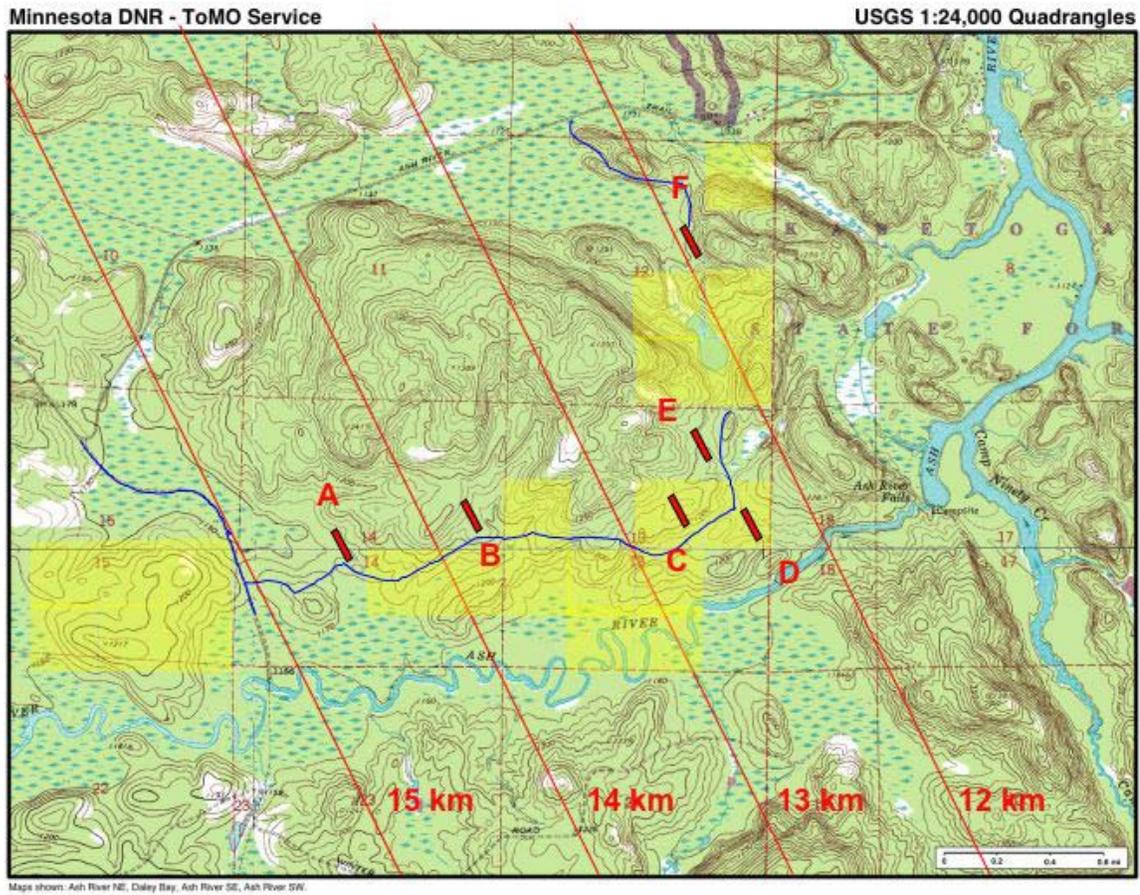

Fig. 8.2: The USGS topographic map for the Ash River Trail sites. The rectangles show a 200 m by 40 m laboratory footprint. The yellow shaded land near the laboratory sites belongs to Boise Cascade. Other land in this area belongs to the State of Minnesota.

| Location | Description | Latitude | Longitude | L (km) | T(km) | Angle (mr) | Ownership |
|---|---|---|---|---|---|---|---|
| A | SENW Sec. 14 | 48.375° | 92.869° | 811.4 | 14.37 | 17.6 | State |
| B | SENE Sec. 14 | 48.377° | 92.857° | 811.2 | 13.51 | 16.7 | State |
| C | SWNE Sec. 13 | 48.378° | 92.841° | 810.7 | 12.46 | 15.4 | BCC |
| D | SENE Sec. 13 | 48.377° | 92.836° | 810.5 | 12.19 | 15.0 | BCC |
| E | NWNE Sec. 13 | 48.381° | 92.840° | 811.0 | 12.26 | 15.1 | State |
| F | SWNE Sec. 12 | 48.391° | 92.840° | 812.0 | 11.81 | 14.5 | State |

Table 8.1: Parameters of Sites Near the Ash River Trail. The angles in the table are the full space angle relative to the beam which is about 4.2 km above ground at Ash River.

All detector locations shown in Figure 8.2 are on relatively flat land with few, if any, obvious rock outcrops. Thus, it is reasonable to believe that all of these sites have at least a few meters of soil cover over bedrock. Core drilling will be required to more completely characterize a chosen location. Most of the locations are forested with small aspen trees. In forestry terms, they are generally described as areas of aspen regeneration.

The access to the Ash River Trail site is via U.S. Highway 53, St. Louis County Highway 129 and then via a private road ~ 1-3 km in length, depending on the specific site that is chosen. Highway 129 has some weight restrictions that will necessitate some load rearrangements for ~ 45 days each Spring. There is an existing 7.2 kV, 3 phase power line that runs essentially along the highway. The local power company estimates that 500 kW is readily available with existing facilities; 1 MW or more of power consumption would require an upgrade of the current line. There is an existing fiber optic line along U.S. 53 and along the Ash River Trail.

The site would require installation of utilities along the access road. Domestic water would likely come from one or more wells, which might also be used to fill a storage tank for fire protection water if required (foam fire suppression systems are probably preferred for PVC and liquid scintillator). Domestic sewage would require either a septic system or a holding tank with periodic disposal.

The settlement of Ash River (U.S. Mail address: Orr MN 55771) is located at the end of the Ash River Trail, about 2 km east of the proposed detector site. This area has several motels and restaurants, although much of the activity is seasonal. (See www.ashriver.com for a listing of hotels and restaurants.) There is a new gas station and convenience store at the intersection of the Ash River Trail and U.S. 53, about 12 km from the laboratory site.

At this time, the University of Minnesota is taking preliminary steps towards land acquisition and environmental review of the Ash River sites. Although the University of Minnesota has authority to determine zoning and permitting with respect to its property within Minnesota, minimal land use controversy will facilitate the laboratory construction.

### 8.3. ES&H Issues

Recent exchanges [1] with the Fermilab Environment, Safety and Health Section have indicated that NOvA will need a DOE Environmental Assessment much like the one [2] done for NuMI and MINOS, and that three NOvA components will require special consideration. The first of these is the impact of a major discharge of liquid scintillator to the environment. While a credible scenario resulting in the discharge of the full 23,885 tons of liquid scintillator is difficult to imagine, the Fermilab ES&H Section advises that we design for containment of the full inventory. This will impact our building design in the next section.

The other two special ES&H issues for NOvA are the flammability of liquid scintillator and rigid PVC and the implied fire protection requirements. The Fermilab Fire Protection engineer has studied both liquid scintillator and PVC [3, 4]. The PVC was found not to ignite or become flammable or drip material when it was subjected to various ignition sources, even when it was covered with the liquid scintillator. The BC-517L tested could be ignited with a torch but was difficult to ignite with a low energy flame even when the liquid was pre-heated to $150\,^0F$. Additional testing will be done with the liquid scintillator under pressure to simulate the conditions of melting PVC forming pinholes and spraying the liquid scintillator onto surrounding surfaces.

The Fermilab Fire Protection Engineer advises that we plan a zoned fire suppression system of either dry chemical or non-alcohol foam. Fire fighting with normal water sprinklers is ineffective since the liquid scintillator has a density of 0.86 g/cc. In addition the runoff from fire fighting with water would have to be held in containment due to environmental concerns.

### 8.4. Building

The 30 kiloton NOvA Far Detector requires a detector enclosure ~170 m long by ~22 m wide by ~22 m high. This is a substan-tial structure, so we commissioned two design studies to get a handle on the costs and cost drivers for such large buildings. The first study was sponsored by the University of Minnesota and was performed by CNA Consulting Engineers with subcontracts to Dunham Associates and to Miller-Dunwiddie Architects [4]. This CNA study focused on a cut and cover approach deep in bedrock with a 10-meter overburden to cover a "worst-case scenario" of a possible required cosmic ray shield. As outlined in Chapter 10, we do not believe such an overburden is required.

The second study was done by the Fermilab Facilities Engineering Services Section [5] and focused instead on zero overburden. The Fermilab design was for buildings at any depth but with an above ground portion similar to experimental laboratory buildings at Fermilab. The minimum case has an excavation just down to bedrock to ensure the 30 kilotons is sitting on a solid surface. Bedrock at most of the sites considered above is expected to be under only 10 to 15 feet of soil till.

While the two building design studies had different goals, they did agree with each other in cost at the 20% level when the Fermilab design at the surface was compared to a similar surface design subsection of the Minnesota design done by Miller-Dunwiddie. In addition, both designs had common assumptions about the general site, for example including modest costs for short roads connecting to existing roads and modest cost to bring in nearby power. The Ash River Sites are a close match to these assumptions.

The secondary containment issue discussed in Section 8.4 led us to a building design with the floor level ~ 9 meters below grade. The building is ~ 5 meters wider than the detector and ~ 25 meters longer than the detector as shown in Figures 8.3 and 8.4. The detector therefore sits in a concrete bathtub which is sized to hold the entire inventory of liquid scintillator. This containment design is similar to that used for MiniBooNE at Fermilab. In addition we would paint the inside of the bathtub with epoxy-based paint to ensure the liquid scintillator cannot leave the building. This copies the recent retrofit efforts at the Gran Sasso laboratory. Interior grating covered gutters will direct small scintillator spills to a special sump. All floors and walkways would slope gently towards the gutters which in turn slope towards the sump. The sump(s) would have ejector pumps that transfer the spilled liquid into a separate holding tank.

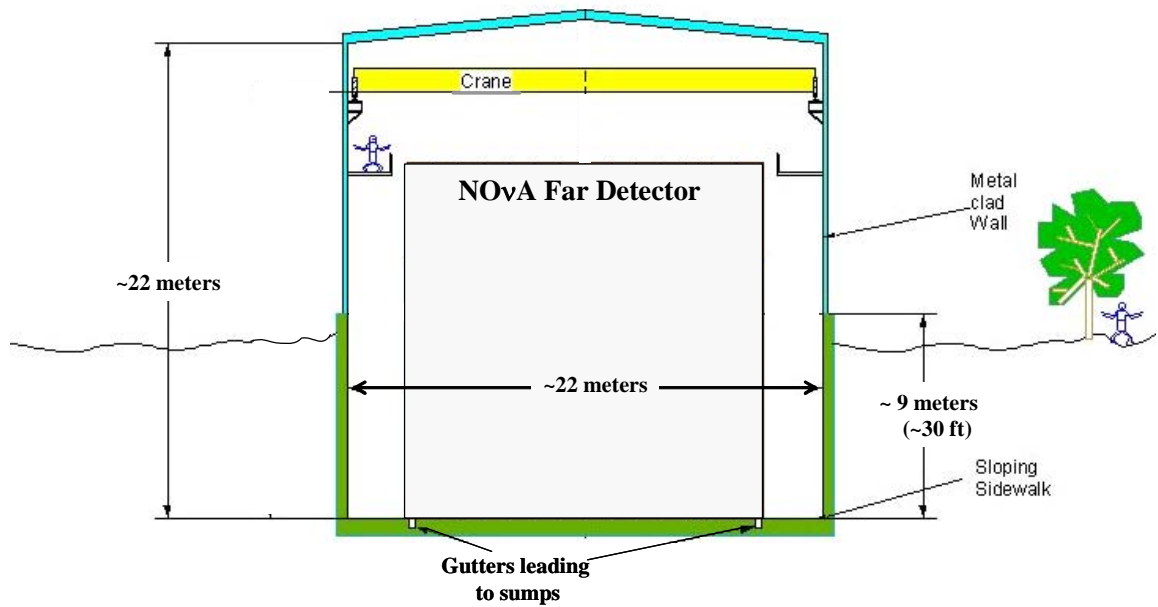

Fig. 8.3: Neutrino beam view of the NOvA Far Detector building. The green shaded portion is the concrete bathtub.

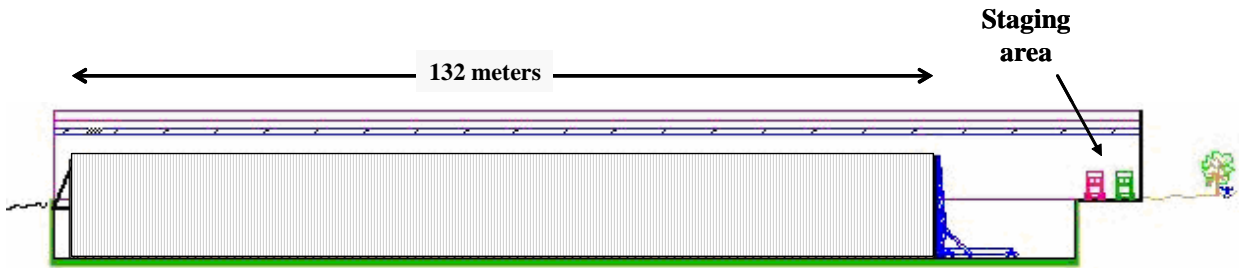

Fig. 8.4: Longitudinal cross section of the Far Detector building. The neutrino beam comes from the left. The green shaded area is the concrete bathtub. The staging area at grade level is shown. The block raiser described in Chapter 5 is shown in its position after the detector is completed.

For the detector design in this proposal, we anticipate constructing a 20-year life, metal-sided, metal-roofed building, similar to the usual experimental area buildings located at accelerator laboratories. The interior building walls would be sheet metal clad to deflect any potential scintillator leaks into the concrete bathtub. The building would have an additional 10-meter long staging and assembly area at grade at one end so that semi-trailers delivering PVC modules could be moved inside for unloading. This staging and assembly area does not require the full 22 m height. Pre-mixed liquid scintillator delivery and storage could be handled in the main building or in another low-roof section attached to the middle of one side of the main building with additional appropriate secondary containment.

The building meets the horizontal wind stress loads, snow loads, and heating and cooling loads required in northern Minnesota. This area sits on the Canadian Shield and is seismically stable, so no special earthquake design features are required.

The building would be insulated, heated and cooled to ~$20 \pm 10\ ^0C$ year round. This level of temperature control ensures that we avoid liquid scintillator oil temperature damage discussed in Chapter 6. A backup emergency heating source will be immune from electrical power failures (e.g., propane with a pilot light instead of electronic thermostat controls and electronic ignition systems). The building would be outfitted with a 5-ton building crane on a ~22-meter bridge. Catwalks below the crane would allow access to the top of the detector along its full length and could double as fall protection. The building would have several mobile scissor-lifts for access along the sides of the detector. Other custom assembly fixtures are described in Chapter 5. A small control room and a small technician work area would be included inside the main structure.

## Chapter 8 References

# 9. The NOνA Near Detector

## 9.1. Introduction

NOνA proposes to construct a Near Detector on the Fermilab site at a distance of about 1 km from the NuMI target in the NuMI access tunnel upstream of the MINOS access shaft. The primary Near Detector design requirement is that it should be as similar as possible to the Far Detector in material and segmentation. This requirement ensures that the efficiencies for signal and background events are identical and ideally will allow us to understand the $\nu_e$ charged current and $\nu$ neutral current beam spectra seen in the Near Detector as a measure of the expected backgrounds to $\nu_\mu \rightarrow \nu_e$ oscillation signals in the Far Detector.

This chapter describes a design based on the same PVC extrusions, the same PVC cell size, the same liquid scintillator, the same wavelength shifting fiber, and the same electronics readout as the Far Detector. Our design is influenced by the physical limitations at the Near site. The space restrictions in appropriate underground Near sites in the NuMI tunnels dictate a small Near Detector and the access to these underground sites through the MINOS shaft dictate a modular design.

A modular Near Detector has other advantages, and in particular, we propose to operate it in a Fermilab test beam and also in the MINOS Surface Building as venues to understand our detector response before the NOνA Far Detector is completed. The test beam can determine the absolute and relative response and energy calibration of the NOνA design. Running the Near Detector in the MINOS Surface Building at Fermilab allows us to easily study low energy neutrino interactions without the overhead of underground access and space restrictions.

## 9.2. Near Detector Location

The NuMI tunnels have several sites that could accommodate a Near Detector of similar construction to the Far Detector. Figure 9.1 shows the layout of the MINOS near-detector hall access tunnel. Starting at the Absorber Hall, on the left side of the figure, the tunnel makes a sharp turn to the west just downstream of the absorber. It continues parallel to the neutrino beam direction at a distance of ~14 meters from the beam axis for a distance of ~250 meters. Then it bends back east to enter the MINOS near detector hall, which is on the beam axis. This access tunnel geometry makes a wide range of off-axis angles accessible for a NOνA Near Detector. The range of sites is shown by three possible near detector locations in Figure 9.1 and in Table 9.1: just upstream of the MINOS near detector, just upstream of the vertical MINOS access shaft, and a third location just downstream of the NuMI hadron absorber. Chapter 10 discusses possible sites and concludes that it may be advantageous to move the Near Detector among several sites approximately midway between Site 1 and Site 2 (at ~ 12-17 mrad off-axis). Underground mobility of the detector will be a design requirement.

| Site | Number of milliradians off-axis |
|------|-------------------------------|
| 1 | ~ 4 |
| 2 | ~21 |
| 3 | ~26 |

Table 9.1: Off-axis angles of the three underground sites in Figure 9.1 as measured from the average pion decay location in the medium energy NuMI configuration, ~ 200 m downstream of NuMI Horn 1.

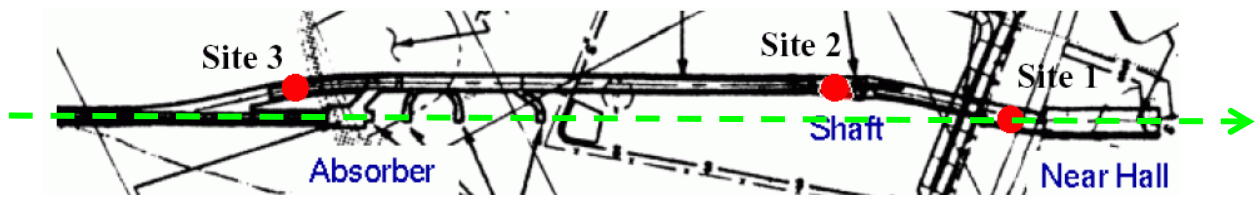

Fig. 9.1: Plan view of the NuMI access tunnel upstream of the MINOS near detector hall. The projection of the beam axis is from left to right along the dotted green line. The beam heads down at 58 mrad relative to the surface.

To reach these NuMI access tunnel sites, a near detector will have to be lowered underground via the MINOS shaft. Figure 9.2 shows a picture from the bottom of the shaft. The Shaft has a D-shaped cross section that is roughly a semicircle with a radius of about 3.3 meters. A 15 ton crane at the top of the shaft provides an additional constraint.

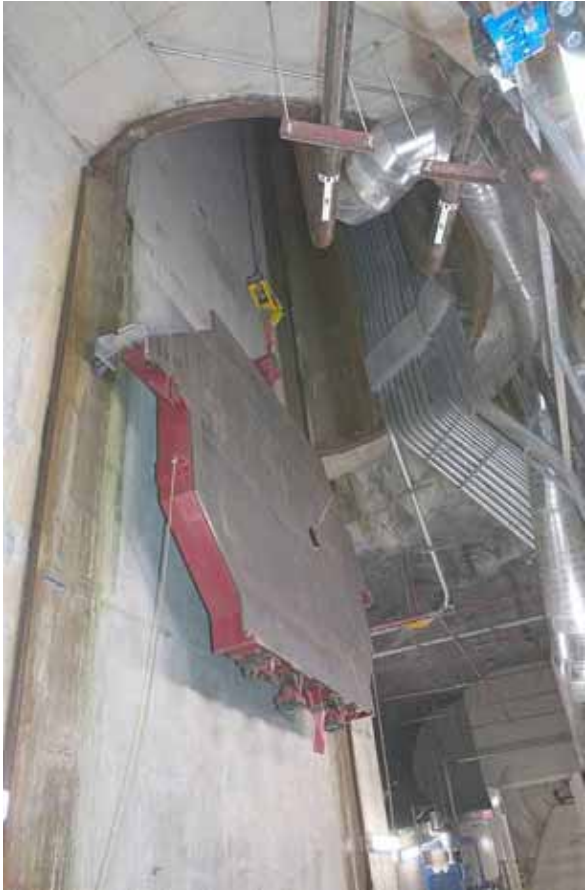

Fig. 9.2: View from the bottom of the MINOS shaft as a MINOS near detector plane comes down the shaft. The D-shaped shaft cross section is evident. The MINOS module shown is ~4.5 m wide by ~3.5 m high by ~0.2 m thick (including the red strong-back frame).

The transverse dimensions of the NuMI tunnels in all these locations are similar to those of Site 2, shown in Fig. 9.3. Each location has approximately 3.5 meters of useable width and about 5.0 meters of usable height. This width leaves about 1 meter for an access walkway around any object placed in the tunnel.

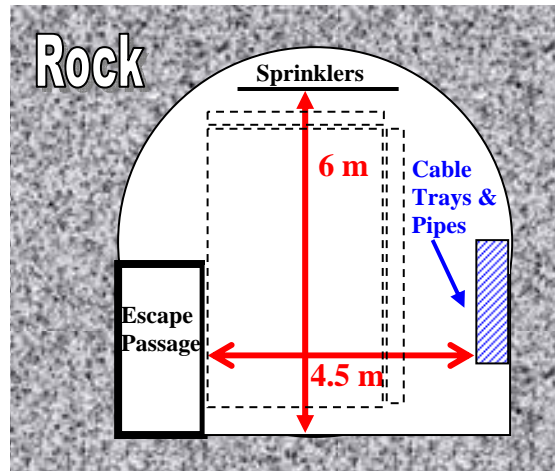

Fig. 9.3: A cross-section view of the access tunnel near Site 2 (see Fig. 9.1). The dotted outline shows the NOνA Near Detector cross section with the fiber manifolds sticking out the top and right side of the device.

### 9.3. Near Detector Design

With these space restrictions we have designed a NOνA Near Detector that is 3.5 m wide and 5.0 m high, indicated as the dotted outline in Figure 9.3. The active area is 3.25 m wide by 4.57 m high and the fiber manifolds plus electronics take up the additional space on the top and on one side of the detector.

The first 8 meters of the detector is composed of the exact same extrusion cells as in the Far Detector design. It is split into three logical parts: an upstream veto region, a fiducial event region, and a shower containment region. Figure 9.4 displays this longitudinal detector structure. The 4.75 m long shower containment length is chosen to fully contain electron showers from charged current $\nu_e$ interactions of a few GeV. The 8 meters of active detector sections are followed by a muon catcher composed of 1.0 meter of steel interspersed with additional planes of liquid scintillator cells. The length of the muon catcher is chosen to so that it plus the shower containment region will contain muons from charged current $\nu_\mu$ interactions of a few GeV.

The fiducial region is further divided transverse to the beam direction with a central 2.5 m by 3.25 m area designated as the fiducial area. This is illustrated in Figure 9.5. The border area is designed to contain the transverse size of electron showers in the few GeV region. The border is further subdivided with the outer 19 cm logically designated as an area where less than 5% of the total energy deposition will be allowed. Our simulations indicate that 96% of good $\nu_e$ events pass these criteria.

Altogether there are 130 planes of liquid scintillator cells, 65 planes with horizontal cells and 65 planes with vertical cells. The total mass of the detector is 262 tons with 145 tons totally active. The fiducial volume has a mass of 20.4 tons.

The detector would be constructed in modular packages 8 planes thick. Each module will be 10.6 tons when full of liquid and about 1.6 tons empty. The modules reasonably fit the MINOS access shaft constraints and could be moved full or empty. The Near Detector parameters are summarized in Table 9.2.

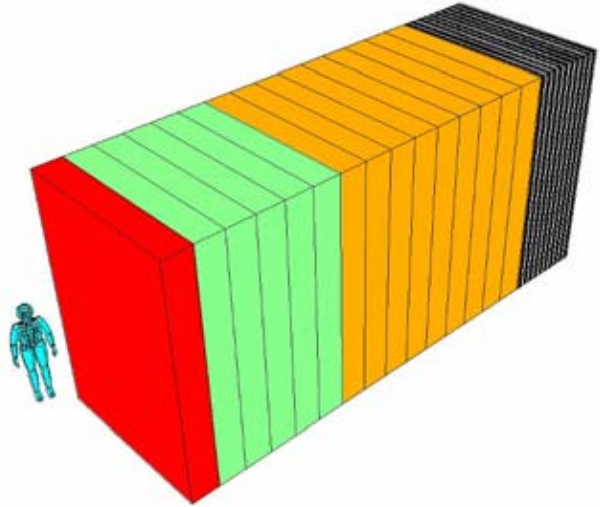

Fig. 9.4: The NOvA Near Detector. The beam comes from the lower left in this diagram. Each modular piece consists of 8 planes of extrusions, 4 vertical interleaved with 4 horizontal planes. The upstream module is a veto region (red), the next 5 modules are the fiducial region (green), and these are followed by a 9 module shower containment region (orange). All parts of these three sections are fully active liquid scintillator cells identical to the Far Detector and the colored areas just represent a logical assignment. Downstream of this active region is a 1.7 m muon catcher region of steel interspersed with 10 active planes of liquid scintillator (black and white).

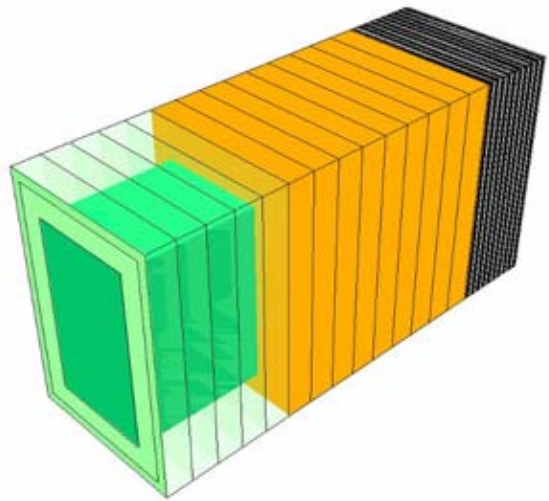

Fig. 9.5: The NOvA Near Detector with the front veto region removed, showing the detector fiducial region (dark green). The fiducial region is surrounded by a border area (lighter shades of green) to contain the transverse size of electron showers in the few GeV region. We would veto events with more than 5% of total energy in the outermost border region (lightest green).

| Near Detector Parameter | Parameter Value |
|---|---|
| Total mass | 262 metric tons |
| Active detector mass | 145 metric tons |
| Fiducial mass | 20.4 metric tons |
| Extrusion cells, liquid scintillator, waveshifting fiber, APD readout | Identical to the NOνA Far Detector |
| Number of channels | 12,480 |
| total Liquid Scintillator | 41,000 gallons |
| Detector<br>  Width (m and # of cells),<br>  Height (m and # of cells,<br>  length (m) | 3.5 m, 80 cells<br>4.8 m, 112 cells<br>9.58 m |
| Total active planes | 130 planes<br>   65 horizontal &<br>   65 vertical |
| Basic modular piece in the active section<br>  # planes<br>  Thickness of module<br>  Empty weight<br>  Full weight | <br>8 planes<br>52.8 cm<br>1,417 kg<br>9,600 kg |
| Veto region,<br>  # of active planes | 8 planes |
| Fiducial region,<br>  # of active planes | 40 planes |
| Shower Containment region,<br>  # of active planes | 72 planes |
| Muon catcher<br>  Steel (m/section,<br>     # of sections)<br>  # of active planes | <br>0.1 m,<br>10 sections<br>10 planes |
| Muon catcher mass<br>  Steel<br>  Scintillator planes | <br>117.5 metric tons<br>11.1 metric tons |

Table 9.2: NOνA Near Detector Parameters.

## 9.3. Near Detector Event Rates

At a site midway between Sites 1 and 2 in Figure 9.1, the event rates in the 20.4 ton fiducial mass will be about 0.09 event per $10^{13}$ protons on the NuMI target. The rate drops about a factor of three near Site 3 and increases about a factor of three near Site 1. The maximum beam from a single Main Injector (MI) pulse is expected [1] to be $6 \times 10^{13}$ protons, so we would get about 0.5 events per MI spill. As we will see in Chapter 10, about two-thirds of these events would be from neutrinos with energies below 5 GeV. We would collect about 6.5 million such events in one year with 6.5 $\times 10^{20}$ p.o.t (see Chapter 11).

The rate of events in the whole active detector is larger. Since the total active mass in 145 tons, we would see a rate of 3.8 events per MI pulse of $6 \times 10^{13}$ protons. Assuming a 500 ns time bin in our electronics and a 10 microsecond spill [2], that would imply 9% of our events would have two or more overlapping events in the active detector. We therefore expect to include an additional 6 planes interspersed throughout the 120 active planes, each with ganged fast MINOS-style Near Detector electronics to identify the presence of more than one event in a spill. Such spills could then be cleanly rejected in an unbiased manner.

## 9.4. Test Beam Program

Given the modular form of the Near Detector, it can be moved to various sites relatively easily. We plan a program to expose the detector to a charged-particle test beam. Using selectable beam momentum settings and a particle identification system, a full response matrix can be measured. Response to cosmic ray muons will be studied simultaneously and compared to these beam interactions. Both the Far Detector and the Near Detector will always see a high rate of out-of-time cosmic ray tracks, which will provide a stable source of muons to monitor the detector response. The collected test beam data will also be used to tune Monte Carlo simulations of the detector response and to aid in developing the most efficient pattern recognition algorithms.

NOνA does not have any unusual demands for the performance of a test beam. However, the beam should have a momentum range from well below 1 GeV up to 5 GeV/c, with the absolute momentum known to a few percent, and an integrated particle identification system. The Fermilab Meson Test Beam Facility in MTest could be used even though it has rather low beam rates at these low energies [3]. Initially we imagine testing prototype Near Detector modules in MTest, but eventually we could put the entire NOνA Near Detector in MTest as shown in Figure 9.5. Since the shower containment and muon catcher regions would be a little thin at higher energies, we would add modules from the prototype Near Detector discussed in Chapter 15 for these measurements.

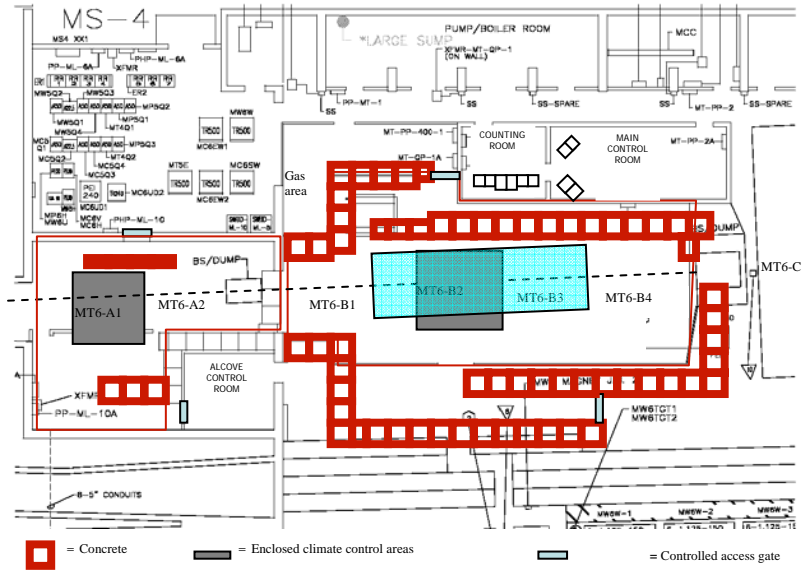

Fig. 9.5: The Fermilab MTest experimental area with the NOνA Near Detector (in blue) superimposed.

Over the life of the experiment we may need access to the test beam for several periods of a few months each. We would rely on Fermilab support for MTest beam line operation, instrumentation, and monitoring.

The Near Detector described here is essentially sampling the Far Detector in the upper right corner as seen by the neutrino beam as shown in Figure 9.6. We could replace the fibers in the Near Detector with 15.7 meter long fibers (coiled up outside the liquid cells) and calibrate / study a different part of our Far Detector. Other positions could be calibrated in the same manner.

### 9.5. Tests of the Near Detector in the MINOS Surface Building at Fermilab

We also plan to put the NOνA Near Detector in the MINOS Surface Building and look at extremely off-axis neutrinos from the NuMI beam. The surface building is about 75 mrad off-axis and the NuMI beam runs parallel to the axis of the building's highbay. The NOνA Near Detector fits easily in the highbay area but would not block access to the MINOS shaft for other users. This is shown in Figure 9.6.

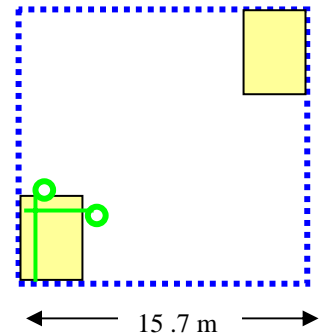

Fig. 9.6: The Far Detector front face (dotted blue line) with two configurations of the Near Detector superimposed. For the lower left version, the extra wavelength shifting fiber gets coiled up outside the Near Detector as represented by the green circles.

The predicted $\nu_\mu$ spectrum in the MINOS Surface Building is shown in Figure 9.7. The main feature is a $\nu_\mu$ beam strongly peaked near 2.8 GeV. These neutrinos are from kaon decays in the NuMI beam [4]. In addition there is a nice $\nu_e$ spectrum which peaks at 1.8 GeV as shown in Figure 9.8.

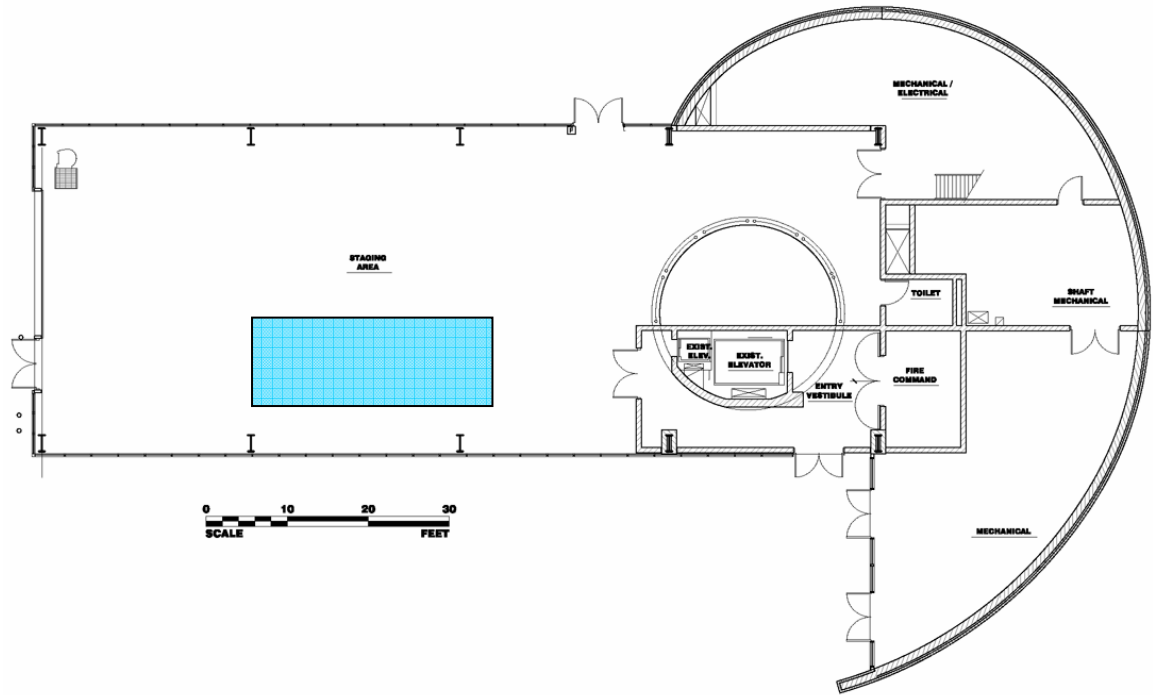

Fig. 9.6: The 9.2 m long NOvA Near Detector (in blue) shown to scale inside the MINOS Surface Building. The NuMI beam runs parallel to the main axis of the building from left to right (but ~105m below the building and ~14m towards the bottom of the figure as shown). The beam is also heading down at 58 mrad relative to the surface.

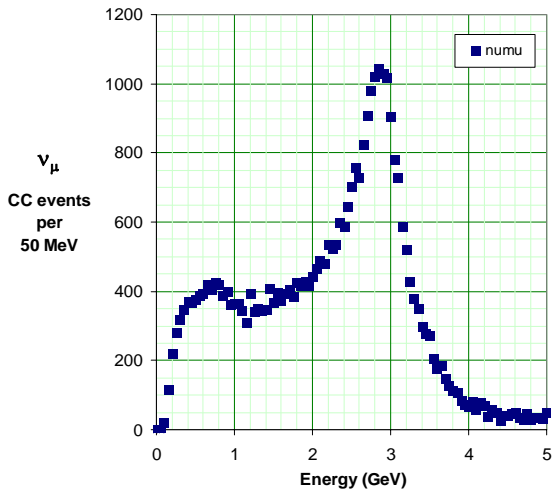

Fig. 9.7: Charged Current $\nu_\mu$ event spectra vs. energy for neutrino events in the 20.4 ton fiducial mass of the NOvA Near Detector placed the MINOS Surface Building for $6.5 \times 10^{20}$ p.o.t.

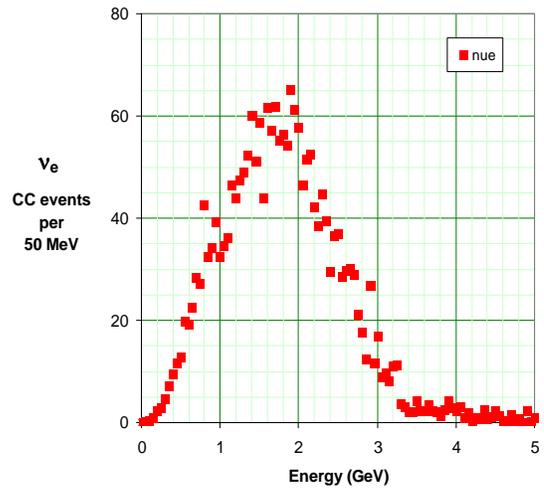

Fig. 9.8: Charged Current $\nu_e$ event spectra vs. energy for neutrino events in the 20.4 ton fiducial mass of the NOvA Near Detector placed the MINOS Surface Building for $6.5 \times 10^{20}$ p.o.t.

The $\nu_e$ over $\nu_\mu$ ratio is 10 - 15% in the 1 – 2 GeV range of Figures 9.7 and 9.8., making the target population of $\nu_e$ rather accessible in any event sample.

The event rates at 75 mrad off-axis are down by a factor of 150 relative to the underground off-axis sites in the NuMI tunnels of Figure 9.1. The total event rate in our 20.4 ton fiducial volume will The be only about 1 event every 5 minutes with the Main Injector operating at 6 x $10^{13}$ p.o.t. per pulse. Still, integrated over a year with 6.5 x $10^{20}$ p.o.t. delivered to NuMI, we would see about 45,000 charged current neutrino events. The event rates are summarized in Table 9.3.

| Event Type | Event Rates |
|---|---|
| Total CC events per MI pulse | 0.0044 |
| MI pulses between CC events | ~ 230 |
| Average time between events | 5 - 6 minutes |
| | |
| | Event Samples |
| CC neutrino events per year | ~45,000 |
| $\nu_\mu$ charged current events per year in the kaon peak between 2.4 and 3.2 GeV | ~ 13,000 |
| $\nu_e$ charged current events per year | ~ 2,200 |

Table 9.3: Event rates for 6 x $10^{13}$ p.o.t. per MI pulse and Event samples for 6.5 x $10^{20}$ integrated p.o.t. delivered to NuMI. These rates and samples are those expected in the 20.4 ton fiducial volume of the NOνA Near Detector at the MINOS Surface Building.

There are clear advantages to starting with the NOνA Near Detector in this position.
- The building already exists and can easily hold the detector as shown in Figure 9.6.
- We can initially avoid the complications of underground access.
- The event rates are low enough that initially we would not need special fast electronics to reject multiple events in a single MI spill.
- The event rates are high enough that substantial numbers can be collected in a year of running.
- In particular, reasonably large samples of $\nu_e$ charged current events will be available to aid in developing more efficient pattern recognition algorithms.
- We can gain experience with a fully active detector running on the surface.

We propose to start immediately with an R&D prototype NOνA Near Detector taking data as soon as possible in the MINOS Surface Building. This is discussed in Chapter 15. During the construction project we would build the final Near Detector from identical objects to those in the NOνA Far Detector.

## Chapter 9 References

[1] draft Fermilab Proton Plan, November 2004, see
http://www.fnal.gov/directorate/program_planning/Nov2004PACPublic/PACagendaNov2004OPEN.htm
[2] The spill length is 9.78 microseconds for the case of no anti-proton operation discussed in Chapter 11. See the NuMI Technical Design Handbook, Chapter 3, "Design Parameters" at http://www-numi.fnal.gov/numwork/tdh/tdh_index.html
[3] The Fermilab Meson Test Beam Facility is described at http://www-ppd.fnal.gov/MTBF-w , the beam has recently been tuned as low as 4 GeV and the yields were about 100 particles per MI spill with 50% of the particles being electrons, E. Ramberg, private communication.
[4] We thank the MiniBooNE collaboration for pointing out this far off-axis effect. Janet Conrad, private communication.

# 10. Backgrounds and Systematics

## 10.1. Introduction

In Chapter 9 we describe a NOvA Near Detector that is virtually identical to our proposed Far Detector. This ensures that the efficiencies for signal and background events are nearly identical in the two detectors. If there were no other effects, then understanding the un-oscillated beam spectra seen in the Near Detector would be a perfect measure of the expected background to $\nu_\mu \rightarrow \nu_e$ oscillation signals in the Far Detector. Unfortunately this simple relation can break down in several ways, leading to incorrect conclusions about the background in the Far Detector and therefore leading to systematic effects in our search for $\nu_\mu \rightarrow \nu_e$ oscillations.

In this chapter we examine several effects which can alter the Near to Far extrapolation and affect our primary measurement. We have not yet fully simulated these effects, but it is important to recognize each one at this proposal stage and indicate our strategies for dealing with each. We find that moving our Near Detector around to different positions in the NuMI tunnel should allow us to understand all these effects sufficiently well such that the total error on the background in our $\nu_\mu \rightarrow \nu_e$ search will be below 10%. The modular nature of the NOvA Near Detector described in Chapter 9 is therefore an important design aspect of the device, since it will be moved several times during the experiment.

NOvA aims to detect excess $\nu_e$ events in the Far Detector. Our principle backgrounds are beam $\nu_e$, beam $\nu_\mu$, neutral current (NC) $\nu$ events, and cosmic rays, each of which can masquerade as $\nu_e$ oscillations. We use the off-axis beam to obtain a nearly monochromatic neutrino energy spectrum and then one of our principle analysis weapons in the Far Detector is an energy cut to eliminate backgrounds. Systematics influencing our result can therefore be tied to energy-dependent effects.

In this chapter we will consider the following list of effects:

- Energy-dependent backgrounds require that we understand the energy calibration of our Near and Far Detectors. The inherent energy resolution of our detectors sets a scale for the precision of this calibration. We will use test beam and other data to set the absolute energy scale to a few percent.
- Predictions of the neutrino beam spectra at the Near and Far sites involve rather simple kinematics in a Monte Carlo simulation, but the input particle production spectra for this Monte Carlo are only known to about the 20% level. We expect Fermilab E-907 MIPP (Main Injector Particle Production) [1] to improve this knowledge before NOvA takes data. Several members of the NOvA Collaboration are also members of the MIPP Collaboration and will have first-hand experience with these improved data.
- The neutrino beam spectra at the Near and Far Detectors cannot be identical because the Near Detector sees a line source from decays only a few hundred meters away while the Far Detector sees a point source from 810 km. The off-axis angle of our detectors makes this situation different from that in MINOS. We will optimize the location of our Near Detector to minimize this effect. The optimization requires moving the Near Detector to different positions for each background.
- A potential MiniBooNE confirmation of the LSND result for short baselines may mean that our Near Detector will see a distorted $\nu_e$ component to the NuMI beam. The NOvA Far Detector would not see the same effect due to its long baseline. If the LSND signal were confirmed, we would have to respond by moving our Near Detector around to different off-axis angles so that we could disentangle the NuMI beam spectra effects from the short baseline oscillation effects.
- The Far Detector $\nu_e$ background from $\nu_\mu$ CC events masquerading as $\nu_e$ comes from the <u>oscillated</u> $\nu_\mu$ spectrum. Therefore any measurement of this background with the un-oscillated beam in the Near Detector will not be quite correct. We can estimate this effect by studying $\nu_\mu$ events with an identified muon as a function of the observed event energy. Different $\nu_\mu$ CC spectra can be seen by the NOvA Near Detector at different off-axis positions.

- To first order the present rather poor knowledge of low energy neutrino cross sections [2] does not matter to NOvA since the same unknown cross section is seen in both the Near and Far Detectors.

  However, since the neutrino beam spectra are not identical at the NOvA near and far sites, some differences in background levels can result from different energy dependences and from different NOvA efficiencies for detection of the different neutrino interaction processes. To second order we can minimize these effects by selecting the best Near site to measure each background.

  Knowledge of the low energy neutrino cross sections will be much improved before the bulk of NOvA data is collected, since Fermilab E-938 MINERvA [3] is designed to attack this very problem. Several members of the NOvA Collaboration are also members of the MINERvA Collaboration and will have direct knowledge of these improved cross section data at a detailed level. Benchmarking our Near Detector measurements against these improved cross section data will help us understand our detector response to neutrinos.

- Our Near Detector will be underground and shielded from cosmic rays, while our Far Detector will be on the surface and unshielded from cosmic ray events that occur within the neutrino beam time window (about 100 seconds per year of running). Monte Carlo calculations indicate this should not be a problem, but we will run our Near Detector on the surface to check this Monte Carlo simulation.

## 10.2. Energy Calibration of the NOvA Detectors

In Chapter 9 we described our plans to calibrate the NOvA Near Detector in a test beam to determine the absolute response and energy calibration of both the Near and Far detectors. As described in Chapter 12, we expect our detector energy resolution to be $\Delta E/E$ (sigma) $\sim 0.10 / \sqrt{E}$ for $\nu_e$ CC events. For a 2 GeV $\nu_e$ event at the peak energy of our oscillation signal, we expect to measure the event energy to about 7%. This 7% resolution does set a rough practical scale to our requirements for understanding shapes of the neutrino energy spectra in both detectors.

Momentum tagged electrons in a test beam will allow us to measure our energy resolution and absolute energy scale directly. With a detector electromagnetic resolution of order $0.10 / \sqrt{E}$, the absolute scale can be determined to a few percent providing the test beam momentum resolution is not the dominant effect. The Fermilab MTest test beam has a momentum bite of a few percent, and the last string of dipoles is instrumented with a tracking system which allows momentum tagging at the $\Delta p/p = 0.25\%$ level. In practice multiple scattering off material in the beam at low energies will limit this tagging ability. Understanding the absolute momentum scale in MTest requires a field map of the final 5 dipoles and a precision shunt resistor on the magnet power supply to monitor that field. Overall, a 1 - 2% absolute calibration of the beam should be attainable, and in turn we should establish the absolute energy scale of our Near Detector to 2 – 3%.

Since high $y$ $\nu_e$ events can contain multiple charged pions with a substantial fraction of the energy, we will also want to use the MTest charged pion beam to understand our detector response. For $\nu_\mu$ events, the NOvA detector energy resolution comes from the total pulse height of the muon track, since the muon range is subject to straggling at the 2-3% level. Running with muons in MTest will allow us to study this resolution directly.

We intend to carry the absolute test beam calibration of the NOvA Near Detector to the Far Detector and through the life of the experiment by constantly monitoring each detector's pulse height response to cosmic ray muons in individual cells. Cross calibration with cosmic ray muons in the test beam will initiate this energy scale tracking.

The extreme off-axis neutrino flux in the MINOS Surface Building provides another calibration path. Figures 9.7 and 9.8 in Chapter 9 illustrate the neutrino CC flux at ~75 mrad off-axis. The $\nu_\mu$ CC peak at 3 GeV is dominated by neutrinos from K decays, and the kinematics of $K \rightarrow \mu \nu_\mu$ vs. $K_{e3}$ decays will allow us to cross correlate the $\nu_\mu$ and $\nu_e$ energy distributions. The 3 GeV peak energy measurement is dominated by muon energy and that provides an absolute calibration of the electron energy scale near 2 GeV. The relative branching ratios for the two decays provide another cross check on the spectra. This measurement will require a longer Near Detector than the one described in Chapter 9 and we would supplement the detector with parts of the prototype Near Detector described in Chapter 15.

## 10.3. Particle Production at 120 GeV and the Neutrino Beam Spectra

Existing particle production data have errors of order 20% and this translates into a ~20% uncertainty [1] in the neutrino flux produced in NuMI with the Medium Energy configuration NOνA proposes to use. The MIPP experiment intends [4] to run with 120 GeV protons to directly measure the flux from the MINOS target. These data should reduce the particle production uncertainties to the level of a few percent. With these data the NOνA neutrino fluxes should then be predicted to about 5% when combined with NuMI horn magnetic field measurements.

The MIPP measurements are clearly important to NOνA since an improved beam Monte Carlo will help us understand any Near to Far differences in the neutrino spectra. We hope that MIPP can get the appropriate MINOS target data as planned during 2005.

## 10.4. Near Detector Location and the Neutrino Beam Spectra

NOνA plans to use the NuMI beam in the medium energy configuration. In this section we compare the neutrino beam spectra at several possible Near Detector locations with the spectra at the Far Detector site. Figure 10.1 reproduces Figure 9.1 to again show the possible Near Detector locations. Figures 10.2 and 10.3 show the charged current (CC) $\nu_\mu$ event rates for several locations in the NuMI halls along with the unoscillated $\nu_\mu$ CC rate expected in the Far Detector. The actual Far Detector $\nu_\mu$ flux will be quite different from Figure 10.2 due to $\nu_\mu \rightarrow \nu_\tau$ oscillations, but the figure does show the shape of the total neutrino flux and particularly the flux of the neutral currents (NC) which are a background source for NOνA. Matching the un-oscillated Far Detector $\nu_\mu$ CC spectrum shape in the Near Detector will be important for our understanding of the NC backgrounds.

There are two main differences between the muon neutrino spectra at these sites and the spectrum at the far detector. The peak is broader at the near site than at the far site, and the "high energy tail" is a larger fraction of the total event sample in the near detector. The energy spectrum is widened at the near detector because there is a broad range of decay positions of the parent pions, so there is no single "off axis angle" seen at one position. At the Far Detector, the range of decay locations has a negligible effect on the off-axis angle. At the Near Detector, the high-energy tail is fractionally higher because these events come from the high-energy pions that decay farthest downstream in the decay pipe. Those high-energy decays are significantly closer to the Near Detector than the decays of the pions that give events in the peak of the distribution and they are at a larger off-axis angle. The prominent kaon decay peak seen in Chapter 9 for the MINOS Surface Building site is also beginning to be visible at Sites 2 and 3 around 8 – 10 GeV.

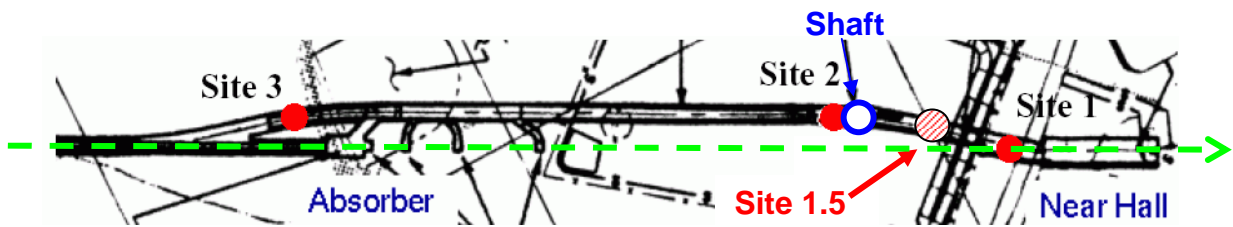

Fig. 10.1: The NuMI access tunnel upstream of the MINOS near detector hall. The beam direction is from left to right as shown by the dashed green line. Our preferred Near Detector sites are in the range between Site 1.5 and Site 2. Site 1.5 is at an off-axis angle of ~ 12 mrad as measured from the average pion decay location in the medium energy NuMI configuration, ~ 200 m downstream of NuMI Horn 1. Site 2 is at ~ 21 mrad.

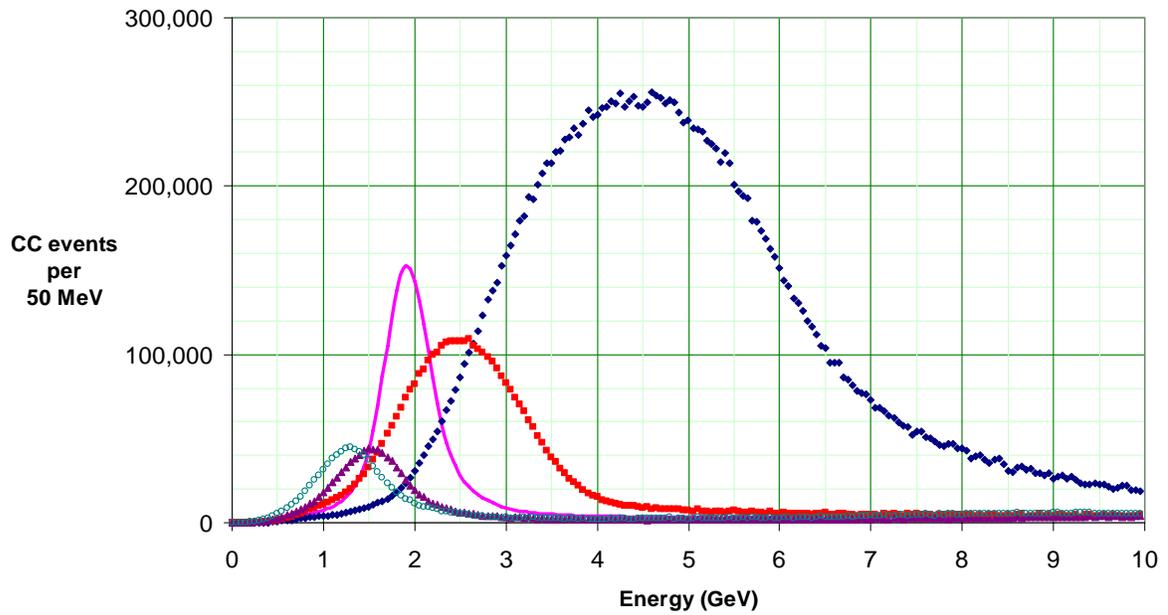

Fig. 10.2: $\nu_\mu$ CC spectra for the various Near Sites [Site 1(blue diamonds), Site 1.5(red squares), Site 2 (purple triangles), Site 3 (open green circles)] for one year of running at 6.5 $10^{20}$ pot. The un-oscillated Far Detector $\nu_\mu$ spectrum for one year of running (times an arbitrary scale factor of 800) is shown as the solid pink line. These spectra are for the NuMI medium energy configuration.

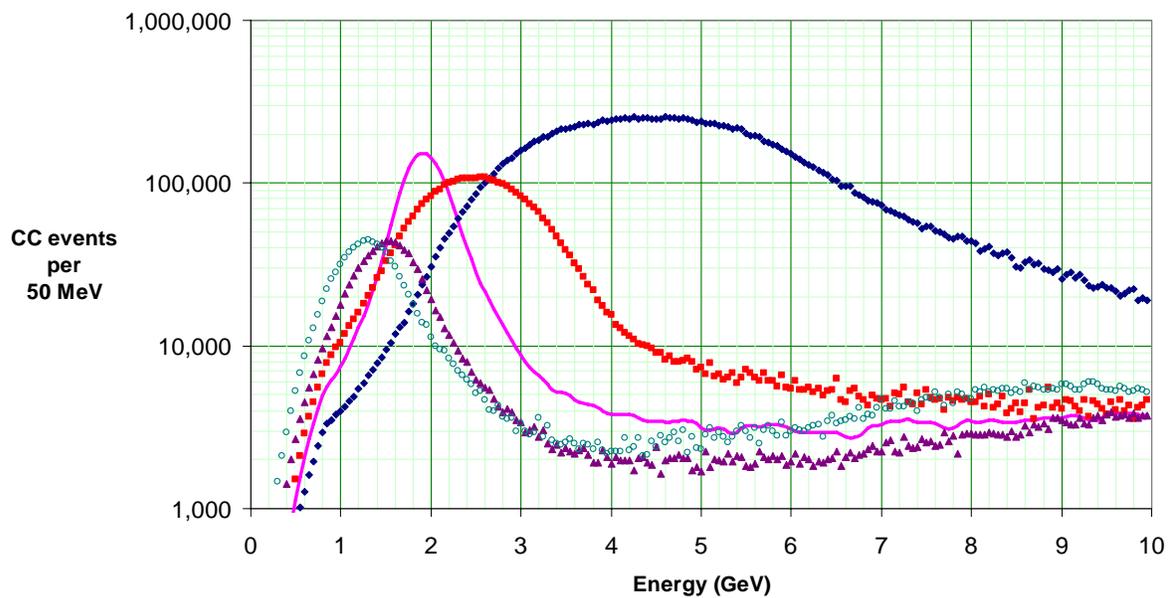

Fig. 10.3: Same $\nu_\mu$ CC spectra as shown in Figure 10.2 but with a logarithmic vertical scale to better display the high energy tails of the distributions.

Based on the total ν flux spectrum comparison alone, the preferred Near Detector location would be approximately midway between Site 1.5 and Site 2, where the best match is made to the NC background seen at the Far Detector.

Figures 10.4 through 10.7 show the beam $\nu_e$ spectra for these same sites, again for the NuMI medium energy configuration. In each figure the beam $\nu_e$ spectrum for the Far Detector is superimposed with an arbitrary normalization so that the two distributions agree at ~2 GeV. Based on these beam $\nu_e$ spectra comparison alone, the preferred Near Detector location would be Site 1.5. Over the 1.5 – 2.5 GeV energy range, the Near and Far distributions agree to within ~7% in every bin.

One cannot optimize for both the electron and total neutrino fluxes at the same time. The electron neutrinos come predominantly from the muon decays farther downstream in the decay pipe while the muon neutrinos, which make up 99% of the total flux, originate from somewhat farther upstream. Site 1.5 gives electron neutrino spectra reasonably similar to those at the Far Detector. A site midway between Site 1.5 and Site 2 would be a better match to the total neutrino spectrum at the Far Detector site. The solution is to move our Near Detector to the appropriate site for each background study.

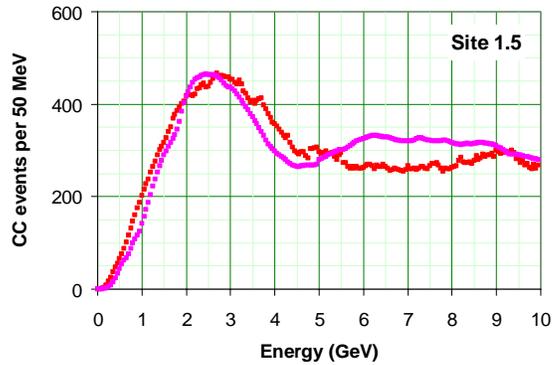

Fig. 10.5: The beam $\nu_e$ event rates for one year of data in the Near Detector located at Site 1.5 (red squares). The Far Detector beam $\nu_e$ distribution (pink line) is also shown assuming no oscillation, but has been normalized to have the same value at ~2 GeV.

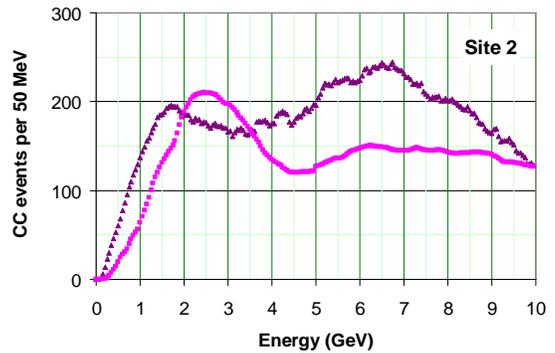

Fig. 10.6: The beam $\nu_e$ event rates for one year of data in the Near Detector located at Site 2 (purple triangles). The Far Detector beam $\nu_e$ distribution (pink line) is also shown assuming no oscillation, but has been normalized to have the same value at ~2 GeV.

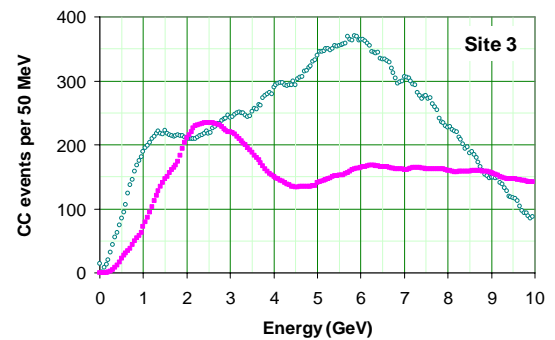

Fig. 10.7: The beam $\nu_e$ event rates for one year of data in the Near Detector located at Site 3 (open green circles). The Far Detector beam $\nu_e$ distribution (pink line) is also shown assuming no oscillation, but has been normalized to have the same value at ~2 GeV.

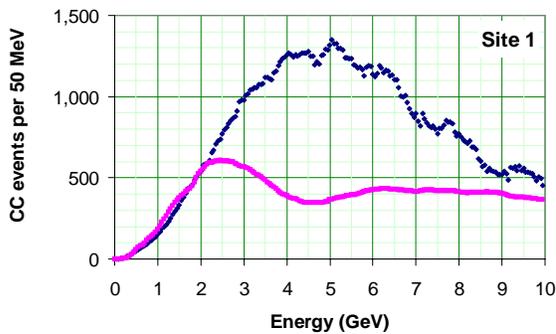

Fig. 10.4: The beam $\nu_e$ event rates for one year of data in the Near Detector located at Site 1 (solid blue diamonds). The Far Detector beam $\nu_e$ distribution is also shown (solid pink line) assuming no oscillation, but has been normalized so that the distributions have the same value at ~2 GeV. Both distributions assume the NuMI medium energy configuration.

## 10.5. The Effect of a Possible MiniBooNE Confirmation of LSND

A MiniBooNE confirmation of the LSND result for short baselines would mean that our Near Detector will see a distorted $\nu_e$ component to the NuMI beam. LSND reported [5] an anti-$\nu_\mu \rightarrow$ anti-$\nu_e$ oscillation probability of 2.6 x $10^{-3}$. In the naive oscillation framework given by

$$P_{ab} = \sin^2(2\theta_{ab}) \sin^2[1.27 (\Delta m^2) (L/E)], \quad (1)$$

the allowed LSND parameter space is shown in Figure 10.8.

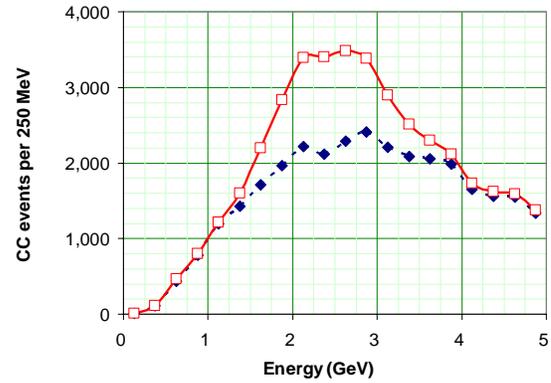

Fig. 10.9: The NOνA Near Detector $\nu_e$ charged current spectrum with (open red squares) and without (solid blue diamonds) the effect of an LSND short baseline oscillation with $\Delta m^2 = 2.5$ eV$^2$ and $\sin^2(2\theta_{\mu e}) = 2.6$ x $10^{-3}$. These spectra are for one year of running at 6.5 x $10^{20}$ pot in the NuMI medium energy beam. No detector resolution effects, $\nu_e$ CC efficiencies, or NC backgrounds are included here.

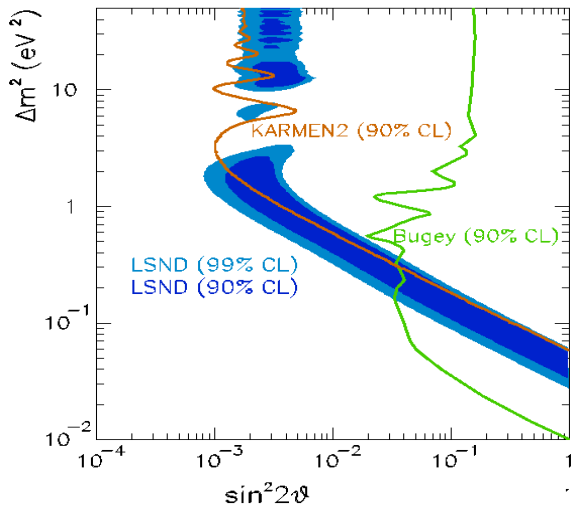

Fig. 10.8: The final LSND allowed region (90, 99%CL), together with the final KARMEN2 90%CL excluded region (Feldman-Cousins approach), and the 90%CL Bugey excluded region. This image comes from the MiniBooNE Public Plots web area.

Figure 10.9 shows a 60% effect in the Near Detector $\nu_e$ spectrum for the case of an LSND signal at 2.5 eV$^2$ and $\sin^2(2\theta_{\mu e}) = 2.6$ x $10^{-3}$ (center of the LSND range, but excluded by KARMEN2 at 90% confidence level). Figure 10.9 shows an extreme case, but Figure 10.10 shows the effect persists for all values of $\Delta m^2$ and $\sin^2(2\theta_{\mu e})$ consistent with the LSND, KARMEN2, and Bugey results. There is always an effect at the level of 20% or more in the measured "beam $\nu_e$" spectrum amounting to ~100 events per 50 MeV bin over a wide energy range.

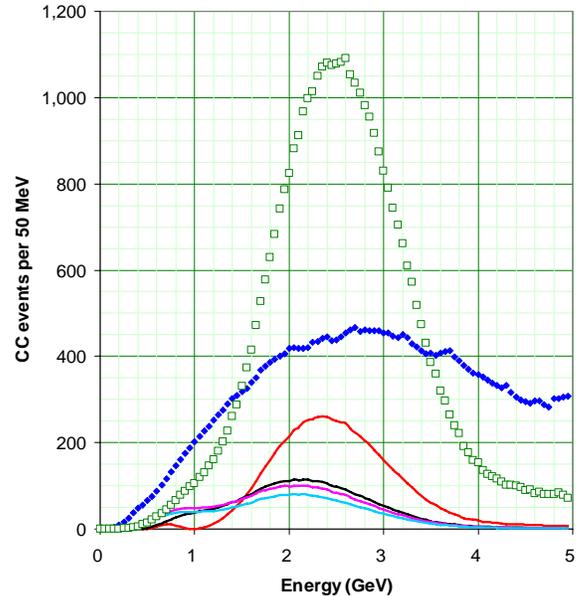

Fig. 10.10: The NOνA Near Detector $\nu_e$ CC spectrum (solid blue diamonds) at Site 1.5 compared with a short baseline oscillation effect at different values of $\Delta m^2$ (four solid lines for 2.5, 1.0, 0.4, and 0.25 eV$^2$) consistent with the allowed LSND parameter space in Figure 10.8. The four lines have ($\Delta m^2$, $\sin^2(2\theta_{\mu e})$ ) parameters of (2.5, 2.6 x$10^{-3}$, red line), (1.0, 4 x$10^{-3}$, black line), (0.4, 2 x$10^{-2}$, pink line), and (0.25, 4 x$10^{-2}$, light blue line). The parent $\nu_\mu$ CC spectrum for the oscillation is shown at 1% of its value (open green squares). No detector resolution effects, $\nu_e$ CC efficiencies, or NC misidentification backgrounds are included here.

The effect in the NOνA Near Detector depends on $\Delta m^2$ and L/E, but it is also a strong function of the parent $\nu_\mu$ spectrum in the off-axis near beam. The Near and Far NOνA detectors would see different effects from this short baseline oscillation. For a large $\Delta m^2 = 2.5$ eV$^2$, the Far Detector sees a small effect as the second factor in Equation (1) just averages to 0.5 and the overall value for $P_{ab}$ goes to 0.0013. For small $\Delta m^2 \sim 0.3$ eV$^2$, the Far Detector sees an order of magnitude larger effect since the first factor in Equation (1) approaches 0.04 and so $P_{ab}$ approaches 0.02. Meanwhile the effect seen in the Near Detector is dominated by the second factor in the equation and the particular values of $\Delta m^2$ and L/ E. In general, we would always see a substantial effect in the Near Detector and incorrectly extrapolate the beam $\nu_e$ spectrum to the Far Detector. Depending on the oscillation parameters we could extrapolate too high a background or too low a background as summarized in Table 10.1.

| LSND parameters | | Near Detector excess at 2 GeV | Far Detector excess at 2 GeV |
|---|---|---|---|
| $\Delta m^2$ (eV$^2$) | $\sin^2(2\theta_{\mu e})$ | | |
| 2.5 | 2.6 x 10$^{-3}$ | 52% | 46% |
| 1.0 | 4.0 x 10$^{-3}$ | 26% | 70% |
| 0.4 | 20.0 x 10$^{-3}$ | 24% | 350% |
| 0.25 | 40.0 x 10$^{-3}$ | 19% | 700% |

Table 10.1: Expected excess beam $\nu_e$ events for both the Near and Far NOνA Detectors for several LSND parameters consistent with Figure 10.8 as a percentage of the beam $\nu_e$'s observed in the absence of an LSND effect. The Near percentages are from Figure 10.10. The Far percentages are relative to the beam Monte Carlo prediction (in the absence of an LSND effect) of 2.85 x 10$^{-3}$ for the $\nu_e$ to $\nu_\mu$ ratio at the far site.

If the LSND signal were confirmed, we would likely have to respond by moving our Near Detector around to a wide variety of different off-axis angles (Sites 1 -3) to exploit the different parent $\nu_\mu$ spectra and different L/E distributions. These data sets would allow us to disentangle the NuMI beam spectra effects from the short baseline oscillation effects. In Chapter 13 we turn this LSND "background" argument around and ask what NOνA could contribute to measurements in this sector if MiniBooNE confirms the LSND effect.

## 10.6. Backgrounds from $\nu_\mu$ Charged Currents

In the analysis described in Chapter 12, in order for the $\nu_\mu$ CC events to be misidentified as NC events, there has to be a track identified as an electron (most likely an asymmetrically decaying $\pi^0$) *and* the muon has to be missed. The cases where a muon itself is misidentified as an electron are rare due to the good $\mu$-e separation in our detector. High $y$ events (where $y$ is defined as the fractional neutrino energy loss) form the majority of the $\nu_\mu$ CC background.

To determine the fraction of $\nu_\mu$ charged current events that would pass all analysis cuts, one can measure that fraction for events with identified muons, and then predict the number of times that the muon is undetected. This procedure works in the limit that the nature of the hadronic system in a neutrino charged current interaction is dependent only on the hadronic energy of the system, and not on the neutrino energy.

$\nu_\mu$ CC events, to a very good approximation, are characterized by a flat $y$ distribution near high $y$. Thus, for a specific neutrino energy, the distribution of these events with longer muon range which satisfy our $\nu_e$ signal criteria should be flat when plotted as a function of muon range (= 1-$y$). The contribution to the background from $\nu_\mu$ CC events with shorter muon range (dominated by unidentified muons) can then be obtained by integrating the extrapolation of the observed distribution. In reality, the $y$ flatness expectation is altered by the fact that our selection criteria for the $\nu_e$ signal interacts somewhat with the energy of the muon. By allowing a slope in this distribution and its extrapolation, these effects can be incorporated. We have tried this procedure in simulations and find that we can extract the actual number of Near Detector $\nu_\mu$ CC background events to about ±30%. The simulation indicates that this Near Detector background extraction procedure will translate into about a 15% error in our prediction of the Far Detector background.

In addition, our energy spectrum is not monochromatic. This background from $\nu_\mu$ CC events masquerading as $\nu_e$ comes from the <u>oscillated</u> $\nu_\mu$ spectrum (see Figure 4.5). Therefore a measurement of this background with the un-oscillated beam in the Near Detector will not be quite correct. We can get a handle on this effect by repeat-ing the described extrapolation procedure for the different $\nu_\mu$ energy spectra available at different near sites as illustrated in Figure 10.2. Again, this means moving the NOvA Near Detector around to different underground sites.

## 10.7. Neutrino Cross Section Uncertainties

Assuming identical Near and Far Detectors, the present imprecise knowledge of low energy neutrino cross sections [2] does not matter to first order to NOvA since the same unknown cross section is seen in both the Near and Far Detectors. The predictions for the Far Detector background could still have second order effects due neutrino cross sections if we do not choose our Near site(s) carefully. Measuring these neutrino cross sections with the NOvA Near Detector will be an excellent benchmark of our understanding of both the Far and Near Detectors.

Neutrino interactions in this energy regime are classified as four different kinds of processes: Quasi-elastic (QE), Resonance, Coherent, and Deep Inelastic Scattering (DIS). Each process can be either neutral current (NC) via Z-exchange or charged current (CC) via W-exchange.

In the QE process, a nucleon is knocked out of the nucleus and the final state lepton is a muon or electron for CC events and a $\nu_\mu$ or $\nu_e$ for NC events. In Resonant processes a $\Delta$ resonance is created, which then decays to a proton + pion, or a neutron + pion. The DIS process produces multiple pions.
Figure 10.11 shows the current status of measurements for these processes in the CC channels [2]. At NOvA neutrino energies, these CC processes are all about equal in magnitude and each known only to about 20 - 30%. These CC cross sections are changing within our narrow off-axis energy band as indicated in Figure 10.11.

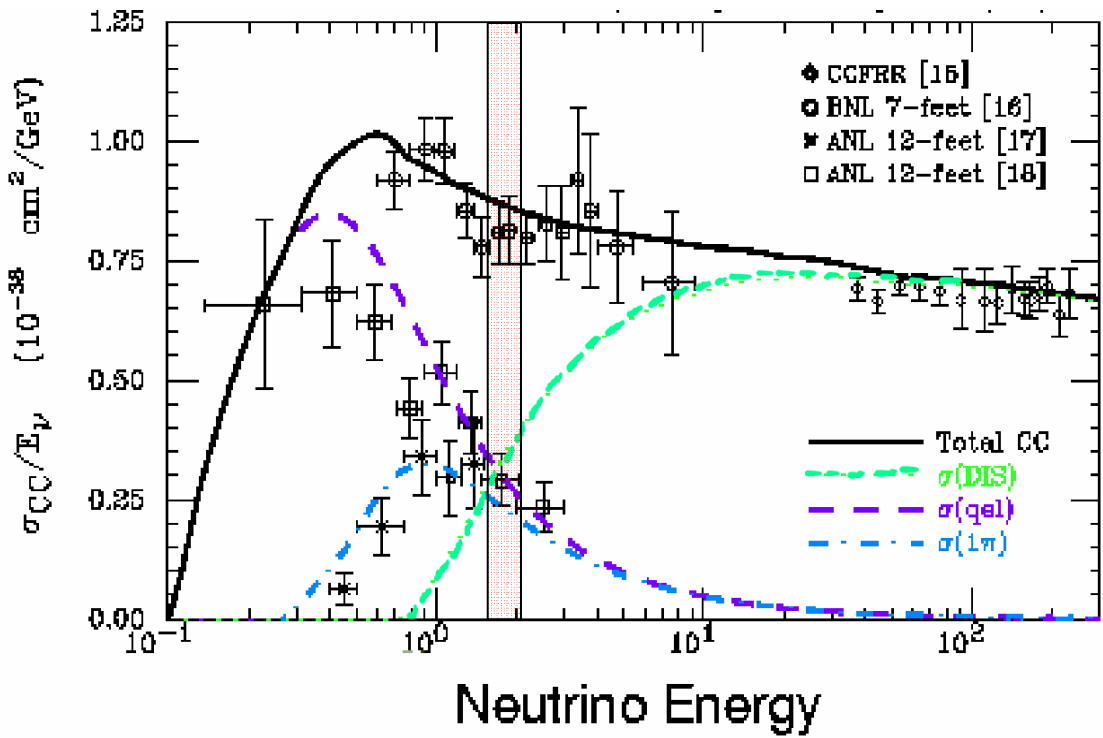

Fig. 10.11: A compilation of low energy charged current neutrino cross sections from G. Zeller [2]. The transparent red band indicates the peak energy of NOvA oscillated $\nu_e$ events.

The NC cross sections are even more poorly constrained by existing data, often with only one experiment contributing any data at all, so that a figure like Figure 10.11 is not even possible. Based on the few existing data, it appears that the NC cross sections are known to about 50% at best. NC Resonance and NC DIS events in the NOνA detector can fake $\nu_e$ CC events if a higher energy NC neutrino interaction creates a $\pi^0$ which gets misidentified as a 2 GeV electron. This feed-down from higher energies means we are interested in the NC cross sections well above 2 GeV. The parent neutrino energy spectrum for NC $\pi^0$ events which fake a beam $\nu_e$ at 2 GeV is roughly flat in energy.

In Coherent processes the neutrino scatters off the nucleus as a whole, and the only final state particle produced (besides the lepton) is a single charged pion for the CC process, or a single neutral pion for the NC process. Coherent CC events are not a problem for NOνA since they should not fake our $\nu_e$ oscillation signal. Coherent NC interactions have the same properties as the more familiar NC processes discussed above in that the observed energy is typically significantly less than the energy of the parent neutrino energy initiating the interaction. Figure 10.12 shows a compilation of both CC and NC coherent pion production cross-section measurements. These cross sections are known only to about 50% and their absolute values are roughly 20% of the NC Resonance plus DIS processes.

All these data will be substantially improved during the next few years. K2K, with 1.3 GeV neutrino data, and MiniBooNE, with 0.8 GeV neutrino data, each have high statistics samples and will contribute improved low energy cross section data. MINOS is optimized for $\nu_\mu$ detection [6] and can probably only add to the $\nu_\mu$ CC data. The MINERνA experiment [3] aims to measure all of the relevant CC cross sections to the 5% level and the NC cross sections to the 20% level before NOνA begins taking data. MINERνA will make the greatest difference to NOνA, particularly because MINERνA will run in the NuMI low energy beam and collect data on neutrino cross sections in the 1.5 – 5 GeV range.

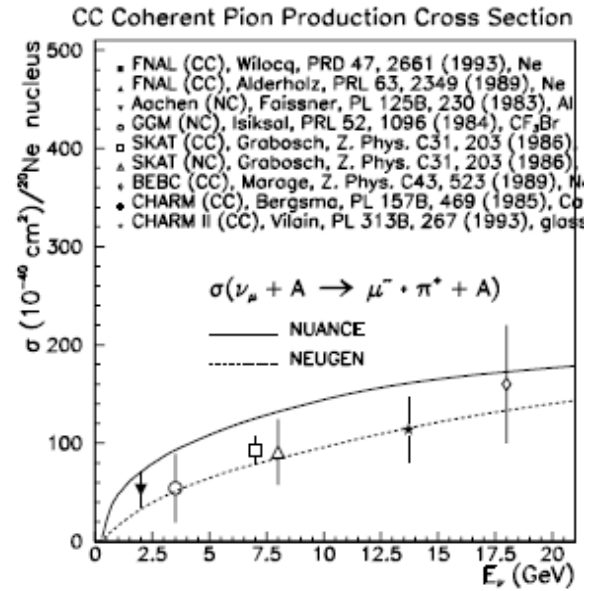

Fig. 10.12: Compilation of coherent pion production cross-section measurements from G. Zeller [2]. Both CC and NC pion production data are shown. NUANCE and NEUGEN are Monte Carlos which model these cross sections in different ways.

MINERνA data on NC above 2 GeV will be invaluable to NOνA simulations. In addition, we will be able to compare NOνA measurements of CC and NC QE, Resonance, Coherent, and DIS interactions to the precise MINERνA results and understand the NOνA detector response to each process.

In Chapter 13 we discuss measurements NOνA should be able to contribute to the low energy neutrino cross section picture.

**10.8. Summary of Beam Backgrounds**

Many of the uncertainties discussed in this chapter will be greatly reduced by the time NOνA runs: MIPP data will constrain the beam flux calculations, MiniBooNE will illuminate the LSND question, and MINERνA will measure the basic neutrino cross sections. We will calibrate our Near Detector and study its properties in a test beam and in the MINOS Surface Building. Still, the NOνA Collaboration and the laboratory need to be aware of these various effects and that has been the motivation for this discussion.

Ignoring NC effects, the NOvA Near Detector at Site 1.5 sees a "real" beam $\nu_e$ distribution ("real" defined as $\nu_e$ from decays in the secondary beam off the NuMI target) that can be ~ 7% different from the Far Detector as shown in Figure 10.5. The Near and Far Detectors both see additional "fake" beam $\nu_e$ distributions from contamination by NC and Coherent NC events with $\pi^0$s, and in fact the two detectors will see somewhat different NC effects since they see somewhat different parent energy spectra. All these effects can conspire to introduce systematics into our determination of the $\nu_e$ background for any $\nu_e$ appearance signal in the Far Detector.

A NOvA Far Detector $\nu_e$ appearance signal energy distribution will have a very different shape from the beam $\nu_e$ energy distribution ("real" or "fake"), peaking near 2 GeV as shown in Figure 4.5. We exploit that difference in our analysis to obtain large rejection factors for the various backgrounds, and this is discussed in Chapter 12. Our simulations indicate we will see a 19.5 event background to a NOvA $\nu_e$ appearance signal measurement. This background is composed of 61% from "real" beam $\nu_e$ events (11.9 events), of 2.5% from $\nu_\mu$ CC backgrounds (0.5 events), and of 36.5% from NC backgrounds (7.1 events).

From a simulation point of view, the first of these backgrounds ("real" beam $\nu_e$) will be known to ~ 7% from our matching of the Near Detector at Site 1.5 to the far site. This implies an uncertainty of 0.8 event on the 11.9 events. The second ($\nu_\mu$ CC) background will be known to ~ 15% from our extrapolation procedure described in section 10.6. This implies an uncertainty of 0.08 events on the 0.5 events. The last background (NC) will be known to ~ 5% from the kinematics of the flux prediction for the unoscillated neutrino spectrum in the Far Detector. This implies an uncertainty of about 0.4 events on the 7.1 events. Assuming uncorrelated errors for these three processes would indicate an overall uncertainty at the level of about 0.9 events (5%) for the 19.2 event background discussed in Chapter 12.

Measurements and understanding of these backgrounds with our Near Detector will require careful work and data taking in several off-axis positions.

## 10.9. Cosmic Ray Backgrounds

The cosmic ray background will be strongly suppressed in NOvA by the very low duty cycle of the accelerator beam (~10 µs spill every ~2 seconds), directionality of this incident neutrino beam (pointing from Fermilab) and its relatively high energy (1.5-2 GeV).

Our preliminary estimates and simulations, described below, indicate that this background should not be a problem. Furthermore, this background can be measured with very high precision during the off-beam time. It is also our intention to test our estimates during prototype testing.

The atmosphere behaves as a 10-interaction length, 25-radiation length calorimeter for the incident primary cosmic rays. The results of interactions in the atmosphere are extensive air showers, with the following components persisting to the surface: penetrating muons with average energy ~ 4 GeV, showering electrons and photons with average energies in the range of tens of MeV, and some hadrons, primarily neutrons, with average energies of hundreds of MeV. To estimate the effects of these secondary particles on operation of the NOvA Far Detector we assume the detector as described in Chapter 5 and a live-time of the detector of 100 seconds per year (~$10^7$ spills per year, each 10 µs long). We discuss next the manifestation of each component on the detector separately.

*10.9.1. Cosmic Ray Muons:* The muon flux at the surface of the Earth is approximately 120 $\cos^2\theta$ m$^{-2}$ s$^{-1}$ sr$^{-1}$, where $\theta$ is the zenith angle. This flux yields an average of 8 muon trajectories inside the detector per 10 µs spill-gate and a total of $8 \times 10^7$ muons per year in the Far Detector during the active spill. Each of our 500 ns electronic time slices (described in Chapter 7) will contain an average 0.4 muons over the ~ 2000 m$^2$ area of the detector. These muons provide an essential calibration and alignment tool. The muons have a median energy of 4 GeV, and 10% to 20% originate in the same air shower, appearing as in-time multiple tracks. Using an expression for the integral flux as a function of energy and zenith angle [7], we estimate that 50% of the muons will stop in the detector. Muons themselves clearly cannot simulate our signal, which could only happen through their interactions in the detector.

*10.9.2. Cosmic Ray Electrons and Photons:* A significant flux of electrons and photons from the extensive air showers survives at ground level. The net flux is about 50% the muon flux, but their average energy is less than 100 MeV [8] as shown in Figure 10.13. The electrons and photons will generally produce small showers that penetrate short distances (less than 1 m typically) into the top of the detector. Only ~2% have energies above 1 GeV and are capable of producing a significant shower or "splash" at the top of the detector, causing multiple hits in the scintillator strips.

*10.9.3. Cosmic Ray Hadrons:* A small component of hadrons survives to ground level. Neutrons and protons are nearly equal components. Protons, having an electric charge, will be detected as background events with high efficiency. The neutrons have an interaction length of ~1.5 m and their interactions are therefore a potential source of background. Their trajectories are much more vertical than the muons, with average angle ~20° from the zenith, and their median energy is ~100 – 200 MeV. Figure 10.14 shows the integral flux of neutrons incident on the top of the detector calculated from the measured differential flux [9]. We estimate that $1.0 \times 10^5$ neutrons with energies above 2 GeV will interact in the detector per year within the neutrino spill gate; they will be concentrated near the top of the detector.

Even though neutrons at ground level are always accompanied by muons or electrons, this fact is not very useful as a veto for neutrons entering the detector because of large typical spatial separation. Using the standard cosmic ray code CORSIKA [10], we found that only 4% (10%) of the neutrons have an accompanying muon within 50 (100) m, providing no satisfactory veto power for the proposed detector dimensions.

In the few GeV energy region, ~20% of the inelastic neutron interactions produce a single pion, which, in principle, might simulate an electron track. 98% of all CC events have a track within 25° of the neutrino beam direction and thus a pion from a neutron interaction must be emitted at an angle at least ~60° to provide a possible background to a beam neutrino event. From kinematics, the maximum possible energy of a pion to be emitted at 60° is 1.5 GeV, just at the edge of possible acceptance. With the addition of a topology requirement that the track should be electron-like, we estimate that background from neutron interactions will be at the level of <1 event/year.

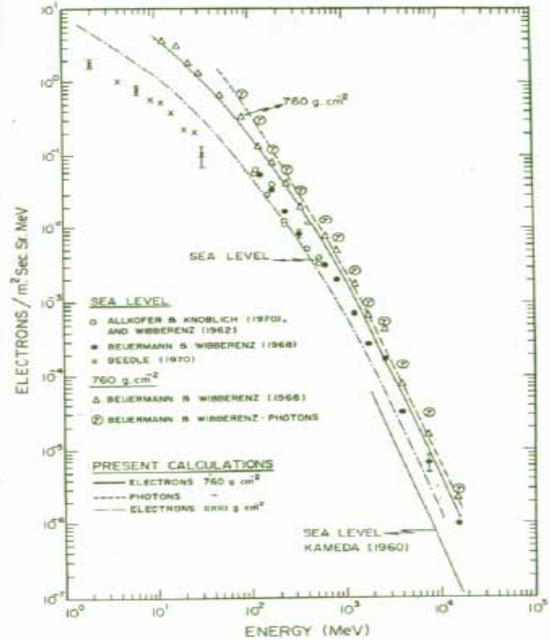

Fig. 10.13: Observed and calculated differential energy spectra of electrons and photons from ground based measurements. This figure comes from reference [8].

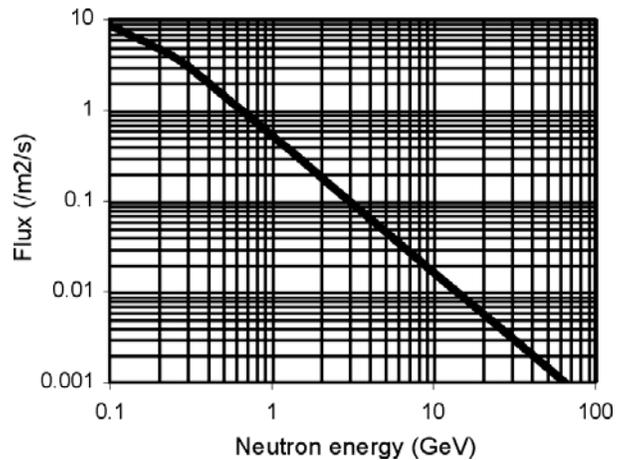

Fig. 10.14: Integral flux of neutrons at ground level

A GEANT simulation of neutrons incident on the detector was performed to determine this number more precisely. In the simulation, neutrons with energies ranging from 1-5 GeV, uniform in azimuth, and fixed zenith angles were incident on the detector. The results were analyzed against the standard selection criteria that are used for signal selection to determine the probability of selecting a cosmic ray neutron event as a signal. Figure 10.15 shows how the selection probability increases as zenith angle increases, as expected, and is still small, even at horizontal incident angles.

Convoluting this result with the measured flux [9], which is steeply falling with both energy and zenith angle, and the detector geometric acceptance, gives an expected number of events shown in figure 10.16. The total number of events selected in a 5 year exposure is 0.44 events.

Off-spill cosmic ray data in NOvA will precisely measure the cosmic ray backgrounds.

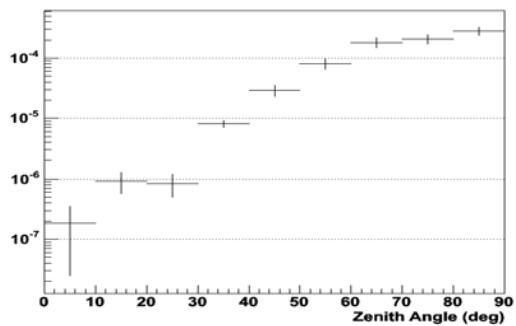

Fig. 10.15: Selection probability for neutrons with energy in the range of 1-5 GeV as a function of zenith angle.

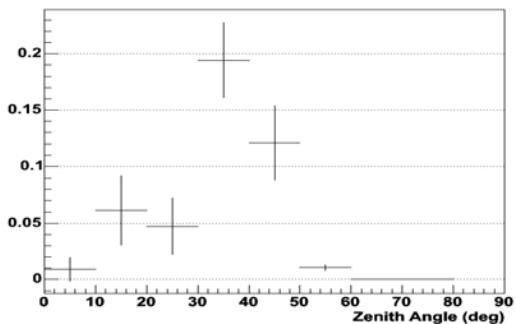

Fig. 10.16: Number of selected events in a 5 year exposure due to cosmic ray neutrons as a function of zenith angle.

# 11. Proton Beam

## 11.1. Introduction

This proposal assumes a 5 year run with the 30 kiloton (kT) NOvA Far Detector and beam intensity in the Main Injector such that the NuMI target receives $6.5 \times 10^{20}$ protons on target per year (pot/yr). The laboratory and the NOvA Collaboration have mutually agreed [1] that NOvA should use this number to illustrate the reach of the experiment in a pre-Proton Driver era. The laboratory and the NOvA Collaboration also have agreed that this number of pot/yr should not be viewed as a promise of delivered beam. We will review the details of this pot/yr calculation in this chapter and assess the probability of not reaching the full beam intensity.

With a Proton Driver at Fermilab the beam intensity expected in the NOvA experiment would increase and we have mutually agreed that $25.0 \times 10^{20}$ pot/yr on the NuMI target should be used to illustrate the reach of the experiment in a Proton Driver era. We also discuss the details of this assumption here.

## 11.2. The Proton Plan

Following the work of a "Proton Committee" [2], the laboratory is assembling a "Proton Plan" for the years 2005-2008. The November 9, 2004 draft Proton Plan [3] indicates the laboratory would be in Phase III of that program beginning in 2008, about one year before NOvA begins data taking. Table 6 of the November 2004 Proton Plan shows the following parameters:

- Booster Batch Size of $5.5 \times 10^{12}$ protons.
- Main Injector (MI) loading with 2 Antiproton Source + 9 NuMI batches. This means slip-stacking is implemented for both programs. (Barrier bucket stacking has also been discussed recently).
- MI Cycle Time of 2.2 seconds.
- MI Intensity of $6.0 \times 10^{13}$ protons.
- Booster maximum rate of 8.3 Hz limited by radiation levels.
- NuMI annual rate of $3.4 \times 10^{20}$ protons on the NuMI target.
- Booster Neutrino Beam annual rate of $2.2 \times 10^{20}$ protons.

Phase III includes 10% down time for NuMI for collider shot setup and a 5% reduction to NuMI to allow for antiproton transfers from the Accumulator to the Recycler. Phase III also includes a 90% operational efficiency factor for NuMI and another 90% efficiency applied to slip stacking for NuMI.

## 11.3. Proton Plan in the Post-Collider Era

Recent direction from the Department of Energy indicates that the Fermilab Proton Plan can be updated to show the end of the Fermilab Collider Program in 2009, about the same time as NOvA would begin operating. We envision several changes to the parameters listed in Section 11.2 above in this post-collider era.

Without a collider program, there is no need for anti-proton production batches in the Main Injector, so 11/9 (a factor of 1.22) more beam could be available to NuMI. Without a collider program, the 10% downtime for NuMI for collider shot setup and the 5% reduction to NuMI for antiproton transfers from the accumulator to the Recycler can be recovered (in total this is another factor of 1.176).

Without a collider program, the Recycler is also available as a proton accumulator. The 2.2 second MI cycle time in Section 11.2 comes from the time to load 11 Booster batches during 12 Booster cycles at 15 Hz (0.8 seconds) plus the MI ramp up and down time of 1.4 seconds. The time to load the Booster batches into the MI can be hidden under the MI cycle time by loading these Booster batches into the Recycler and then taking only one Booster cycle (0.067 seconds) to inject from the Recycler to the MI. This would reduce the effective MI cycle time from 2.2 seconds to 1.467 seconds and give a factor of 1.50 more beam to NuMI.

These three factors imply the NuMI annual rate could be increased by $(1.22)(1.176)(1.50)(3.4 \times 10^{20}$ protons$) = 7.3 \times 10^{20}$ protons. We have agreed with the laboratory to use ~90% of this value ($6.5 \times 10^{20}$ protons per year) in this proposal to illustrate the reach of the NOvA experiment.

As a crosscheck, the total Booster rate in this scenario is now 11 batches in 1.467 seconds or 7.5 Hz. This is comfortably below the Phase III maximum Booster rate of 8.3 Hz limit due to Booster radiation limits anticipated in 2008.

Another required crosscheck is for the number of seconds the MI would operate in one year. The total annual NuMI rate of $7.3 \times 10^{20}$ protons divided by the Phase III MI intensity of $6.0 \times 10^{13}$ per pulse gives $1.22 \times 10^7$ MI pulses in a year. At a cycle time of 1.467 seconds, this means $2.0 \times 10^7$ seconds of MI operations per year or 63% of the available seconds in a year. This can be compared to typical laboratory assumptions of 40 weeks of operations per year with each week averaging 120 hours of beam time yielding $1.73 \times 10^7$ seconds of operations per year. 46 weeks of operations would be required for $7.3 \times 10^{20}$ protons, or 41 weeks for $6.5 \times 10^{20}$.

## 11.4. Possible Limitations to the NOvA assumption of $6.5 \times 10^{20}$ pot per year

The Proton Plan [3] and the factors multiplied together in Section 11.3 could be overly optimistic, so we examine these assumptions in this section. Main Injector proton losses in the acceleration cycle may be a limiting factor as operations approach $6 \times 10^{13}$ protons per cycle. Part of the Proton Plan includes increasing aperture restrictions, reducing beam tails, and adding a collimation system to the MI. MI slip stacking or barrier bucket stacking may not succeed completely when extended from 2 booster batches in pbar production to 11 batches for NuMI. As noted in Section 11.2, the Proton Plan already assumes a 90% efficiency for NuMI operations and another 90% efficiency applied to slip stacking for NuMI.

In Section 11.3 we have assumed that the Recycler can be used as a proton accumulator including slip stacking in the Recycler as was originally assumed to occur in the MI. The laboratory [4] believes this use of the Recycler is a reasonable assumption since the Recycler looks almost identical (same aperture, same energy, …) to the MI at 8 GeV. Using the Recycler would involve some expense. Protons from the Booster currently cannot be injected directly into the Recycler, so the transfer line from the Booster to the MI would have to be redirected. This would not involve civil construction. The Recycler would also require additional RF to handle the increased beam and slip stacking.

So in both of the above cases, the tricky part is the slip stacking of <u>many</u> Booster batches. The Proton Plan does note [5] that

> "slip stacking has been successfully developed for antiproton production, and development for NuMI will be demonstrated by early 2006."

As a measure of the risk that NuMI slip / barrier stacking does not materialize, one could assume only Phase II of the Proton Plan is realized. The parameters of Phase II are as follows:

- Booster Batch Size of $5.3 \times 10^{12}$ protons.
- Main Injector (MI) Loading with 2 Antiproton Source + 5 NuMI batches. This means slip-stacking is implemented only for antiproton production.
- MI Cycle Time of 2.0 seconds.
- MI Intensity of $3.7 \times 10^{13}$ protons.
- Booster maximum rate of 7.5 Hz limited by radiation levels.
- NuMI annual rate of $2.2 \times 10^{20}$ protons on the NuMI target.
- Booster Neutrino Beam annual rate of $2.8 \times 10^{20}$ protons.

Without a collider program, we next apply similar scaling factors to the Phase II NuMI protons following the arguments made in Section 11.2 for Phase III:

- Now it's a factor of $7/5 = 1.40$ since the 2 Booster batches destined for the antiproton source can be redirected to NuMI.
- The same factor of 1.176 applies for recovering collider shot setup time and Accumulator to Recycler transfer time.
- The Recycler could still be used as a proton accumulator. The 2.0 second MI cycle time comes partially from the time to load 7 Booster batches during 8 Booster cycles at 15 Hz (0.467 seconds). Of this time, 0.400 seconds can be hidden under the MI cycle time by loading these Booster batches into the Recycler and then taking perhaps only one Booster cycle (0.067 seconds) to inject from the Recycler to the MI. This would reduce the effective MI cycle time from 2.0 seconds to 1.467 seconds, giving a factor of 1.36 more beam to NuMI.

The net result is $(1.4)(1.176)(1.36)(2.2 \times 10^{20}$ protons$) = 4.9 \times 10^{20}$ protons to NuMI. This is 67% of the annual rate calculated in Section 11.3. We believe this represents the most conservative assumption for NOvA in a pre-Proton Driver era.

We note that a contingency to this 67% assumption exists in a possible upgrade to the MI RF system to shorten the MI ramp rate to as little as 1.0 seconds. This scheme will be discussed below in section 11.6.

## 11.5. Possible Limitations from the NuMI Beamline

The NuMI components were designed for $4 \times 10^{13}$ protons per MI pulse and the discussion in Section 11.3 would increase that instantaneous rate by a factor of 1.50. The NuMI components were designed for a 1.87 second cycle time and the discussion in Section 11.3 would decrease the cycle time to 1.467 seconds, a factor of 1.27. For some components the two factors get multiplied together to give an overall increase of 1.90. In this section we examine these components to see if any of them will limit our assumption of $6.5 \times 10^{20}$ pot per year. These issues are under study and this section gives a first look at the situation.

The beam windows in the system would not be over stressed by these higher intensity and cycle time numbers.

The NuMI target was designed with a safety factor of 1.8 against stress from instantaneous beam, so we would push nearer to the limit with a factor 1.5 more instantaneous beam. However, NOvA proposes to use the medium energy configuration of the NuMI beam and in this case the target is outside of the horn (vs. inside Horn 1 for MINOS in the low energy configuration). This makes solutions somewhat simpler if there is need to reduce target stress or deal with radiation damage. For example it would be possible to replace the target more often or possible to make a moving target so that no one section accumulated too much radiation damage.

The horns themselves would see additional heating from the factor of 1.50 in instantaneous rate, but they can take this load as long as the heat is removed by the water cooling system. The air cooled stripline to the horn would be right on the edge of needing water cooling in the last ten-foot section. Such water cooling is possible and might cost of order $300K.

The Target Pile Air System also would be pushed hard since the capacity of the water chiller is only ~15% oversized. The minimum upgrade would be to add an additional chiller to the existing Air System, gaining about a factor of two.

The Decay Pipe cooling would have to be looked at in more detail. For the Proton Driver case (next section), studies so far have not been able to prove or to disprove that the water cooling piping is adequate. There are cooling lines every $30^0$ around the circumference of the pipe, so stress can increase due to differential heating between the hot spots and the cooling lines $15^0$ away. This Decay Pipe is a vacuum vessel, so one simple fallback here would be to remove the vacuum vessel stress (and attendant heightened design requirements) by filling the pipe with helium at atmospheric pressure. This would reduce the neutrino flux by a few percent.

The absorber could handle the increased instantaneous rate since the aluminum components can easily remove the heat. The current aluminum design allows cooling to one of three modules to fail and the absorber is still adequately cooled by air convection to the adjacent modules. However, the absorber water system cooling capacity would probably have to be upgraded. The absorber was designed with redundant cooling lines, so one solution is to use the original lines plus the redundant ones.

For NuMI running at $4 \times 10^{20}$ pot per year, conservative designs were adopted to ensure that the groundwater radiation concentrations would be well below the regulatory limits[6]. Beginning in 2005, measurements of radiation levels from NuMI running will be available and allow extrapolation to the case of 1.63 (= $6.5 \times 10^{20}$ pot / $4 \times 10^{20}$ pot) times as many pot/year being discussed here. Measurable levels of $^3$H or $^{22}$Na in the groundwater monitoring wells around NuMI are not expected at $4 \times 10^{20}$ pot/year and extrapolations by a factor of 1.63 after initial NuMI running should indicate negligible levels relative to the regulatory limit in these wells. Similarly, measurements during the initial years of NuMI running will be made of the levels of radionuclides in the water pumped from the NuMI tunnel and released to the surface waters. These levels are expected to be at least a factor of 20 (twenty) below the surface water limits, so that a factor of 1.63 increase should not be a problem.

Overall the conclusion is that the NuMI components with additional cooling could handle the 6 x

$10^{13}$ protons in MI pulses and the shortened cycle time of 1.467 seconds as outlined in Section 11.3 for the post-collider era. There are work-arounds for each component in case further study indicates some need further help.

The more conservative beam assumption discussed in Section 11.4 with 70% of 6.5 x $10^{20}$ pot per year delivered to NOνA is handled more easily by NuMI components. This reduced beam scenario has the same cycle time of 1.467 seconds but with only 3.7 x $10^{13}$ protons per MI pulse. The instantaneous rates are down by 40% and within the original design envelope for the components.

## 11.6. NOνA and the NuMI Beam Line with a Proton Driver

Over the last year, the idea of building a new 8 GeV Proton Driver has become a centerpiece of the recommendations of the Fermilab Long Range Planning Committee [7]. The 8 GeV proton linac plan under discussion would increase the MI intensity and shorten the MI ramp time, realizing a 2 MW proton source. This would benefit NOνA as discussed below:

- The MI intensity per pulse would be increased to 15 x $10^{13}$ protons. In this era, slip stacking is not part of the plan and the linac's small beam emittance is used to increase the MI intensity. Compared to the NuMI annual rates calculated in Section 11.3, the Proton Driver would increase the NuMI proton beam by a factor of (15 x $10^{13}$ / 6 x $10^{13}$) = 2.5.
- The MI cycle time would be reduced to 1.37 seconds for the MI ramp (now free of the Booster clock cycle that meant the minimum was 21/15 sec) plus 0.1 seconds to fill the MI from the Proton Driver. Compared to the rates calculated in Section 11.3, the shorter cycle time would increase the NuMI proton beam by a factor of (1.467 / 1.470) = no effect.
- The laboratory is considering adding a MI RF and MI power supply upgrade to the Proton Driver scheme to decrease the MI cycle + fill time further to as little as 1.00 seconds. This would increase the NuMI proton beam by a factor of (1.467 / 1.00 ) = 1.467. Decreasing the MI cycle time in this way would carry a substantial additional cost of 10-20% to a Proton Driver project.

Overall the Proton Driver could increase the annual NuMI beam intensity to (2.5)(1.0)(1.47)(6.86 x $10^{20}$ protons) = 25.2 x $10^{20}$ protons.

As a conservative fall-back position, we think of an annualized NuMI intensity using only the Proton Driver itself and realizing (2.5)(6.86 x $10^{20}$ protons) = 17.2 x $10^{20}$ protons. This is 68% of the annual rate calculated above. We believe this represents the most conservative assumption for NOνA in the Proton Driver era. We note that in the event a Proton Driver is not approved, decreasing the MI cycle time is still an option and still provides additional reach to a Fermilab neutrino program.

All changes discussed in this section impact the stability and lifetime of the target, horns and other systems in the NuMI beam line. These issues are still under study and upgrades to the target and beam line are anticipated.

## Chapter 11 References

# 12. Simulations of NOνA Performance

## 12.1. Introduction

We have simulated the signals and backgrounds for $\nu_\mu \to \nu_e$ oscillations using relevant parts of the MINOS experiment software, the NEUGEN3 neutrino interaction generator and the GEANT3 detector simulation. The steps in the simulation were
1) Generation of the event interaction.
2) Calculation of the detector response to the generated particles.
3) Reconstruction, i.e. track finding and fitting. A quadratic fit is made to each track using the pulse height-weighted cell positions in each plane.
4) Calculation of various parameters associated with each track.
5) Assignment of particle identity to each track (e, μ, p, γ, or hadron).
6) Calculation of the interaction vertex.
7) Preliminary identification of events with
   a) A measured energy within 25% of the nominal off-axis energy.
   b) No significant energy deposition near the detector boundaries.
   c) An electron candidate, which starts near the vertex and has no gaps near the vertex.
   d) No μ or γ in the event.
8) Separation of signal and background events using a maximum likelihood analysis with the following variables
   a) Total measured energy
   b) Fraction of total energy carried by the electron
   c) Mean pulse height near the origin of the electron
   d) Pulse height per plane for the electron
   e) Number of hits per plane for the electron
   f) Energy upstream of the vertex
   g) Curvature of the electron
   h) Missing transverse momentum
   i) Fraction of total electron energy contained in the first half of the electron track
   j) rms deviation of electron hits from the fitted track
   k) number of tracks identified as hadrons in the event

The maximum likelihood optimization was done by maximizing a figure of merit (FoM) defined as the signal divided by the square root of the background, assuming that the oscillation is given by the formula

$$P(\nu_\mu \to \nu_e) = 0.5 \sin^2(2\theta_{13}) \sin^2\left(\frac{1.27 \Delta m_{32}^2 L}{E}\right), \quad (1)$$

where $\Delta m_{32}^2 = 0.0025$ eV$^2$, L = 810 km, and the energy spectrum is given by the NuMI medium energy beam. The matter, solar, and CP effects are not included in Eq. 1, but are incorporated in the discussion of the physics potential of NOνA in Chapter 13.

## 12.2. Detector Optimization

*12.2.1. Cell Dimensions*: Results for a few combinations of cell widths and depths are shown in Table 12.1. It appears that widening the transverse dimension of the cell from 3.8 cm to 5.4 cm causes a significant decrease in the FoM. However lengthening the longitudinal dimension of the cell from 4.5 cm to 6 cm appears to have little effect. A slightly different set of simulations indicate a reduction in the FoM for cells longer than 6 cm. Such cells would also require a thickening of the cell walls for structural reasons. Since lengthening the cell reduces the cost per unit mass of the detector, we have chosen 6 cm long cells rather than the 4.5 cm long cells described in Appendix B of the previous version of the proposal. The cell width of 3.8 cm has been retained.

| Cell width | Cell depth | Relative FoM | Electron energy resolution |
|---|---|---|---|
| 3.8 cm | 4.5 cm | 1.00 | 10.0% |
| 3.8 cm | 6.0 cm | 1.02 | 10.7% |
| 5.4 cm | 4.5 cm | 0.90 | 9.9% |

Table 12.1. Simulation results for various cell dimensions.

*12.2.2. Detector Off-Axis Transverse Location:* Table 12.2 shows results for various transverse detector locations for both neutrinos and antineutrinos. The choice of transverse location depends on the physics goals of the experiment and this topic is discussed in detail in Chapter 13.

| Off-axis distance | $\nu$ or $\bar{\nu}$ | Number of signal events | Number of background events | Figure of merit |
|---|---|---|---|---|
| 8 km | $\nu$ | 284.5 | 61.2 | 36.4 |
| 10 km | $\nu$ | 227.4 | 39.0 | 36.4 |
| 12 km | $\nu$ | 142.4 | 19.5 | 32.2 |
| 14 km | $\nu$ | 90.5 | 12.9 | 25.2 |
| 8 km | $\bar{\nu}$ | 147.2 | 32.2 | 25.9 |
| 10 km | $\bar{\nu}$ | 109.0 | 18.5 | 25.3 |
| 12 km | $\bar{\nu}$ | 71.8 | 12.1 | 20.6 |
| 14 km | $\bar{\nu}$ | 49.3 | 8.4 | 17.0 |

Table 12.2: Number of $\nu_\mu \to \nu_e$ oscillation signal and background events and FoM for different off-axis detector locations. The numbers are for a 5-year run at $6.5 \times 10^{20}$ pot/year and do not include matter, solar, and CP effects, which are included in the discussion of physics potential in Chapter 13. They assume that $\sin^2(2\theta_{13}) = 0.10$, $\sin^2(2\theta_{23}) = 1.0$, and $\Delta m^2_{32} = 0.0025$ eV$^2$. The number of signal events is proportional to $\sin^2(2\theta_{13})$, but the number of background events is essentially independent of it. The variation with $\Delta m^2_{32}$ is discussed in Chapter 13.

The background listed in Table 12.2 is typically about two-thirds from beam $\nu_e$'s produced from muon and kaon decay and one-third from neutral-current events. The background from $\nu_\mu$ charged-current events is quite small, less than one event. This is shown in the bottom half of Fig. 12.1, which plots the number of each class of background events as a function of the number of accepted signal events generated by changing the cut on the likelihood function. The numbers in Fig. 12.1 are for a 5-year neutrino run with NOvA situated at 12 km off axis and the other conditions as in Table 12.2. The top half of Fig. 12.1 shows the resulting FoM as a function of the number of accepted signal events.

The accepted fraction of $\nu_\mu$ charged-current events is approximately $4 \times 10^{-4}$ and the accepted fraction of neutral-current events is approximately $2 \times 10^{-3}$. The efficiency for accepting a $\nu_e$ event from $\nu_\mu \to \nu_e$ oscillations is approximately 24%.

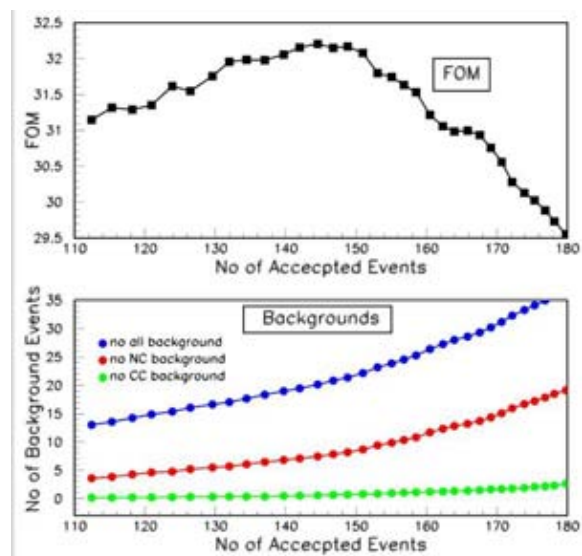

Fig. 12.1: The numbers of $\nu_\mu \to \nu_e$ oscillation background events and the FoM as a function of the number of accepted signal events generated by varying the cut on the likelihood function. The top half of the figure shows the FoM. The bottom half shows the number of background events. The green (bottom) curve shows the number of misidentified $\nu_\mu$ charged-current events; the red (middle) curve shows the number of misidentified neutral-current events; and the blue (top) curve shows the total number of background events including the number of beam $\nu_e$ events.

## 12.3. Detector Performance
The simulations described in the preceding sections allow us to study the NOvA detector performance. We discuss here those features that are most relevant for the physics of highest current interest.

*12.3.1. Energy resolution:* There are several areas where energy resolution helps in improving quality of physics. In brief, they are:

a) In reducing the intrinsic beam $\nu_e$ background for the $\nu_\mu \to \nu_e$ appearance analysis. This is the only handle one has on that background.

b) In reducing the neutral-current and $\nu_\mu$ charged-current backgrounds for this analysis; the energy distributions from these two sources generally will not peak at the oscillation maximum and be much broader.

c) In measuring the dominant oscillation mode parameters. This will be discussed in Section 13.6

For the first two, it is sufficient that the energy resolution should be good enough so that there is no appreciable broadening in the measured energy spread of the convolution of the beam energy distribution and the oscillation function. Then, the eventual energy cuts are determined by the natural energy spread of the beam. For the last, the energy resolution should be as good as possible.

The true and measured energy distributions for all $\nu_\mu \to \nu_e$ events are shown in Fig. 12.2. The additional spread in measured energy distribution due to the resolution is hardly perceptible – the rms width of the distribution changes only from 19.2% to 21.4%.

Another useful way of looking at the energy resolution is to look for correlations with the fraction of the total energy that goes into the electron, effectively the (1-$y$) parameter of the interaction. This is shown in Figure 12.3 where the energy information, represented as $\sigma E$ and defined as the difference between the true and measured energies divided by square root of true energy in GeV. The scatter plot of the left shows $\sigma E$ as a function of (1-$y$). Clearly as $y$ approaches zero the number of events increases and the energy resolution improves. The middle plot shows the mean $\sigma E$ as function of (1-$y$) indicating adequate weighing of hadronic and electromagnetic energy deposition. Finally, the plot on the right shows the $\sigma E$ distribution; the rms width of the fitted curve is 8.7%. Restricting that sample to events passing all cuts for $\nu_e$ identification reduces this width to 6.7%.

*12.3.2. Electron / muon separation:* The electrons and muons look quite different in the NOvA detector. The electrons tend to deposit more energy per plane and are more "fuzzy", i.e. have more hits per plane. In addition, electrons, because of their showering nature tend to have a larger rms spread of the accepted hits and also have more gaps, whereas the muon tracks are rather continuous.

These are the principal parameters that distinguish muons from electrons and the separation is excellent. This is illustrated in Fig. 12.4, which shows the first two of these parameters, i.e. average pulse height per plane and the average number of hits per plane, both for electrons and muons. The muon hits per plane distribution is significantly broadened by the finite angle of the muon with respect to the beam direction, i.e. crossing of two cells in one plane. Once that is corrected the distribution will be even narrower.

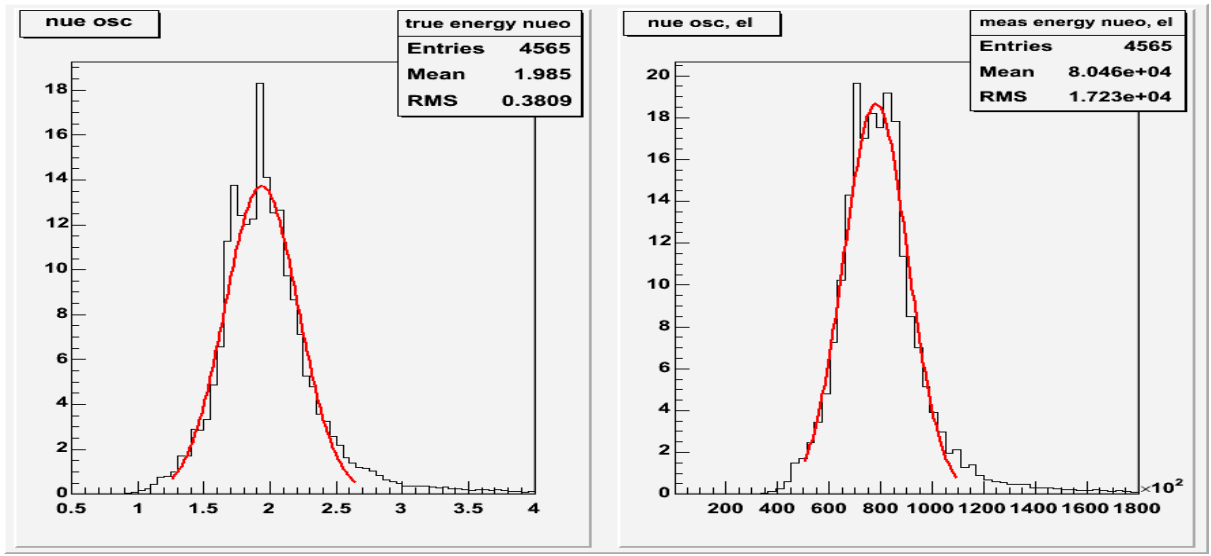

Fig. 12.2: True (left) and measured (right) energy distributions for all events with a reconstructed electron track. The units are GeV for the true energy and attenuation-corrected photoelectrons for the measured energy.

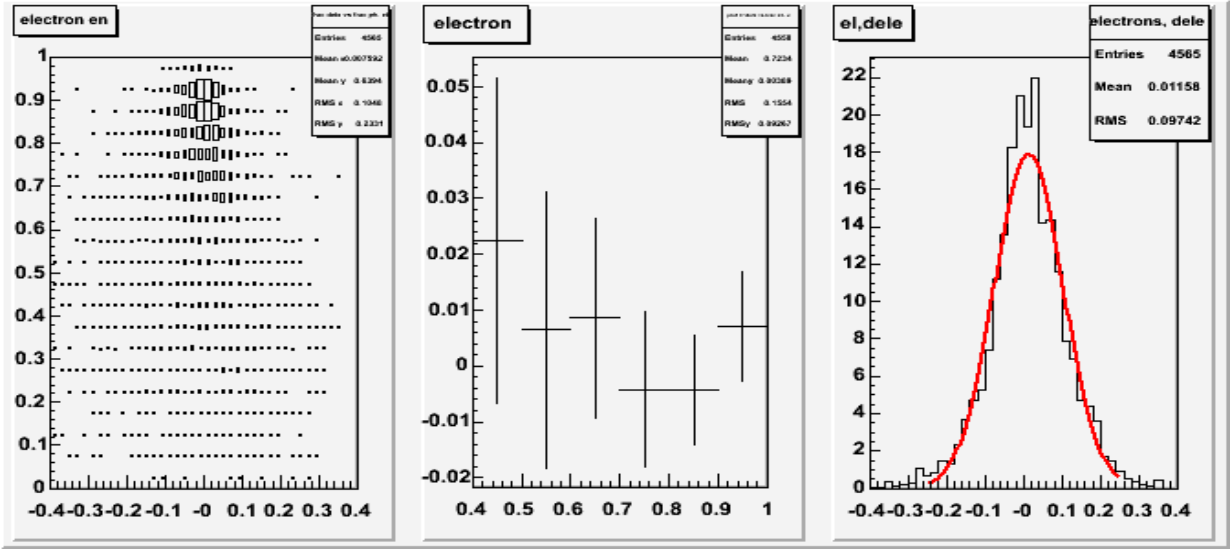

Fig. 12.3: Left: Measured minus true energy difference divided by square root of true energy in GeV, $\sigma E$, on the horizontal axis as function of (1-$y$) on the vertical axis. Middle: average $\sigma E$ on the vertical axis as a function of (1-y) on the horizontal axis. Right: Distribution of $\sigma E$.

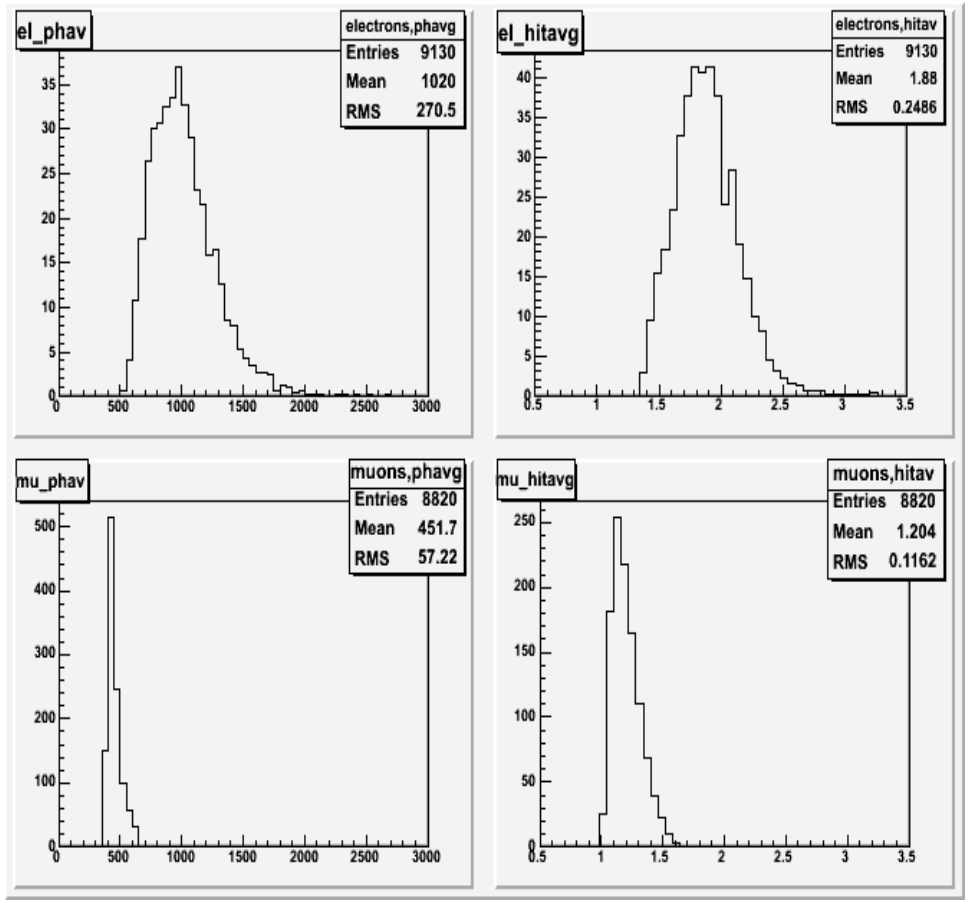

Fig. 12.4: Distributions of average pulse height/plane (left) and average number of hits per plane (right) for electrons in electron charged-current events (top) and muons in muon charged-current events (bottom).

Figure 12.5 further illustrates NOvA performance in separating electrons and muons. Forty thousand $\nu_\mu$ charged-current events and 40,000 $\nu_e$ charged-current events were generated. Events outside the fiducial volume, events with energies clearly too high or too low to be of interest, and events in which no tracks were found were eliminated. The $\nu_e$ events were required to have a found electron with an average number of hits per plane greater than 1.4, and the $\nu_\mu$ events were required to have a found muon with an average pulse height per plane less than 550. The remaining events are separated into two bins by event total energy, and for each energy band the average pulse height per plane is plotted versus the average number of hits per plane in Fig. 12.5.

These distributions indicate that the electron/muon separation is very clean. Thus the main mechanism for muon charged-current events looking like electron charged-current events would be production of $\pi^0$'s, which then simulate electrons.

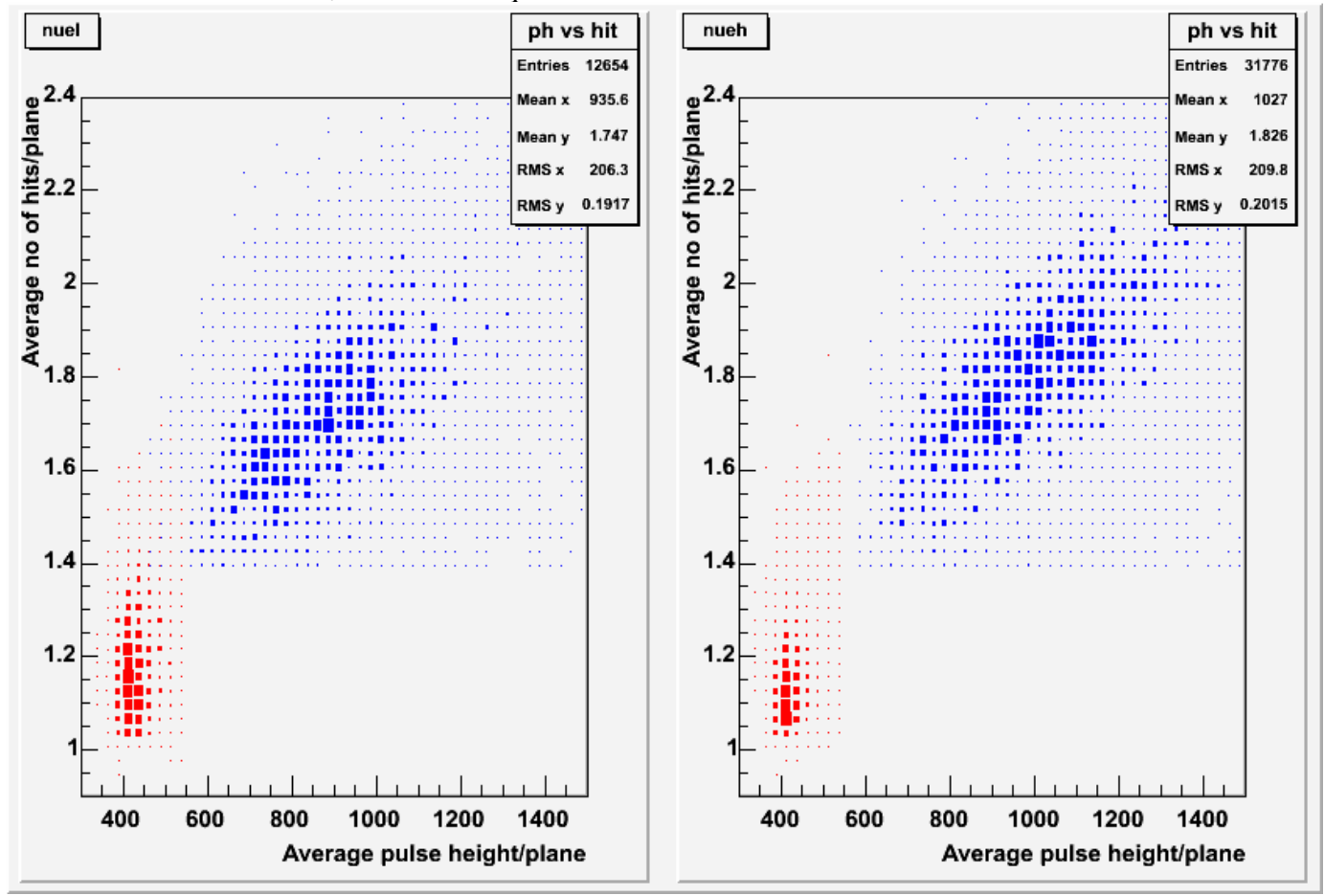

Fig. 12.5: Average pulse height per plane versus average number of hits per plane plotted for low energy events (left) and high energy events (right). Electrons are in blue; muons are in red. See text for additional details.

## 12.4. Typical Events

Figures 12.6-12.9 present a few typical events that illustrate the performance of the detector, using one example of each category of events: passing $\nu_e$ charged-current, failing $\nu_e$ charged-current, passing neutral current background and passing $\nu_\mu$ charged-current background. These events are typical in so far that no special effort has been made to select them. They illustrate the most important characteristics of the different categories of events: the passing $\nu_e$ events tend to be rather clean without much extraneous pulse height and with most of the energy in the electron. On the other hand the failing $\nu_e$ events have most often a low energy electron. Both NC and $\nu_\mu$ CC background events tend to have an energetic $\pi^0$ that is called an electron and the muon from the $\nu_\mu$ CC background events has rather low energy. In addition these background events tend to be somewhat "messier".

Each figure has the *x-y* view on top and the *y-z* view on the bottom. The indicated color code represents the relative pulse height of the hits. The lines represent the trajectories of the final state particles with the following color code: charged leptons in red, charged pions in blue, protons in black, and $\pi^0$ in green. The length of the colored trajectory is proportional to the energy of the particle but is not its expected length in the detector.

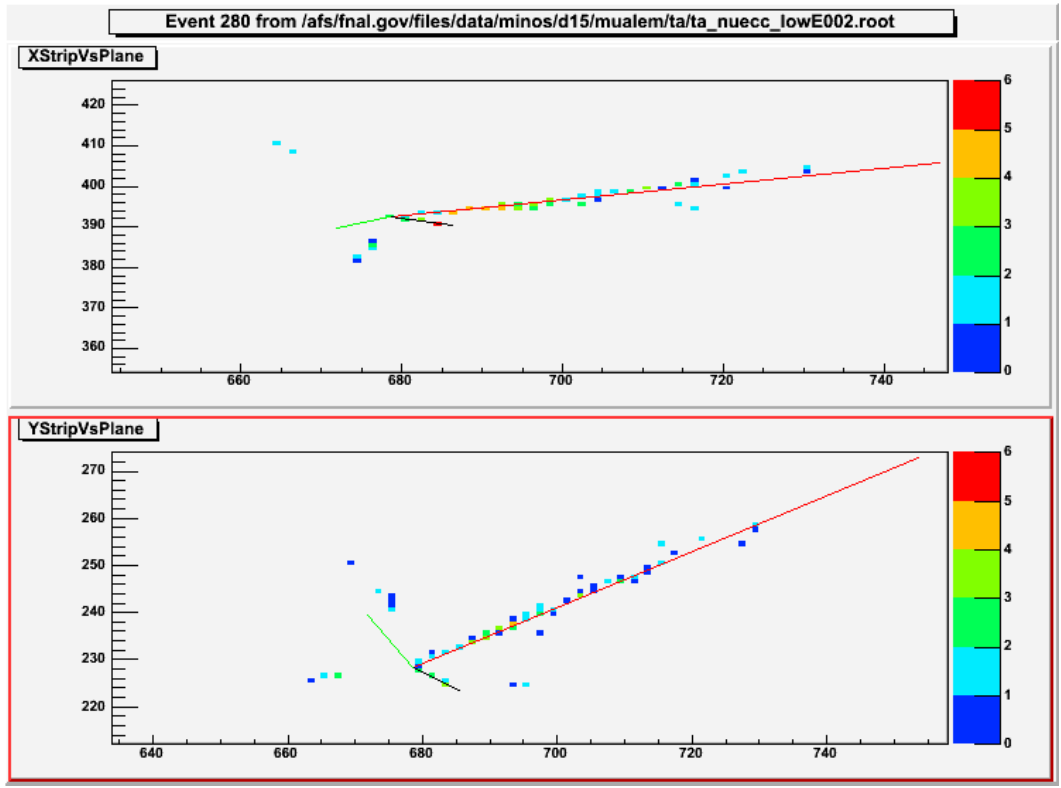

Fig. 12.6: An accepted $\nu_e$ charged-current event : $\nu_e A \rightarrow pe\pi^0$ , $E_\nu$ = 1.65 GeV. See text for explanation of the codes.

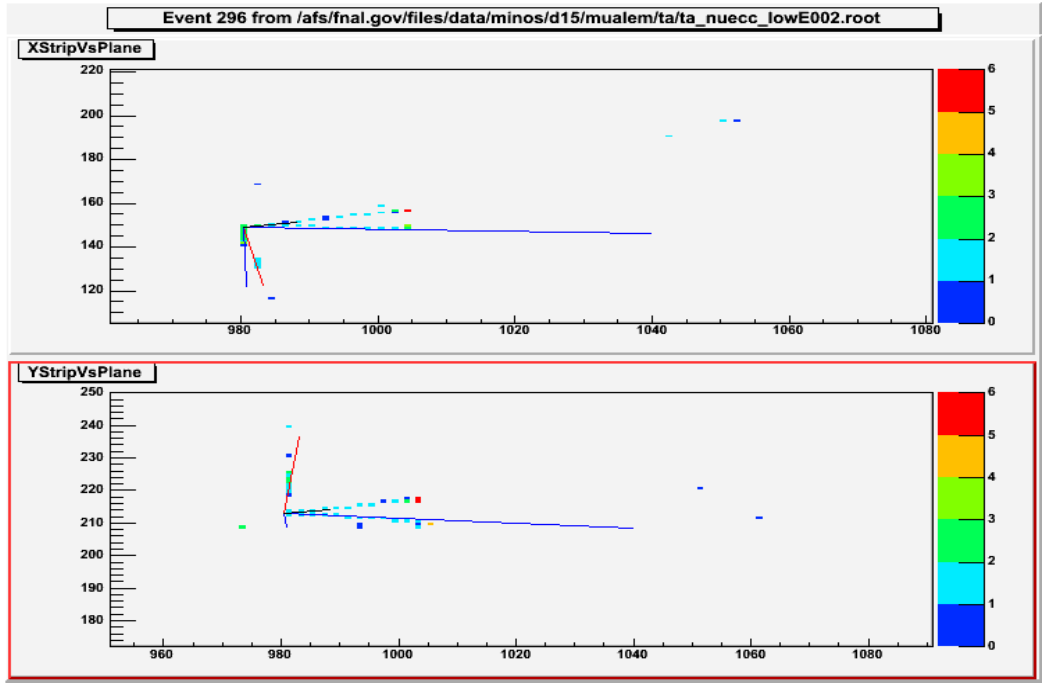

Fig. 12.7: A failing $\nu_e$ charged-current event : $\nu_e A \to pe\pi^+\pi^-$, $E_\nu$ = 1.87 GeV. See text for explanation of the codes.

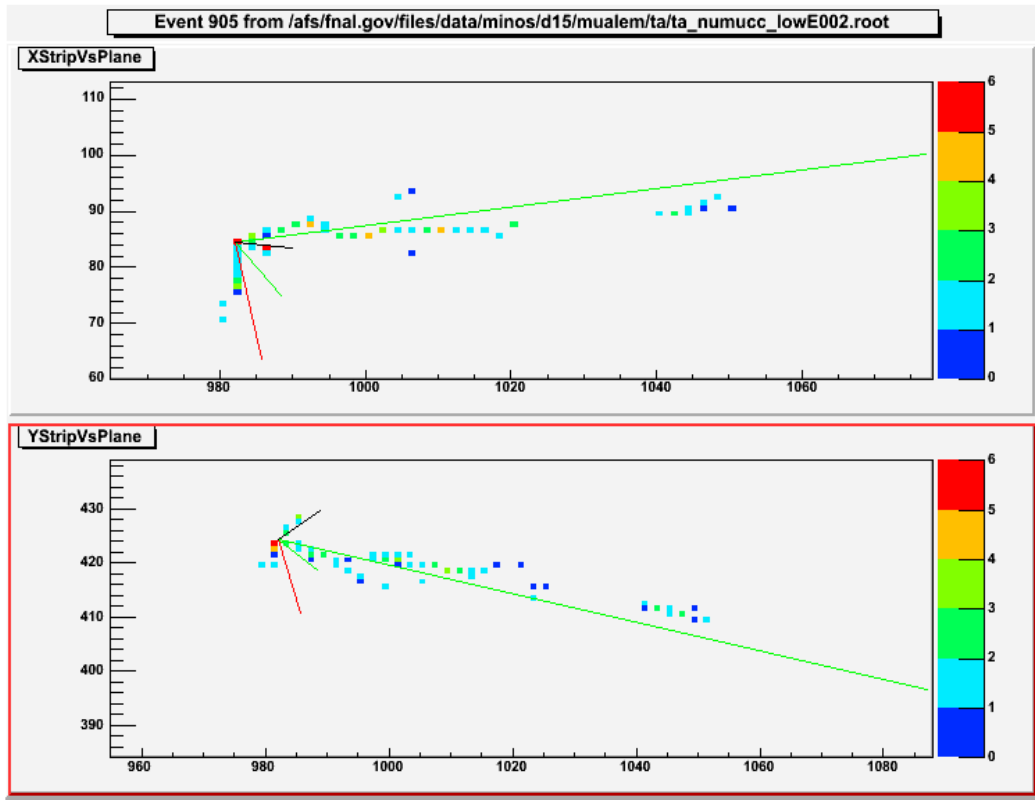

Fig. 12.8: A background $\nu_\mu$ charged current event: $\nu_\mu A \to p\mu\pi^0\pi^0$, $E_\nu$ = 1.70 GeV. See text for explanation of the codes.

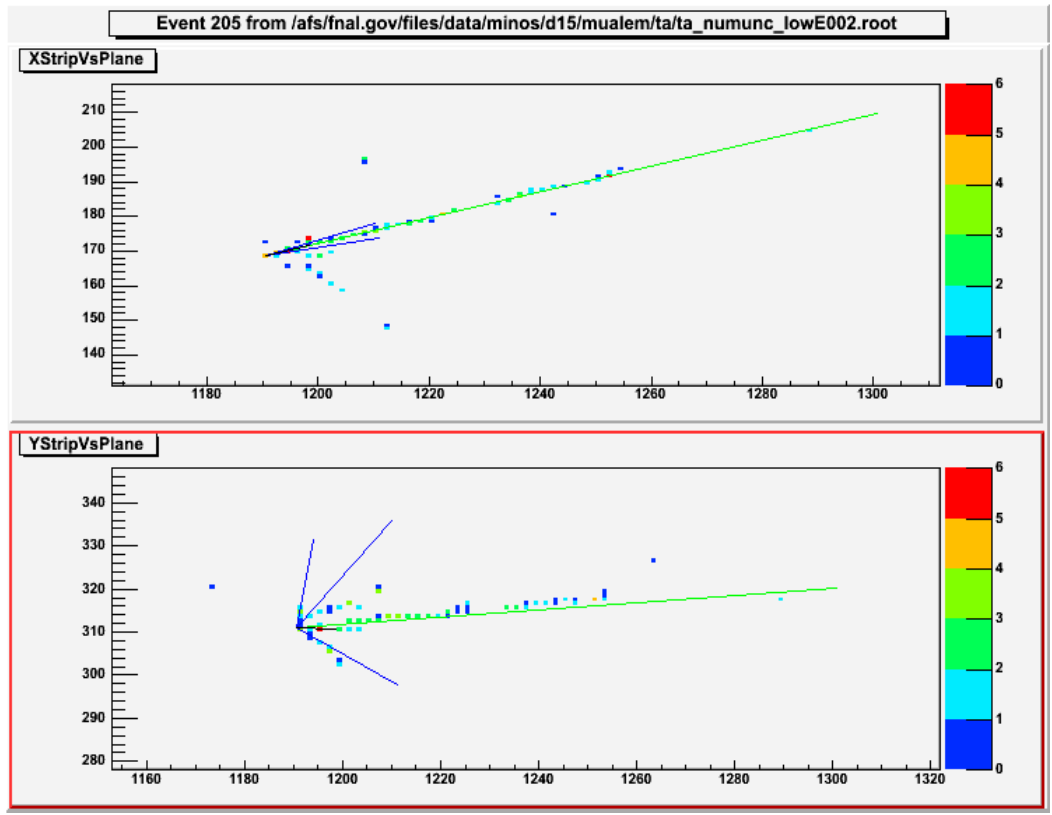

Fig. 12.9: A background neutral-current event: $\nu A \rightarrow p\nu\pi^+\pi^-\pi^-\pi^0$, $E_\nu = 4.95$ GeV. See text for explanation of the codes.

## 12.5 Prospects

The combination of the high level of segmentation and the "totally active" nature of NOνA yields a large amount of information for each event. While the simulations described here have attempted to use most of that information, it is unlikely that we have already found the optimum ways of increasing the efficiency for signal and the rejection of background. Thus, the results presented here should be considered as a lower bound on the ultimate performance of the detector

# 13. Physics Potential of NOνA

## 13.1. Introduction

Assuming that light sterile neutrinos either do not exist or do not mix with active neutrinos, there are currently three parameters of neutrino oscillations about which we have no information or only upper limits: $\sin^2(2\theta_{13})$, the sign of $\Delta m^2_{32}$ (i.e., whether the solar oscillation doublet has a higher or lower mass than the third state which mixes in the atmospheric oscillations), and the CP-violating phase δ.

The goal of NOvA will be to acquire information on all three of these parameters. However, provided that $\theta_{13}$ is in the range accessible to conventional neutrino beams, the unique contribution of the NuMI neutrino program will be the resolution of the mass hierarchy. This can only be done by experiments that measure the matter effect due to $\nu_e$'s traveling long distances through the earth. Planned future experiments in both Japan [1] and Europe [2] are concentrating on baselines that are too short for this purpose.

The determination of whether the solar neutrino doublet is at a higher or lower mass than the third neutrino mass state is important in its own right, for interpreting neutrinoless double beta decay experiments, and for the eventual measurement of CP violation in the lepton sector. As an example of the last, consider Fig. 13.1, which is taken from the T2K LoI [1]. The T2K collaboration is proposing a very ambitious long-term program to make precision measurements of CP violation by increasing the JPARC proton intensity by a factor of 5 (to 4 MW) and by building a new detector, HyperKamiokande, which will have twenty times the mass of SuperKamiokande. Fig. 13.1. shows the numbers of $\nu_e$ and $\bar{\nu}_e$ appearance events with two years of neutrino and six years of antineutrino running for $\sin^2(2\theta_{13}) = 0.1$. It is clear that without a resolution of the mass hierarchy, there are large areas of the parameter space in which the CP phase cannot be determined with any precision. The JPARC program is relying on the NuMI program for this information. This will be made quantitative in Section 13.5.

Given this unique role for the NOνA

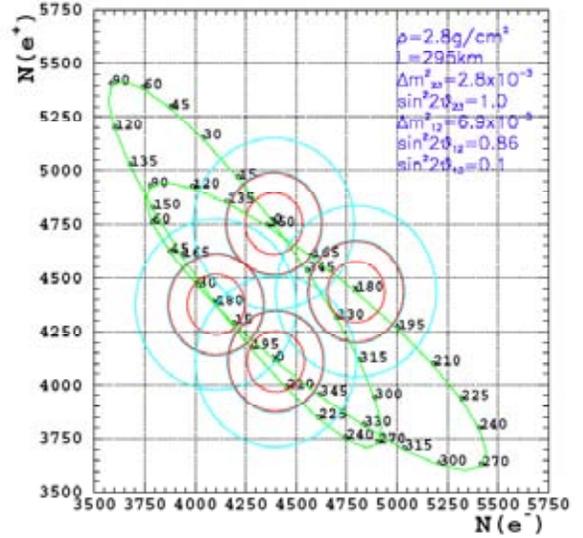

Fig. 13.1: The numbers of $\nu_e$ and $\bar{\nu}_e$ appearance events with two years of neutrino and six years of antineutrino running for $\sin^2(2\theta_{13}) = 0.1$ in an anticipated experiment utilizing an upgraded JPARC proton beam and the HyperKamiokande detector. Each of the two green contours corresponds to the different mass hierarchy and the numbers on the contours are the CP phase in degrees. The red circles correspond to the 90% confidence level contours and the blue circles correspond to three standard deviation contours. The outer circles include errors due to a 2% systematic uncertainty. From the T2K LoI [1].

experiment, we believe it should be designed and sited to optimize this role.

The next section will introduce the problem of optimizing the siting of NOvA. Section 13.3 will discuss the sensitivity of NOvA to the observation of a signal in $\nu_\mu \to \nu_e$ oscillations, which is related to the sensitivity to $\sin^2(2\theta_{13})$. Section 13.4 and 13.5 will explore the sensitivity of NOvA to the neutrino mass ordering and the CP-violating phase δ, respectively. These sections will also show how NOvA fits into a long-range step-by-step program for the measurement of these parameters.

We will conclude this chapter with a discussion of other physics that NOvA can do. Section 13.6

will discuss how NOvA can improve on MINOS's measurements of $\sin^2(2\theta_{23})$ and $\Delta m^2_{32}$. Sections 13.7 and 13.8 will discuss measurements that could be made with the NOvA Near Detector, and Section 13.9 will discuss the NOvA sensitivity to galactic supernova explosions.

## 13.2 The Optimization Problem

There are two aspects to the optimization problem. The first is illustrated in Fig. 13.2, which shows all of the values of the parameters consistent with a (perfectly measured) 2% $\nu_\mu \to \nu_e$ oscillation probability 12 km off axis at an 810 km baseline. There are three parameters, $\sin^2(2\theta_{13})$, shown on the vertical axis, the two possible mass orderings, the normal hierarchy, shown by the solid blue curve and the inverted hierarchy, shown by the dashed red curve, and the CP phase δ, shown as values around the ellipses. The horizontal axis shows the result of a (perfect) measurement of the $\bar{\nu}_\mu \to \bar{\nu}_e$ oscillation probability.[9]

NOvA is capable of making two measurements, the neutrino and the antineutrino oscillation probabilities near the first oscillation maximum. In some cases, these two measurements are capable, in principle, of measuring all three parameters, up to a two-fold ambiguity in the CP phase. For example a neutrino oscillation probability of 2% and an antineutrino oscillation probability of 4% or 1%, determine the mass hierarchy unambiguously. However, a neutrino oscillation probability of 2% and an antineutrino oscillation probability of 2% cannot resolve the inherent ambiguity shown in Fig. 13.2. A third measurement is needed in this case, either from an experiment done elsewhere at a different baseline, or from an additional measurement on the NuMI beamline, for example, on the second oscillation maximum.

---

[9] Figs. 13.2 and 13.3 are drawn assuming that $\sin^2(2\theta_{23}) = 1.0$. If it is less than unity, then there will be a two-fold ambiguity in the value of $\sin^2(2\theta_{13})$ derived from $\nu_\mu \to \nu_e$ oscillations since the "atmospheric scale" oscillation probability is largely proportional to $\sin^2(\theta_{23})$. Since this factor is the almost the same for all $\nu_\mu \to \nu_e$ oscillation experiments, it will not affect the resolution of the mass hierarchy or the determinations of the CP-violating phase δ by these experiments. It will, however, affect the comparison of these experiments to reactor experiments, and may eventually be resolved by the comparison of precise reactor and accelerator oscillation experiments.

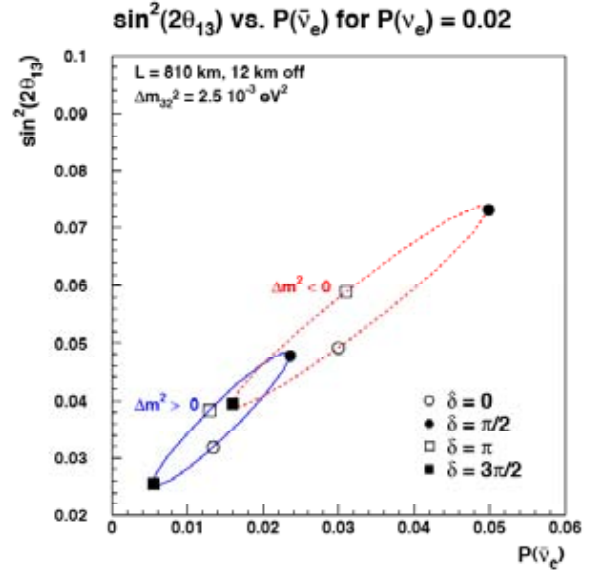

Fig. 13.2: Plot of the possible results of a measurement of a 2% neutrino oscillation probability. See text for an explanation.

Figure 13.3 shows the same information as Fig. 13.2, except for neutrino oscillation probabilities of 0.5%, 1%, 2% (again), and 5%. This figure illustrates that the fraction of possible δ values for which there is an ambiguity increases with decreasing values of $\theta_{13}$.

The second aspect of the optimization problem is illustrated in Fig. 13.4. The figure of merit (FoM) squared and the neutrino asymmetry are plotted as a function of the off-axis transverse angle for $\Delta m^2_{32} = 0.0025$ eV$^2$. The FoM is defined as the signal divided by the square root of the background. It is proportional to the sensitivity (in standard deviations) for seeing an oscillation signal, and the inverse of its square is proportional to amount of detector mass times beam flux required to obtain a given result. The neutrino asymmetry is defined as the neutrino oscillation probability minus the antineutrino probability divided by their sum, due to the matter effect. Thus, it is a measure of how far the two ellipses separate in Figs. 13.2 and 13.3. The ability to resolve the mass hierarchy will depend on both the rate of events as given

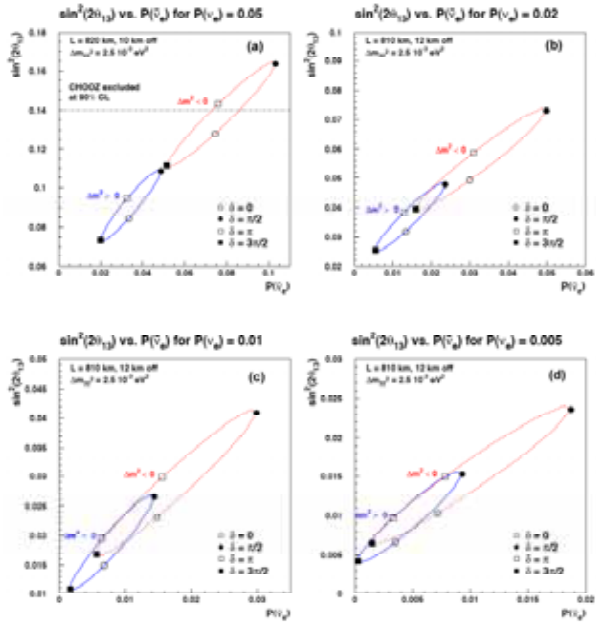

Fig. 13.3: Plots of the possible results of a measurement of a (a) 5%, (b) 2%, (c) 1%, and (d) 0.5% neutrino oscillation probability. See text for an explanation.

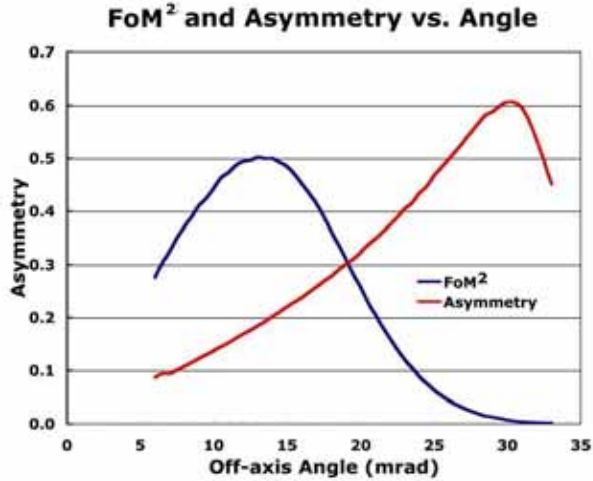

Fig. 13.4: Figure of merit squared (arbitrary units) and neutrino oscillation asymmetry due to the matter effect for $\Delta m^2 = 0.0025$ eV$^2$ versus off-axis angle. See text for an explanation. This figure is for illustrative purposes. It is based on a toy model and may not agree precisely with the simulation data presented in this chapter.

by the FoM and separation given by neutrino asymmetry.

Figure 13.4 shows that the sensitivity to observing the oscillation will not optimize at the same place as the sensitivity to the mass hierarchy. In Section 13.4, we will see that siting NOvA at 12 km off-axis is optimum for resolving the mass hierarchy, and this is what we propose. However, in the next section we will show that optimizing for resolving the mass hierarchy results in only a small loss of sensitivity for seeing the oscillation. Further, this optimization will be approximately correct for each possible future stage of the evolution of the NOvA program, and it is largely insensitive to the value of $\Delta m^2_{32}$ within the range suggested by the latest SuperKamiokande and K2K analyses.[10]

The conclusion of this chapter will be that NOvA is optimized for a long-range program that is capable of resolving the mass hierarchy over most of the range accessible to conventional neutrino beams. In addition, we will show that with the construction of a Proton Driver at Fermilab, NOvA will have a substantial capability to measure CP violation, both alone and in combination with other experiments.

### 13.3. Sensitivity to Observing $\nu_\mu \to \nu_e$ Oscillations

Figures 13.5 and 13.6 show the calculated three standard deviation discovery limit for $\nu_\mu \to \nu_e$ oscillations in terms of the three unknown parameters, assuming $\Delta m^2_{32} = 0.0025$ eV$^2$. The vertical axis represents the fraction of possible $\delta$ values for which a 3-$\sigma$ discovery could be made. At a fraction of 1.0, a 3-$\sigma$ discovery can be made for all values of $\delta$. This sets the $\sin^2(2\theta_{13})$ limit for a certain discovery. At lower values of $\sin^2(2\theta_{13})$, a 3-$\sigma$ discovery is only possible for a range of delta. When there is no value of $\delta$ that gives a 3-$\sigma$ discovery, the fraction is 0.0, and this sets the lower limit for $\sin^2(2\theta_{13})$ at which a discovery is possible. A fraction of 0.5 may be taken as the typical value.

---

[10] See Section 3.2 for a summary of results. Both the SuperKamiokande analysis of high-resolution events and the K2K analysis give a 90% confidence level lower limit for $\Delta m^2_{32}$ of 0.0019 eV$^2$.[3]

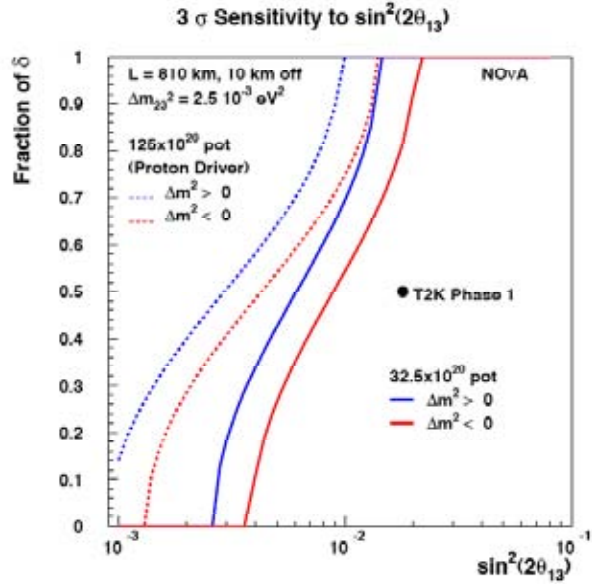

Fig. 13.5: Three standard deviation discovery limits for the observation of $\nu_\mu \to \nu_e$ oscillations for the NOvA detector situated 10 km off the NuMI beamline. See the text for more details.

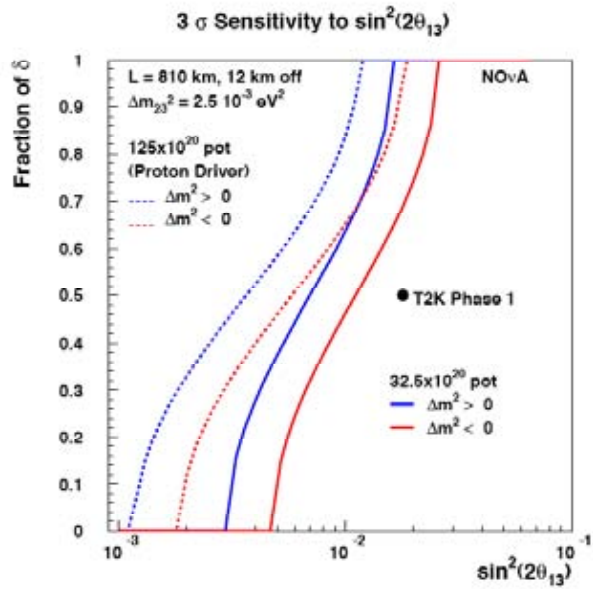

Fig. 13.6: Three standard deviation discovery limits for the observation of $\nu_\mu \to \nu_e$ oscillations for the NOvA detector situated 12 km off the NuMI beamline. See the text for more details.

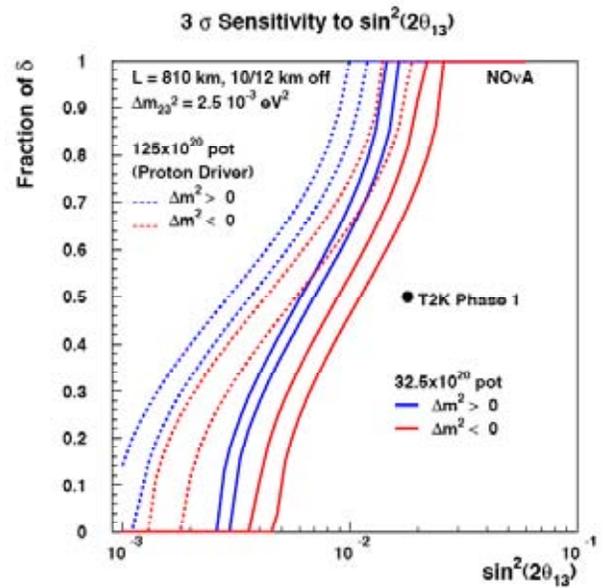

Fig. 13.7: Data from Figs. 13.5 and 13.6 superimposed for comparison purposes.

The curves represent the two possible values of the sign of $\Delta m^2_{32}$ and different assumptions on the number of protons on target (pot) that the experiment might see in a five-year run. (If these figures are being viewed in gray scale, the line to the right for each number of protons represents the inverted mass hierarchy.)

The value of $35.2 \times 10^{20}$ pot represents our estimate of what Fermilab might be able to deliver in a five-year run as discussed in Chapter 11, while $125 \times 10^{20}$ pot represents the expectation with the Booster replaced by a new Proton Driver. A 5% systematic error on the background determination has been included in these and the other calculations presented in this chapter, but as can be seen from Table 12.2, the statistical errors on the backgrounds always dominate. The three standard deviation sensitivity of the T2K phase 1 proposal [1] is also shown in these figures.

Figures 13.5 and 13.6 differ in that the former displays data for the NOvA detector situated 10 km off-axis, while the latter is for 12 km off-axis. There is some loss of sensitivity in going from 10 to 12 km. This is best seen in Fig. 13.7, which superimposes the data from the previous two figures. There is only a minor loss of sensitivity for the normal mass hierarchy, because the larger matter effects at 12 km enhance the neutrino oscillation probability. The loss is somewhat larger, but still relatively small, for the inverted mass hierarchy, where the matter effects suppress the neutrino oscillation probability.

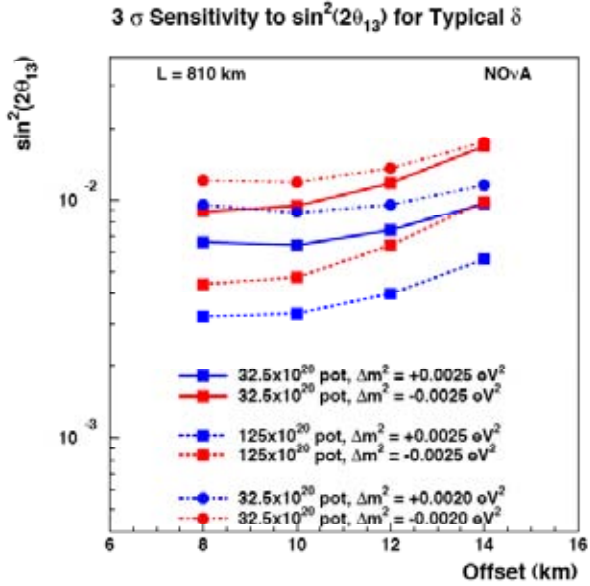

Fig. 13.8: Three standard deviation discovery limits for the observation of $\nu_\mu \to \nu_e$ oscillations for the typical CP phase δ versus the NOvA detector off-axis distance for the integrated fluxes and $\Delta m^2$ values shown.

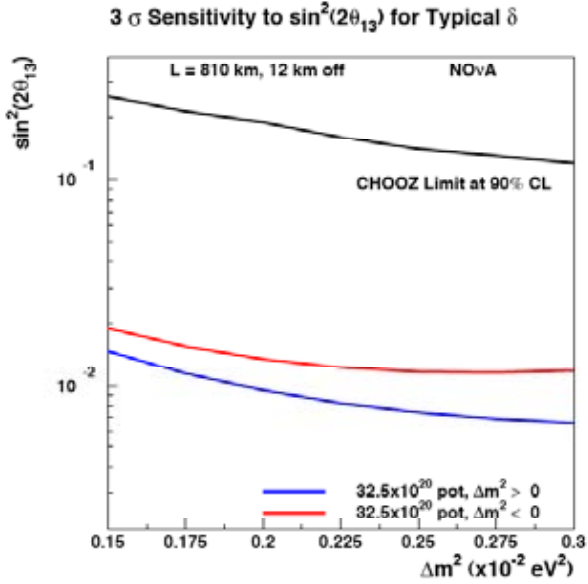

Fig. 13.9: Three standard deviation discovery limits for the observation of $\nu_\mu \to \nu_e$ oscillations for the typical CP phase δ versus $\Delta m^2_{32}$ for the NOvA detector sited 12 km off axis

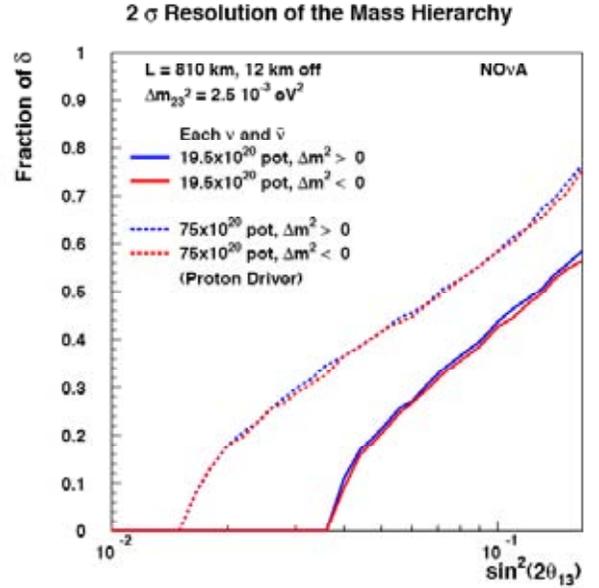

Fig. 13.10: The 95% confidence level resolution of the mass hierarchy versus $\sin^2(2\theta_{13})$ for three years of running each neutrinos and antineutrinos, with and without a proton driver.

Figure 13.8 shows the three standard deviation discovery limits for the typical δ for both $\Delta m^2_{32} = 0.0025$ eV$^2$ and $\Delta m^2_{32} = 0.0020$ eV$^2$ as a function of the off-axis distance. For all cases, the sensitivity maximizes around 8 to 10 km off-axis. Figure 13.8 also shows the loss of sensitivity going from $\Delta m^2_{32} = 0.0025$ eV$^2$ to $\Delta m^2_{32} = 0.0020$ eV$^2$. However, it should be noted that this is not a loss in range, since the CHOOZ limit [4] is correspondingly weaker at 0.0020 eV$^2$. This is further illustrated in Fig. 13.9, which shows the three standard deviation discovery limits for the typical δ for NOvA sited at 12 km off axis as a function of $\Delta m^2_{32}$.

### 13.4. Sensitivity to the Mass Hierarchy

*13.4.1. NOvA Alone:* Figure 13.10 shows the 95% confidence level resolution of the mass hierarchy as a function of $\sin^2(2\theta_{13})$ for the NOvA detector sited at 12 km off-axis. The 95% confidence level has been chosen since the mass hierar-chy is binary, so 20:1 odds should be reasonably convincing. The assumed scenario is that within three years of neutrino running, a three-σ signal is observed for $\nu_e$ appearance, after which the running is switched to antineutrinos for studying the mass hierarchy. Thus, Fig. 13.10 assumes three years of each neutrino and antineutrino running, both with and without a proton driver.

The shapes of the curves are easily understood from Fig. 13.3. There is a limited range of δ values for which two measurements can resolve the mass hierarchy, and this range decreases with decreasing values of $\sin^2(2\theta_{13})$. There is a reasonable region of parameter space in which NOvA could resolve the mass hierarchy before a Proton Driver is available, and a larger region after.

To emphasize the point that only a long baseline experiment can resolve the mass hierarchy, we have calculated the sensitivity of T2K phase 1, if it were to run for three years each on neutrinos and antineutrinos. This is shown in Fig. 13.11. The horizontal scale has been expanded in order to show the T2K sensitivity, which otherwise would be off-scale to the right. The CHOOZ limit for $\Delta m_{32}^2 = 0.0025$ eV$^2$ is also indicated [4]. Points substantially to the right of this limit are largely irrelevant. We emphasize that the results for T2K are our calculations, since the T2K collaboration, quite sensibly, has not proposed this measurement.

Figure 13.12 shows the mass hierarchy resolution sensitivity for all of the simulations in Table 12.2. This figure displays the value of $\sin^2(2\theta_{13})$ for which the δ value is at the limit of first quartile, i.e., the δ value such that 25% of δ values give a lower value of $\sin^2(2\theta_{13})$ and 75% give a higher value. This δ was chosen because the typical δ is in the region of the CHOOZ limit for running before the Proton Driver, and thus less relevant. However, the siting optimization does not depend significantly on which δ value is chosen.

Fig. 13.12 shows that the mass hierarchy resolution optimizes around 12 km off-axis for both $\Delta m_{32}^2 = 0.0025$ and 0.0020 eV$^2$. It appears to optimize between 10 and 12 km for $\Delta m_{32}^2 = 0.0025$ eV$^2$ and between 12 and 14 km for $\Delta m_{32}^2 = 0.0020$ eV$^2$.

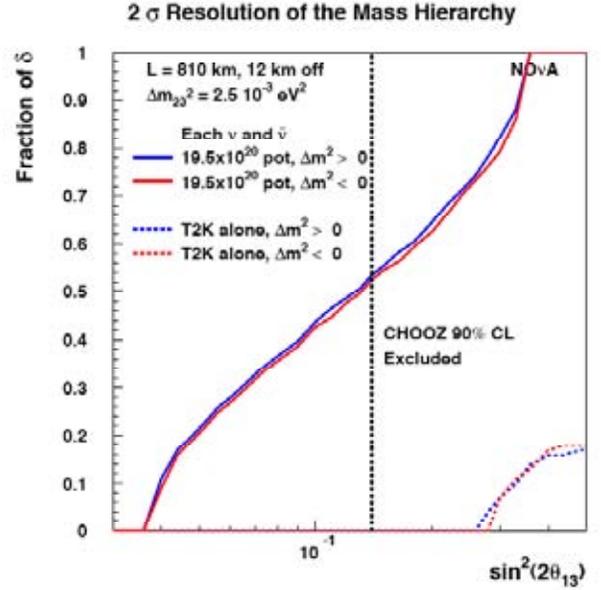

Fig. 13.11: A comparison of NOvA's and T2K's abilities to resolve the mass hierarchy alone.

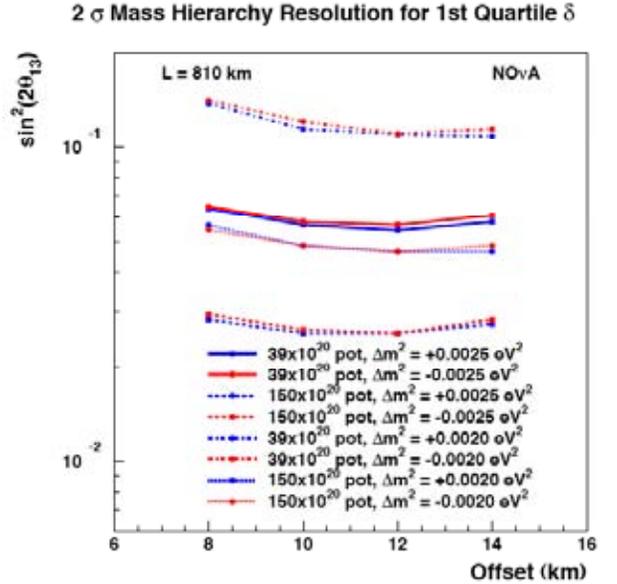

Fig. 13.12: 95% confidence level for resolution of the mass hierarchy for the 1$^{st}$ quartile δ. See the text for additional explanation.

*13.4.2: NOvA in Combination with Another Measurement:* If the neutrino oscillation parameters are such that the mass hierarchy cannot be resolved by NOvA alone, then combining NOvA measurements with the measurement of another detector will be necessary. The most obvious candidate is T2K. Figures 13.13 and 13.14 show these results. Figure 13.13 is for NOvA without the Proton Driver combined with T2K phase 1. Figure 13.14 is for a later time in which NOvA with the Proton Driver can be combined with T2K with an upgraded proton source. For this later case, we have calculated the results assuming either that the T2K detector is SuperKamiokande or HyperKamiokande.

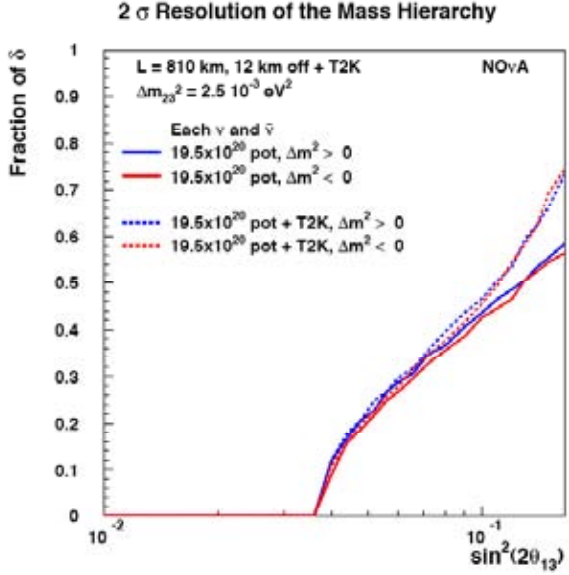 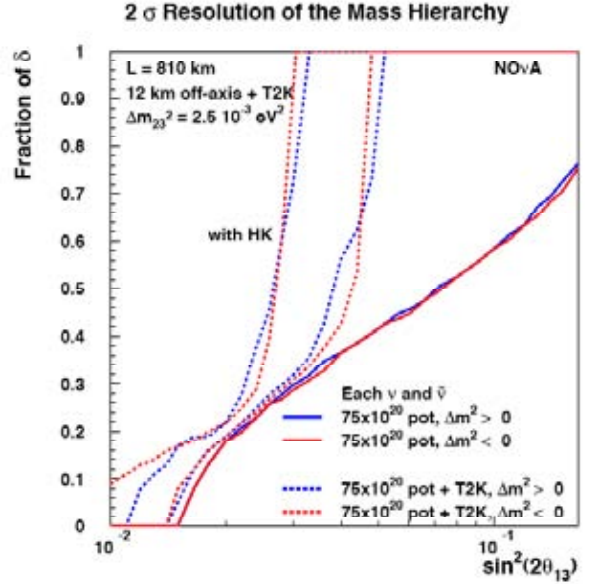

Fig. 13.13: A comparison of the 95% confidence level resolution of the mass hierarchy with NOvA alone (solid curves) and the combination of NOvA and T2K phase 1 data (dashed curves). It is assumed that both NOvA and T2K run three years each on neutrinos and antineutrinos.

Fig. 13.14: A comparison of the 95% confidence level resolution of the mass hierarchy with NOvA alone with the Proton Driver (solid curves) and the combination of NOvA and T2K data with an upgraded proton source (dashed curves). The curves labeled "HK" assume that the T2K detector is HyperKamiokande; the other set of dashed curves assume that it is SuperKamiokande. It is assumed that both NOvA and T2K run three years each on neutrinos and antineutrinos.

The structure of these plots is that the combination with T2K does not have much effect until a critical value of $\sin^2(2\theta_{13})$, after which the mass hierarchy is resolved for all values of δ. The reason for this is fairly easy to understand. We are comparing two distributions that have approximately the same structure due to the CP phase, and differ primarily by a factor of 2.3 in the matter effect. Thus, sufficient statistics to pass the 95% confidence level threshold happens for all values of δ at approximately the same point.

The difference between the critical value of $\sin^2(2\theta_{13})$ for HyperKamiokande is only about 30 to 40% lower than that for SuperKamiokande, even though the former has twenty times the mass of the latter. This is because the statistical precision is limited by the number of events in NOvA.

If comparisons with T2K are insufficient to resolve the mass hierarchy, then an attractive approach would be to do a measurement with an additional detector on the NuMI beamline to measure events at the second oscillation maximum. At the second maximum the matter effect is smaller by a factor of three and the CP violating effects are larger by a factor of three, both for the same reason – the energy is smaller by a factor of three.

There will be sufficient information available at that time that it will be known whether this technique will work and how much detector mass will be required. For the purpose of our calculation, we have adopted the following scenario. After two years of running with the Proton Driver, it is realized that a second off-axis detector will be needed and it is constructed in four years and then runs for an additional six years. Thus, there will be twelve years of NOvA data with a Proton Driver and six years of data with the second detector, both split equally between neutrinos and antineutrinos. It is not clear at this time what technology would be best for the second detector. Liquid scintillator, water Cerenkov, and liquid argon detectors are all reasonable candidates. For the purpose of this calculation, we have just assumed a 50 kT detector with efficiencies equal to those of T2K. The detector is assumed to be sited 30 km off axis at a baseline of 710 km. Since we want to minimize the matter effects in this detector, there is no reason to place it at a longer baseline. The results are shown in Fig. 13.15. The mass hierarchy is resolved for all values of δ for values of $\sin^2(2\theta_{13})$ greater than 0.01 to 0.02.

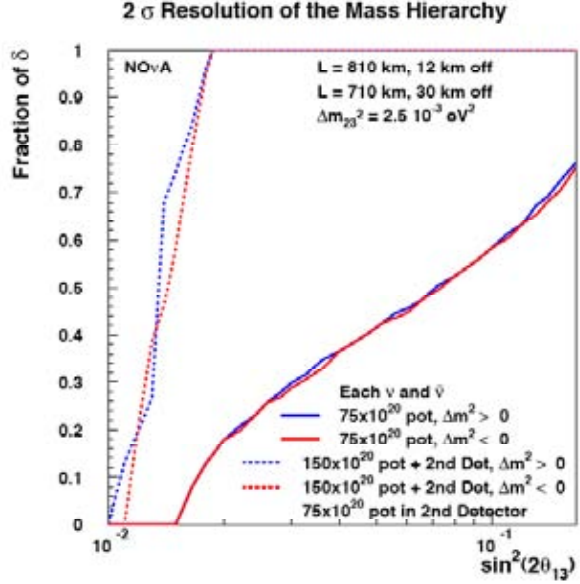

Fig. 13.15: A comparison of the 95% confidence level resolution of the mass hierarchy with NOvA alone (solid curves) and the combination of NOvA and an additional NuMI detector sited to measure the second oscillation maximum (dashed curves). See the text for details of the scenario.

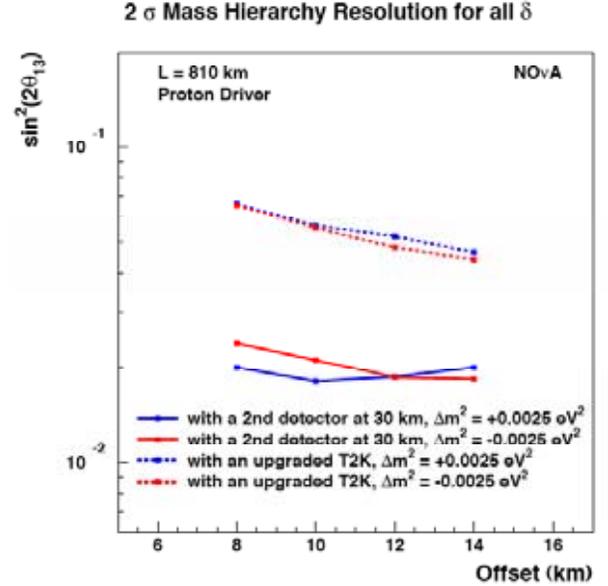

Fig. 13.16: 95% confidence level for resolution of the mass hierarchy for all values of δ for various NOvA off-axis distances. The dashed lines are for a combination of NOvA data with the Proton Driver and T2K data with an upgraded proton source and SuperKamiokande as the T2K detector. The solid lines are for a combination of NOvA data with an additional NuMI detector as discussed in the text.

Figure 13.16 addresses the siting optimization for combinations of NOvA data with T2K data or with that of an additional NuMI detector. It displays the value of $\sin^2(2\theta_{13})$ at which the mass hierarchy is resolved at the 95% confidence level for all values of δ. For the comparison of NOvA data with that from T2K, the optimum off-axis distance appears to be near 14 km. For the comparison of NOvA data with a second off-axis detector, the optimum is shallow and different for the two mass orderings. For the normal ordering, it optimizes at 10 km, while for the inverted ordering, it optimizes at 14 km.

Thus, based on our present knowledge of $\Delta m^2_{32}$, it appears siting NOvA 12 km off axis is reasonable for all stages of the NuMI program. As we get more information on $\Delta m^2_{32}$ from MINOS, we can refine the optimization.

*13.4.3: Summary of the Evolution of the NOvA Program to Resolve the Mass Hierarchy:* Figure 13.17 summarizes the possible evolution of the NOvA program by combining the results shown in Figs. 13.9, 13.13, 13.14, and 13.15. The NOvA program allows the resolution of the mass hierarchy over most of the range in $\theta_{13}$ accessible to conventional neutrino beams. The program is flexible; each stage can be guided by the information obtained in prior stages, and the NOvA detector that we are proposing here remains a key and well-optimized participant throughout the program.

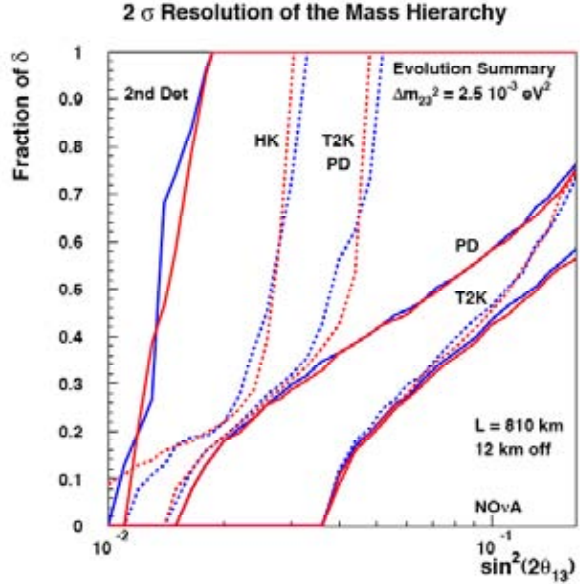

Fig. 13.17: A summary of the data presented in Figs 13.9, 13.13, 13.14, and 13.15.

## 13.5. Sensitivity to CP Violation

*13.5.1: Introduction*: The relationship between the resolution of the mass hierarchy and the observation of CP violation varies from experiment to experiment. Very short baseline experiments, such as the beta beam experiments being planned in Europe [2] have very small matter effects and can measure CP violation phase δ without regard to the determination of the mass hierarchy. Long baseline experiments such as NOνA generally require a resolution of the mass hierarchy to measure the CP phase because maximal CP violation for one mass ordering can have the same or similar neutrino and antineutrino oscillation probabilities as no CP violation for the other mass ordering. An example of this is shown in Fig. 13.3(c). Shorter baseline experiments such as T2K are intermediate between these extremes. This section will explore the capability of NOνA to measure the CP violating phase δ and the power of combinations of NOνA measurements with those of other experiments.

One should keep in mind that CP-violating effects are proportional to the first power of $\theta_{13}$, while CP-conserving effects are, for the most part, proportional to the square of $\theta_{13}$, as can be seen in Fig. 13.3. This has led some to argue that the ability to measure δ is independent, to some extent, of the value of $\sin^2(2\theta_{13})$. We will see that there are regions of $\sin^2(2\theta_{13})$ in which the probability of measurement is flat. We will also see that there can be peaks and dips in the probability as a function of $\sin^2(2\theta_{13})$ due to the complex relationship between CP-violating effects and matter effects.

In order to take this relationship into account, we use the following measure of our ability to measure CP violation: the fraction of possible δ values for which there is a three standard deviation demonstration of CP violation, that is, that δ is neither zero nor π for both mass orderings. Of course, this fraction can never be 100%, since there will always be some range of δ values very close to zero or π. A rough way to convert this measure into a one standard deviation measure of δ is that a small, but non-zero fraction corresponds to 30 degrees, a 25% fraction to 22.5 degrees, a 50% fraction to 15 degrees, and so on.

*13.5.2: Simulation Results:* Neither NOνA nor T2K can demonstrate CP violation even at the two standard deviation level with six years of running without an enhanced proton source. However, both experiments gain some ability with their proposed proton drivers. This is shown in Fig. 13.18, in which both experiments are assumed to have run three years each on neutrinos and antineutrinos and the T2K detector is assumed to be Super-Kamiokande. T2K and NOνA have a similar reach in $\sin^2(2\theta_{13})$, but T2K saturates at a lower fraction of δ due to its inability to resolve the mass hierarchy. Combining measurements from both experiments gives a large gain in both the breadth and precision of the measurement. The sharp rise around $\sin^2(2\theta_{13}) = 0.05$ is due to the resolution of the mass hierarchy, as discussed in Section 13.4.3 and seen in Fig. 13.14.

Fig. 13.19 shows the same information as Fig. 13.18, except that HyperKamiokande is assumed to be the T2K detector. The twenty-fold increase in mass gives it high statistical precision. The role of NOνA is to resolve the mass hierarchy so that the precision can be used, as was discussed in the opening section of this chapter.

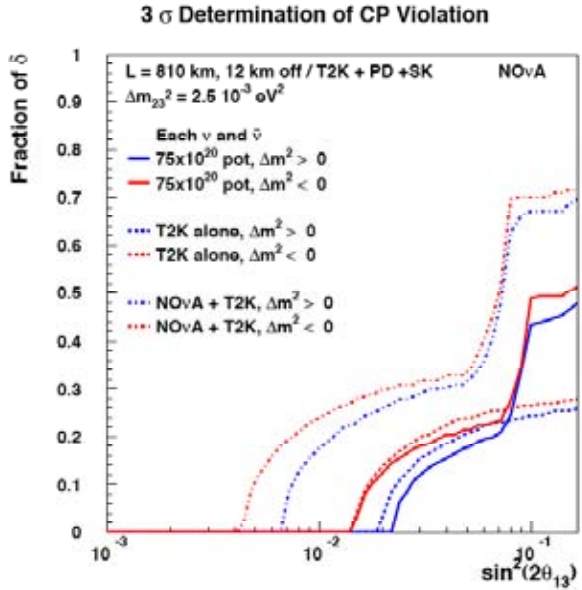

Fig. 13.18: The fraction of δ values for which CP violation can be demonstrated at three standard deviations. A three year run on each of neutrinos and antineutrinos is assumed for NOvA with the Proton Driver and for T2K with an enhanced proton source and SuperKamiokande as the detector.

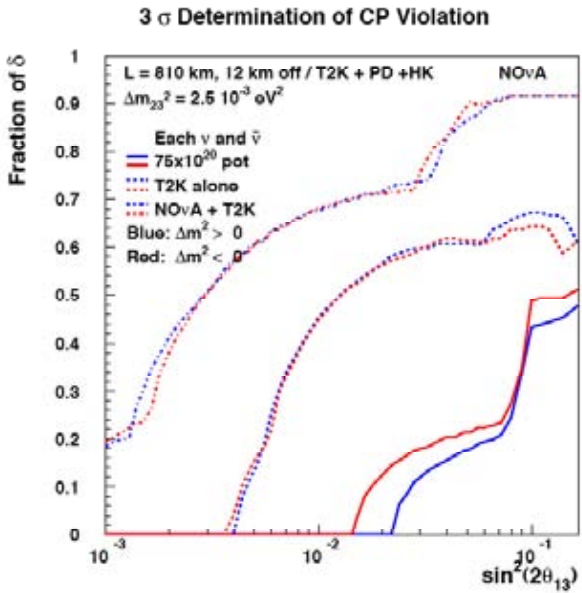

Fig. 13.19: The same as Fig 13.18 except that Hyper-Kamiokande is assumed to be the T2K detector.

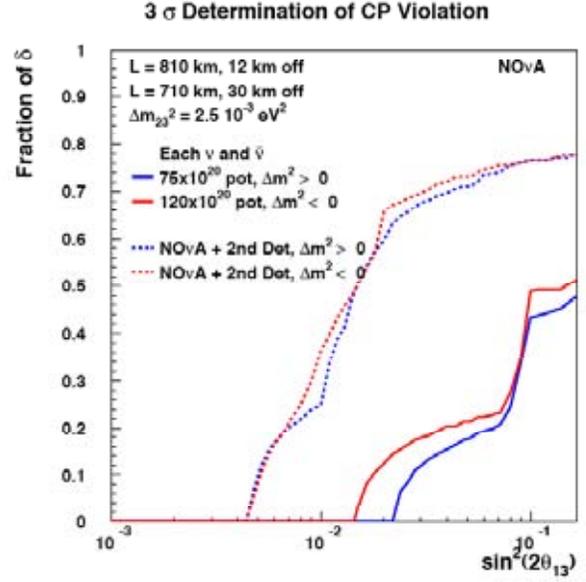

Fig. 13.20: The fraction of δ values for which CP violation can be demonstrated at three standard deviations for NOvA with the Proton Driver and combined with an additional detector on the NuMI beam line, as discussed in the text.

Finally, Fig. 13.20 addresses the CP violation measurements that could be made by a combination of NOvA and the additional detector on the NuMI beamline, running at the second oscillation maximum, which was suggested in Section 13.4.3 to resolve the mass hierarchy in the case of small values of $\sin^2(2\theta_{13})$. This figure shows that there is also a good capability for measuring CP violation at these $\sin^2(2\theta_{13})$ values.

## 13.6. Measurement of the Dominant Mode Oscillation Parameters

One of the most important measurements in neutrino physics today is the precise determination of $\sin(\theta_{23})$. The best current measurement comes from the SuperKamiokande study of atmospherically produced neutrinos [3,5]. This measurement is consistent with maximal mixing, $\sin^2(2\theta_{23}) = 1$, but with a considerable uncertainty. At the 90% confidence level, $\sin^2(2\theta_{23}) > 0.92$, which translates into a rather large range of possible values of $\sin^2(\theta_{23})$, namely $0.36 < \sin^2(\theta_{23}) < 0.64$.

There are three reasons why determining $\sin(\theta_{23})$ is of high interest:

(1) If the mixing is maximal, it might be due to some currently unknown symmetry.

(2) The $\nu_\mu \to \nu_e$ oscillation is mostly proportional to $\sin^2(\theta_{23})\sin^2(2\theta_{13})$ while $\bar{\nu}_e$ disappearance, measured by reactor experiments is proportional to $\sin^2(2\theta_{13})$. Thus, if the mixing is not maximal, there is an ambiguity in comparing accelerator and reactor experiments, or conversely

(3) whether $\theta_{13}$ is greater than or less than $\pi/4$, which measures whether $\nu_e$'s couple more strongly to $\nu_\mu$'s or $\nu_\tau$'s, can probably best be measured by comparing precise accelerator and reactor measurements.

The deviation of $\sin^2(2\theta_{23})$ from unity is measured by the depth of the oscillation dip in the $\nu_\mu$ disappearance spectrum. Thus, precision in this quantity requires good statistics in this region, excellent neutrino energy resolution, and good control of systematics. NOvA offers the possibility of satisfying all of these requirements.

It appears that the best way to meet these requirements is to limit the analysis to totally contained quasielastic events, i.e., those events in which the geometrical pattern of energy deposition is consistent with the presence of only an energetic muon and a possible recoil proton.

We have performed a preliminary study of how well NOvA can use these events to measure $\sin^2(2\theta_{23})$ and $\Delta m^2_{32}$ using a parametric representation of the energy. This procedure is justified by the nature of these events, which are extremely clean as is demonstrated by a typical quasielastic event displayed in Fig. 13.21.

With the exception of energy deposited in the PVC walls, which can be estimated from the trajectories, all of the final state energy should be visible in NOvA. The overall scale of unknown missing energy will be from the boiloff neutrons from the struck nucleon. The typical Fermi momentum is about 250 MeV/*c*, corresponding to a kinetic energy of the nucleon of about 33 MeV, or about 2% of the neutrino energy. Considering the various sources that contribute to the energy resolution, including photoelectron statistics, straggling fluctuations[11], saturation effects in the scintillator, undetected neutron emission, nuclear excitation, and reabsorption and rescattering in the nucleus, we conclude that the energy resolution should be in the 2 to 4% range [6]. The absolute energy scale can be determined from stopping cosmic ray muons and should be understood to better than 2%.

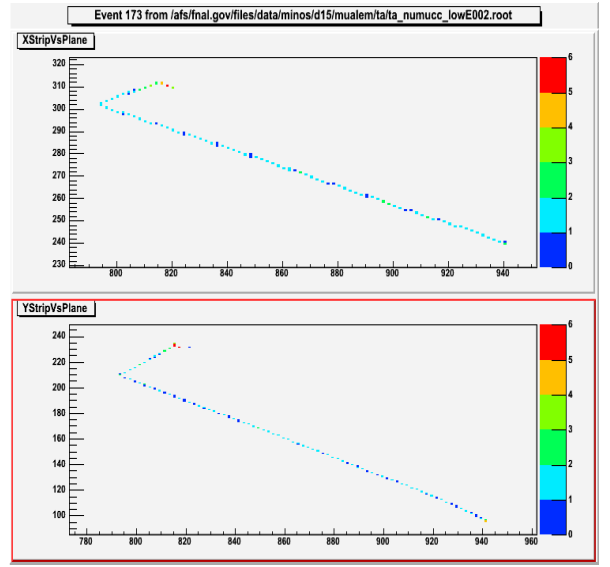

Fig. 13.21: An example of a quasielastic $\nu_\mu$ CC interaction in the NOvA detector. Note the proton scatter near the end of its range. The color of the hits indicates the relative pulse height.

The calculated one and two standard deviation contours are displayed in Figs. 13.22 and 13.23 for assumed values of $\sin^2(2\theta_{23})$ of 0.95, 0.98, and 1.00. Figure 13.22 is for a five-year neutrino run without a Proton Driver and Fig. 13.23 is for the same length run with a Proton Driver. The energy resolution has been assumed to be 2%, but the contours do not change markedly as one increases the resolution to 4%.

Note that the precision of the $\sin^2(2\theta_{23})$ measurement increases as the value of $\sin^2(2\theta_{23})$ approaches unity. For maximal mixing, the error on the measurement of $\sin^2(2\theta_{23})$ is about 0.004 without a Proton Driver and about 0.002 with.

---

[11] The main source of straggling fluctuations are the Landau fluctuations in the energy loss along the muon path. Since we measure all of the energy loss, this effect is relevant only for the straggling fluctuations in the PVC walls.

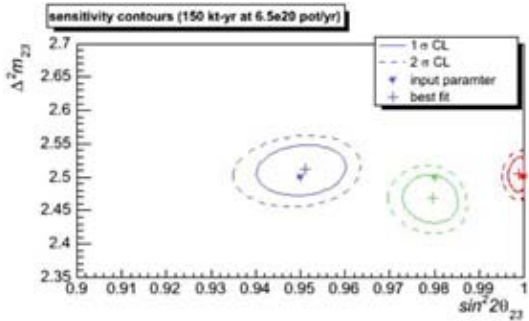

Fig. 13.22: One and two standard deviation contours for the simultaneous measurements of $\Delta m^2_{32}$ and $\sin^2(2\theta_{23})$ for a five-year neutrino run without a Proton Driver.

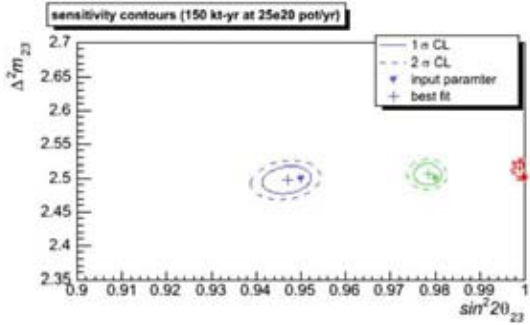

Fig. 13.23: One and two standard deviation contours for the simultaneous measurements of $\Delta m^2_{32}$ and $\sin^2(2\theta_{23})$ for a five-year neutrino run with a Proton Driver.

## 13.7. Short Baseline Neutrino Oscillation Measurements with the NOvA Near Detector

In Chapter 10 we noted that an LSND[7] oscillation would distort our beam $\nu_e$ spectrum. While a short baseline oscillation will complicate our background subtraction process at the Far Site, it also presents an opportunity. If MiniBooNE confirms the LSND signal, NOvA can expect to see hundreds of excess $\nu_e$ events in the Near Detector, providing additional confirmation for a result of immense importance.

At Site 1.5 (see Chapter 10), the $\nu_\mu$ spectrum peaks at about 2.5 GeV and has a width of about 1.5 GeV FWHM. In the extreme case of $\Delta m^2 = 2.5$ eV$^2$, 0.26% of the copious $\nu_\mu$ spectrum would oscillate into $\nu_e$'s. Our "beam $\nu_e$" measurement would be off by about 60% at the peak of the short baseline oscillation and we would overestimate our beam $\nu_e$ background at the far site. For other values of $\Delta m^2$, the effect is less pronounced but still significant, as shown in Fig. 13.24.

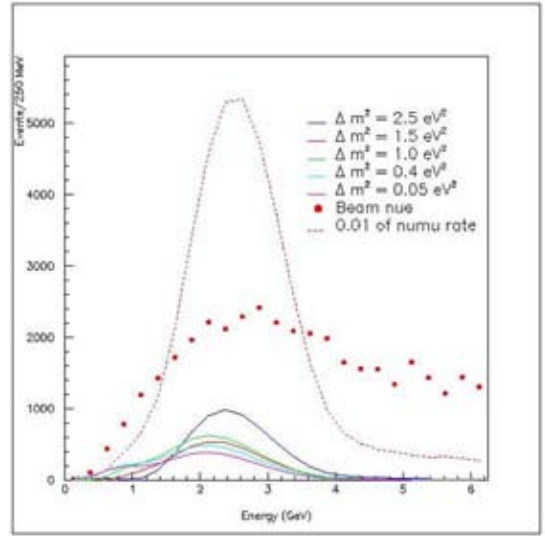

Fig. 13.24: The NOvA Near Detector (at Site 1.5) $\nu_e$ CC spectrum (red dots) compared with a short baseline oscillation effect at different values of $\Delta m^2$ consistent with the allowed LSND parameter space. The parent $\nu_\mu$ CC spectrum for the oscillation is shown at 1/100 of its value. No detector resolution effects or backgrounds are included here. The plot corresponds to 6.5 x 10$^{20}$ pot (about one year) on a near detector with 20.4 tons of fiducial mass.

NOvA is sensitive to effects over a large range of $\Delta m^2$, but can it differentiate one $\Delta m^2$ from another? Figure 13.25 shows the number of events at Site 1.5, after subtraction of the expected beam $\nu_e$ spectrum, for a number of different short baseline oscillation scenarios. We assume an oscillation with parameters of ($\Delta m^2$, $\sin^2 2\theta$) = (1.0 eV$^2$, 0.004) for a data run with 6.5 x 10$^{20}$ pot (about one year) on a near detector with 20.4 tons of fiducial mass and compare it to curves for oscillations with different $\Delta m^2$'s. The error bars are dominated by the statistical error on the expected beam $\nu_e$ events per bin. Systematic errors, backgrounds and detector resolutions are not included, though given the size of the signal we do not expect background to be an issue. It is apparent that some curves can be easily differentiated from others. Our ultimate sensitivity on $\Delta m^2$ and $\sin^2 2\theta$ will depend on how well we can control systematic errors.

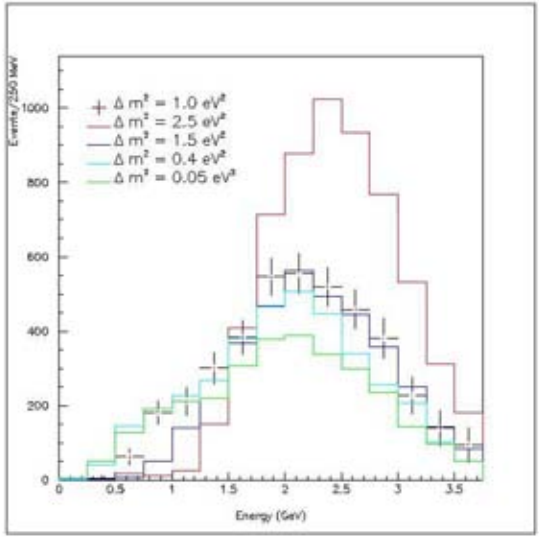 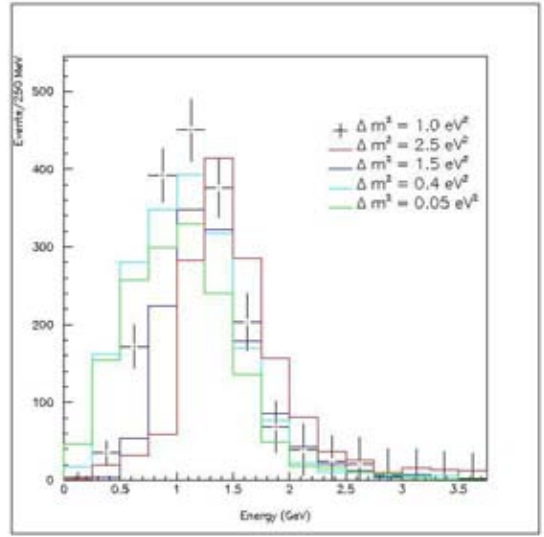

Fig. 13.25: Events after subtraction of the expected beam $\nu_e$ spectrum at Site 1.5 for a oscillation with parameters of ($\Delta m^2$, $\sin^2 2\theta$) = (1.0 eV$^2$, 0.004) for a data run with 6.5 x 10$^{20}$ pot (about one year) on a near detector with 20.4 tons of fiducial mass compared to histograms for other $\Delta m^2$'s. The error bars are dominated by the statistical error on the expected beam $\nu_e$ events per bin. Systematic errors, neutral current background and detector resolutions are not included.

Fig. 13.26: The same as Fig. 13.25, except for a near detector located at Site 3.

Near detectors in different locations could allow for better discrimination of the oscillation parameters. Figure 13.26 repeats the exercise illustrated in Fig. 13.25, but at Site 3 where the energy spectrum peaks at lower energy. Again it is apparent that some curves can easily be discriminated from others. The relative characteristics of the curves change from Fig. 13.25 to Fig. 13.26 due to the different neutrino energy spectra. This could allow for simultaneous fits to the spectra from both locations and result in a more accurate extraction of the oscillation parameters. Again, the ultimate accuracy that can be achieved will depend on systematic errors.

Even if MiniBooNE definitively rules out the LSND oscillation in $\nu_\mu \rightarrow \nu_e$, it is still possible that the oscillation occurs in $\bar{\nu}_\mu \rightarrow \bar{\nu}_e$ where LSND observed their most significant excess. MiniBooNE would now want to run in antineutrino mode, and the APS Joint Study on the Future of Neutrino Physics [8] has strongly recommended that the LSND result be tested with both neutrinos and antineutrinos.

MiniBooNE has recently noted [9] that there are new CP violation models in which the oscillation probability for $\bar{\nu}_\mu \rightarrow \bar{\nu}_e$ s can be three times as large as $\nu_\mu \rightarrow \nu_e$. However, the antineutrino production and interaction cross sections make this test more challenging than for the case of neutrino running. It seems clear that the NOvA Near Detector could also contribute to this antineutrino effort using the off-axis beam in antineutrino mode.

MiniBooNE also notes [9] that they have a large ~30% "wrong sign" background of neutrinos in their antineutrino event samples (to be compared to only about a 2% wrong sign antineutrino background in their neutrino samples). This is presumably due to the difficulty in defocusing the leading positively charged particles produced in their target. This is of course important for non-magnetic detectors like MiniBooNE and NOvA.

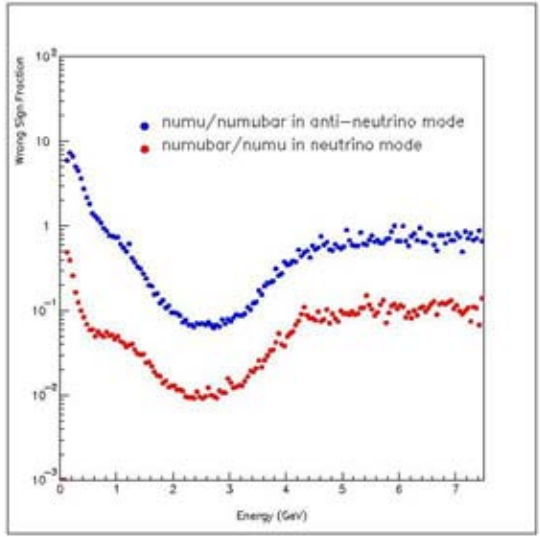

Fig. 13.27: Wrong sign fractions in the neutrino (red) and antineutrino (blue) beams for the NOνA Near Detector at Site 1.5.

The NuMI off-axis beam has different properties than the MiniBooNE beam, as shown in Figure 13.27. Away from the off-axis energy peak, the wrong sign fraction in the antineutrino beam is about 100%, but at the peak it is only about 7%. This lower level of neutrino contamination could allow NOνA to make a clean confirmation of a possible observation by MiniBooNE using anti-neutrinos.

## 13.8. Neutrino Cross Section Measurements with the NOνA Near Detector

The large samples of neutrino events collected in the NOνA Near Detector should allow some "bread and butter" measurements of basic neutrino cross sections much as planned by MiniBooNE and MINERνA. The narrow energy spectrum in the off-axis beam would make these NOνA measurements unique.

Figure 13.28 illustrates the charged current (CC) neutrino energy spectrum that will be seen by NOνA in the Near Detector running for one year in the NuMI medium energy beam at the off-axis Site 1.5 (see Chapter 9). At that time MINERνA [10] could collect an on-axis data sample in the same medium energy beam and will have already collected a large data sample with ~ 7 x $10^{20}$ pot on-axis in the NuMI low energy beam. These three potential data samples are shown together in Fig. 13.28 as a function of neutrino energy.

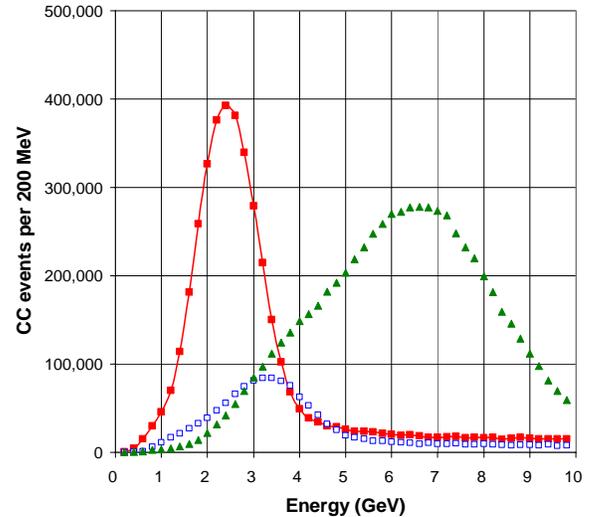

Fig. 13.28: Charged Current $\nu_\mu$ events seen in the NOνA Near Detector at Site 1.5 for 6.5 x $10^{20}$ pot (~1 year of data) in the NuMI medium energy beam (red squares with red line). The CC $\nu_\mu$ events seen by MINERνA in their 4 ton "active target" are shown for 7 x $10^{20}$ pot (~ 3 years of data) in the NuMI low energy beam (open blue squares). The CC $\nu_\mu$ events are also shown for MINERνA for 6.5 x $10^{20}$ pot (~ 1 year of data) in the NuMI medium energy beam (solid green triangles).

It is clear from Figure 13.81 that NOνA would see a high statistics data sample with a narrow energy spectrum peaked at ~2.4 GeV, to be compared with MINERνA's expected data samples peaked at ~3.3 GeV and ~6.5 GeV. The MiniBooNE data [11] peaks at lower energy (~0.7GeV). The K2K SciFi and SciBar data [12] also peak at lower energy (~1.3 GeV).

MINERνA aims to measure CC cross sections to 5% and NC cross sections to 20% with the data illustrated with the dashed curve in Figure 13.28. Both NOνA and MINERνA have totally active detectors. The NOνA Near Detector is less well suited to measuring cross sections since NOνA has a larger granularity and does not have the MINERνA additions designed for complete containment of events.

We have not yet studied the NOνA capabilities in detail, but several strategies are obvious. We could cut harder on the defined fiducial volume to reduce systematic effects from events with energy escaping out the sides or end of the NOvA Near Detector. Since NOvA sees multiple events in each 500 nsec window within the 10 microsecond MI spill (see Chapter 9), we would limit any analysis to events with only one event per 500 nsec window and the NOvA curve in Figure 13.28 already has this efficiency factor included.

Studying neutral current (NC) cross sections with the samples in Figure 13.28 is particularly interesting for the NOvA $\nu_e$ appearance analysis since NC events with $\pi^0$s can fake a $\nu_e$ event. These NC backgrounds feed down from higher neutrino energies, so measuring the NC cross sections above our oscillation $\nu_e$ peak at 2 GeV is most interesting. The NC/CC ratio is about 0.2, so NOvA would have over 700,000 NC events with the "true energy" distribution shown. Like MINERvA, NOvA could study NC $\pi^0$ production cross sections producing by looking for two photon conversions in the detector. We have not yet studied the NOvA efficiency for these specific event types.

## 13.9. Supernova Detection

*13.9.1 Supernova Physics:* At 7:35 AM on February 23, 1987, the first of about 20 neutrinos from a supernova explosion in the Large Magellanic Cloud (LMC) were detected by the IMB, Kamiokande II, and Baksan detectors [13]. Optical observations of the LMC taken at 9:30 AM showed no evidence of the blast, but it was present in observations taken one hour later at 10:30 AM indicating the neutrino signal led the optical signal by at least two hours [14].

Though the number of events was small, these neutrinos generated a great deal of interest in both the astrophysics community by confirming the core-collapse model of supernova explosions, and in the particle physic community by limiting various neutrino properties such as their mass and magnetic moment [15].

During a supernova event, the star radiates 98% of its energy in the form of neutrinos of all flavors. The neutrinos are emitted over a period of roughly 10 seconds with a time constant of roughly 3 seconds. The initial burst contains half the total neutrino signal in the first second. Before they escape the exploding star, neutrinos are trapped and reach thermal equilibrium. The neutrinos from a supernova which are detected by experiments on the earth, have an energy spectrum which peaks at roughly 20 MeV and extends out to roughly 60 MeV. Neutrino detectors capable of detecting neutrinos in this range expect to see roughly 400 events per kiloton of detector mass via the inverse beta decay reaction $\nu_e p \to e^+ n$ from a supernova located at a distance of 10 kiloparsecs (roughly the distance to the galactic center). The galactic supernova rate has been estimated to be roughly $3 \pm 1$ per 100 years [16].

Currently there are several neutrino experiments that are sensitive to supernovas. These include, SuperKamiokande, SNO, KamLAND, MiniBooNE, LVD, and AMANDA. However, several of these (SNO, KamLAND, MiniBooNE, AMANDA) are planning to complete their running over the next several years. In the time period in which NOvA expects to run, there may only be SuperKamiokande, Borexino, and ICECUBE in operation. In this case, NOvA would serve as an important backup to the SuperKamiokande detector as a detector of comparable size that is capable of supernova detection. NOvA would also serve as an important input to the supernova early warning network (SNEWS [17]), which provides astronomers an automatically generated supernova alert. In the case where both SuperKamiokande and NOvA detect a supernova signal, the neutrino flight paths through the earth to the detectors will be different, allowing the matter effect on neutrino oscillations in the earth to be studied [18].

*13.9.2 Detection of Supernova neutrinos:* A supernova explosion at a distance of 10 kpc will produce a total of 9000 neutrino interactions via inverse beta decay in the NOvA detector. Electrons at these energies produce roughly one hit strip per 15 MeV of energy, so that the majority of events between 20 and 40 MeV will have coincident hits in adjacent strips of the detector which exceed 0.5 mip (energy deposited by a minimally ionizing particle). Figure 13.29 shows the number of signal events selected by cuts on the minimum pulse height per hit and the minimum number of hit strips in the detector. Selecting hits with more than two hits which have a pulse height more than 0.5 mip yields an 80% efficiency for detection of supernova neutrinos. Folded with an estimated 50% reconstruction efficiency would yield an estimated signal of 1800 events in the first second of the neutrino burst.

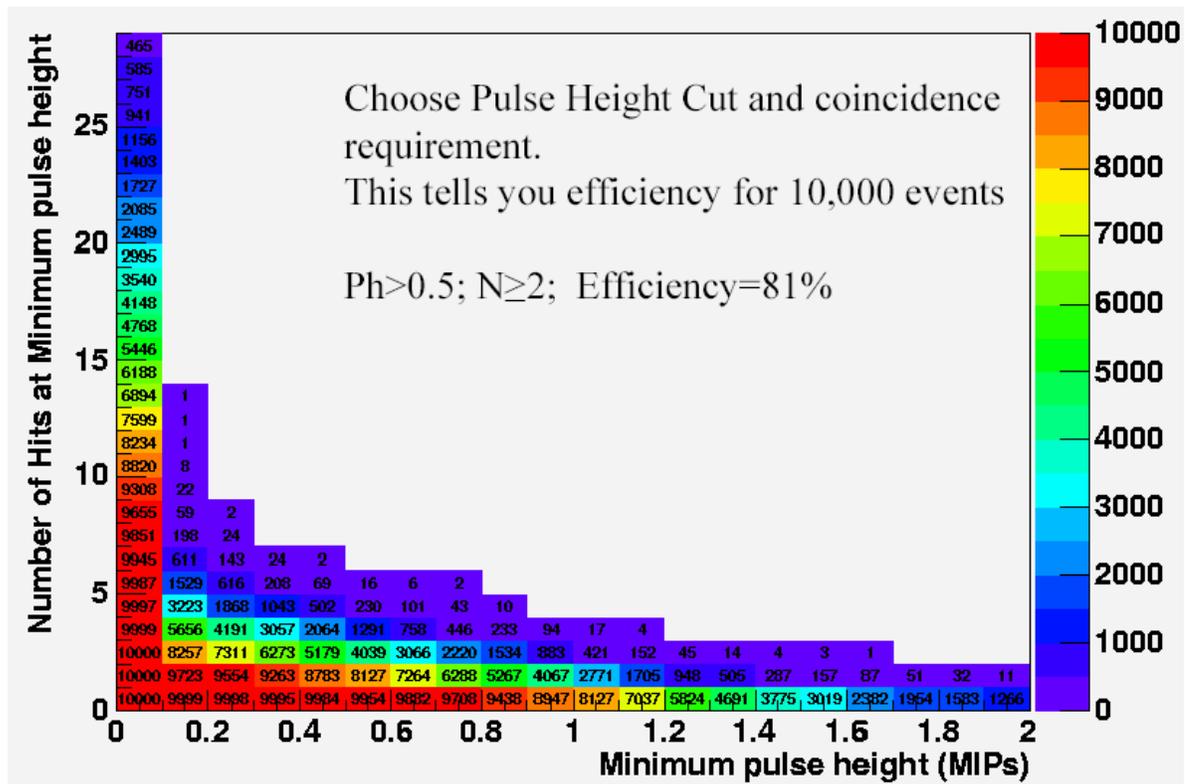

Fig. 13.29: Fraction of supernova neutrino interactions accepted as a function of minimum pulse height seen in a single cell of the detector (horizontal axis) and the minimum number of hits cells (vertical axis). Cuts on >0.5 and >2 accept 80% of the signal events.

Backgrounds to this neutrino signal have several sources from cosmic rays and neutrons. Detailed calculations have yet to be done, but it is possible to estimate the backgrounds based on experience with other surface detectors.

● Backgrounds from natural radioactivity will be small as the maximum energy for these decays is 2 MeV which is well outside the signal region of 20-60 MeV.

● The electromagnetic component of the cosmic ray backgrounds expected in NOvA is roughly 300 kHz at 10 MeV dropping to 150 kHz at 100 MeV. We expect roughly half of these particles to have energies in the 20-60 MeV signal region, and of those, 90% of them can be removed by placing the fiducial boundary for the detector 1.5 m from the detector edge. This reduces the rate from this source to roughly 10-20 kHz.

● The rate of neutrons entering the detector is roughly 30 kHz above 100 MeV. Of these, only 10% survive a 1.5 m fiducial boundary cut, yielding a background rate of roughly 3 kHz.

● The cosmic ray muon rate in the NOvA detector will be roughly 500 kHz. The muons themselves should be very easily vetoed using a 1.5 m veto region around the central part of the detector. Roughly half of the cosmic ray muons will stop in the detector and will eventually produce Michel electrons with energies up to 52 MeV – right in the heart of the supernova neutrino signal region. Assuming that muons can be tracked, one can place a time-dependent veto region around the muon track. Assuming this region is vetoed for 15 μs, the rate of Michel electrons would be reduced by 99.9%, to roughly 0.25 kHz. Likewise, the beta decay of $^{12}$B, which produces positrons below 15 MeV, can be vetoed.

Taking all sources together, we estimate that the background trigger rate will be between 15 and 25 kHz. A supernova signal would appear as an upward fluctuation of this trigger rate of roughly 3.6.

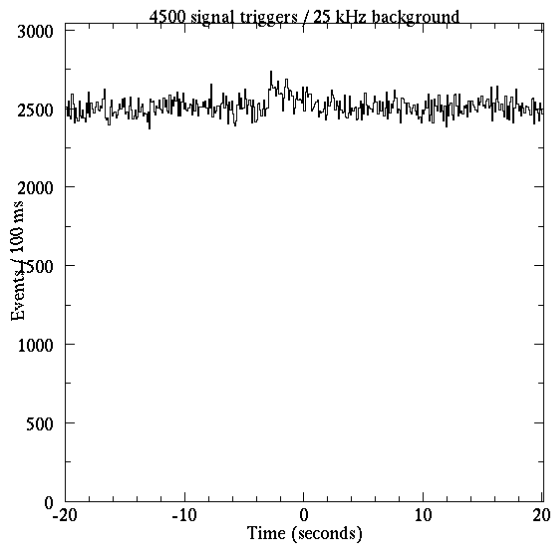

Fig. 13.30: The expected trigger rate as a function of time for a supernova explosion at a distance of 10 kpc. The supernova signal appears at time t=-3 s over a background of 25 kHz.

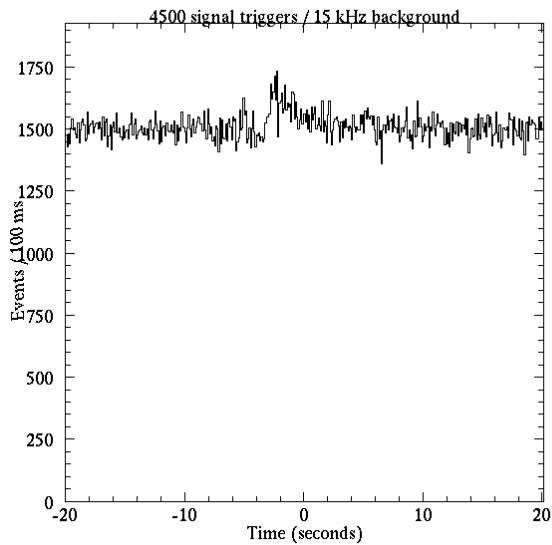

Fig. 13.31: As figure 13.30, but assuming a supernova signal at t=-2 s over a 15 kHz background.

kHz in the first second. Figures 13.31 and 13.31 shows the expected time signature for a supernova signal in the NOvA detector. Assuming 25 kHz of background, the supernova signal is at the edge of detection; at 15 kHz the signal is very clear.

*13.9.3 DAQ Requirements*: Supernova detection would place significant demands on the DAQ. First, the front end electronics would have to operate with very low dead time and the DAQ would be required to operate in a "free running" mode, and not rely solely on a beam trigger signal from Fermilab. The readout system proposed in Chapter 7 meets these requirements. Second, as the search for a supernova signal would have to be done off-line in real time, the DAQ would have to be able to handle very large data rates – roughly 3 Gb/sec. This is also possible by dividing the detector into sections that are read out and analyzed independently.

## Chapter 13 References

# 14. Cost Estimate and Schedule

## 14.1. Introduction

We have developed a cost estimate for a construction project to build the detector described in this proposal. In this chapter we explain briefly the methods used in the cost estimate and contingency analysis. We also discuss some of the important features of the estimates for each of the major elements of the proposal. We present a model for costing operational expenses, both during and after the completion of the construction project. The bottom line is that a construction project would total $ 165 M in FY2004 $, including a 50% contingency. No escalation is included. Additional required R&D is discussed in Chapter 15.

We have developed a schedule for the construction of the NOvA detectors, and that schedule is discussed in this chapter. The bottom line is that a project construction start at the beginning of FY07 leads to a complete detector in July 2011. Data taking with a fraction of the detector begins in October 2009.

## 14.2. Cost Estimate Methodology

In preparing this cost estimate we have primarily followed the principles used in costing and tracking the MINOS Detector construction project. For each major system we have itemized the materials and services (M&S) that must be procured, fabricated or assembled. Each system is itemized to the lowest level that is realistic for the current state of the system design. For each cost estimate we indicate the source of the estimate as a vendor quote, engineer's estimate or physicist's estimate. These sources are used to distinguish the confidence level in each estimate and hence are used in the contingency determination.

For each system we also itemize the labor tasks associated with the construction of each system. The cost of each task is determined by identifying the type of labor and duration of time required to carry out the task. Each type of labor is assigned a labor rate. For the purposes of this preliminary estimate we have used labor rates based on Fermilab salaries and fringe (SWF) for technicians, designers, drafters, engineers and project management personnel. For staff and installers at the far detector site we have used the labor rates currently applicable at the Soudan Underground Laboratory. Labor estimates have been made by either engineers or physicists based on time and motion studies or recent experience with similar tasks.

For each detector system we have included costs for engineering, design, inspection and administration (EDIA) throughout the life of the construction project. At this stage we have done this by estimating the person-years required based on experience from similar scale projects. We have also included costs for project management and ES&H oversight throughout the life of the construction project.

Our cost summary includes an estimate for institutional overhead based on percentages calculated from the actual costs incurred on the MINOS Detector project, namely 28% for SWF, 9% for M&S procurements under $500k, 1.5% on the first $500k of procurements of $500k or larger, and 0% on the remaining amount of the procurement. Additional overhead on large procurements at collaborating institutions must be negotiated with each institution and we have included contingency funds to cover any such agreements that might not be overhead free.

Contingency is estimated on each item or task based on the confidence level of the estimate, or on an analytical calculation based on a plausible variation of the unit cost or labor estimate. Finally we add an additional allowance to bring the overall contingency on the complete project to 50%. This is discussed in Section 14.4.3.

Table 14.1 summarizes the results of our cost estimate for the construction of this experiment. Our Total Project Cost (TPC) estimate is $ 165 M and that number includes a 50% contingency. The TPC includes the cost for the detectors and for other costs associated with the project. All costs are presented in FY 2004 dollars. At this time the cost estimate is in the form of an Excel Workbook and has not been linked

| WBS | Description | Base Cost (K$) | Overhead (K$) | Contingency (K$) | % Contingency | Sub-total (K$) |
|---|---|---|---|---|---|---|
| 1.0 | Far Detector | | | | | |
| 1.1 | Active Detector | | | | | |
| | 1.1.1 PVC Modules + Assembly | 19,513 | 2,184 | 7,085 | 33% | 28,782 |
| | 1.1.2 Liquid Scintillator + handling | 24,063 | 59 | 6,187 | 26% | 30,309 |
| | 1.1.3 Waveshifting Fiber | 13,400 | 8 | 4,022 | 30% | 17,430 |
| | 1.1.4 EDIA | 1,680 | 470 | 860 | 40% | 3,011 |
| 1.2 | Electronics, Trigger and DAQ | 7,853 | 803 | 4,756 | 55% | 13,412 |
| 1.3 | Shipping & Customs Charges | 4,799 | 960 | 1,200 | 21% | 6,958 |
| 1.4 | Installation | 7,530 | 1,963 | 4,048 | 43% | 13,541 |
| | *Far Detector Sub-total* | 78,837 | 6,446 | 28,159 | 33% | 113,442 |
| 2.0 | Near Detector | 1,678 | 470 | 945 | 44% | 3,092 |
| 3.0 | Building and Outfitting | | | | | |
| 3.1 | Site Work | 5,275 | 158 | 4,075 | 75% | 9,509 |
| 3.2 | Building | 11,532 | 346 | 5,345 | 45% | 17,223 |
| 3.3 | Outfitting | 1,262 | 38 | 1,300 | 100% | 2,599 |
| | *Building & Outfitting Sub-total* | 18,070 | 542 | 10,720 | 58% | 29,332 |
| 4.0 | Project Management | 2,985 | 805 | 948 | 25% | 4,738 |
| 5.0 | Additional Contingency | - | - | 14,145 | | 14,145 |
| | due to the early stage of the cost estimate | | | | | |
| TPC | Total Project Cost | 101,570 | 8,263 | 54,916 | 50% | 164,749 |

Table 14.1: Work Breakdown Structure and cost estimate for a NOνA construction project in FY04 $.

to project software like Microsoft Project or Open Plan.

## 14.3. The Detector Cost Estimate

The proposed 30 kiloton NOνA detector is large but uses only a few types of simple components. This simplicity makes the cost estimating exercise straightforward and easy to understand. Most of the mass of the detector is liquid scintillator and the channels are all read out by a single system of electronics. The detector is a monolithic PVC structure assembled from smaller modules constructed in factories at collaborating institutions with small crews in each factory. The detector is assembled in a short time at the far site using a crew of 34 people, about the same size as the crew which assembled the MINOS detector in the Soudan mine.
In this section we briefly discuss each of the major pieces of the estimate.

*14.3.1. Liquid Scintillator Active Detector:* There are three major components to the liquid scintillator active detector. These are the extruded PVC modules with their endcaps and fiber manifolds, the wavelength shifting fibers, and the liquid scintillator.

The detector requires the assembly of ~24,000 PVC modules with fibers. It is the simplicity of this assembly process that makes the liquid scintillator such a cost effective active detector. The time/motion analysis of the module assembly process indicates that a factory staffed with three assemblers and one supervisor can assemble 10 - 12 modules in one shift. At this rate, three assembly factories can produce the modules at the required rate. The cost of setting up these factories is included in WBS 1.1.1.

The modules are filled with liquid scintillator at the far site. The cost estimate assumes the purchase of pre-mixed liquid scintillator (mineral oil, pseudocumene, and fluors), so there will be no mixing on site. Liquid Scintillator handling is included in WBS 1.1.2 with appropriate storage tanks and a piping system to move the liquid to the detector assembly front.

*14.3.2. Front End Electronics, Trigger and DAQ:* The key components in the detector readout are the APD arrays and the custom front-end electronics to read out the APDs. The detector has ~ 762,000 channels to be read out. The custom electronics require the development of two custom ASICs incorporating a pre-amplifier, integrating amplifier, Cockroft-Walton voltage generators, multiplexer and ADC. The current estimate for the overall production cost of the readout is ~$10 per channel.

*14.3.3. Shipping:* For this design the shipping estimates include shipping (via truck) the empty modules from the three factories to the installation site and mixed scintillator oil (via truck) from Texas to the detector site. There are over 300 truckloads of modules and over 1100 truckloads of scintillator to ship. We deliberately kept the modules less than 53 feet in length so that standard trucks could be used without over-length permit fees. Similarly, no single load will be over the standard road weight limits.

*14.3.4. Installation:* We have developed an installation procedure that enables us to determine the number of people that will be required to install the detector and how long it will take. 2 people on the day shift will handle the incoming modules and scintillator oil. 25 people split over two shifts a day will construct and install detector planes, cable the electronics, and fill the completed planes with scintillator. There are an additional 7 support staff. We have built into the estimate three phases of the installation: ramp-up while assembling a full crew, steady state, and ramp down.

The installation cost estimate also includes the design and materials costs of the specialized tools and fixtures required for the installation process. This includes vacuum lifting fixtures, the "block raiser" and assembly tables described in Chapter 5, several scissor lifts for working at heights, and 220 tons of epoxy.

*14.3.5. Near Detector:* The detector described in Chapter 9 is built of identical objects to those in the NOvA Far Detector. The cost estimate assumes the same cost per module or channel for this device as for the Far Detector. Two items are unique to the Near Detector: ~ 5 planes of detector with fast electronics to resolve multiple events occurring in a 500 nsec window, and a steel or aluminum structure which allows installation of the 8-plane modules via the MINOS shaft to the underground enclosure. Cost estimates for these two special items are based on physicist estimates and have a 100% contingency.

## 14.4. Other Costs

*14.4.1. Far Detector Site, Building and Outfitting:* The current cost estimate does not include any land acquisition costs. However, the building cost estimate does include general preparation of the site such as clearing and grading. A one-mile access road is also costed. Additional access roadway costs ~$ 0.75 M per mile.

The building cost estimate has been based on a simple industrial style building with no overburden. To estimate costs we are using an algorithm developed by Fermilab Engineering Services Section (FESS), which allows us to specify the detector dimensions, the desired depth below grade of the detector, as well as an installation staging area. The building estimate includes basic utilities such as electrical distribution, fire protection and HVAC but does not include any detector specific structures or outfitting.

Outfitting costs for the detector within the building include additional HVAC and humidity control, detector specific electrical work, and the building bridge crane. The cost of epoxy paint to coat the inside of the containment bathtub is included here as are modest costs to outfit a control room and technician shop.

*14.4.2. Project Management:* We have estimated the manpower needs and corresponding cost of a project office that would oversee the management and administration of this project. This category of project management includes the Project Manager, a deputy, "Level 1" managers for the detector and the conventional construction, a project scheduler, a budget officer and an administrative assistant. Travel costs for this staff are included here.

*14.4.3. Extra Contingency:* We are at a very early stage of the design of this experiment. Most designs are only conceptual and still require detailed engineering. Our line by line contingency analysis gives a total contingency in the range of ~35% and we feel this is not sufficient. The detector is composed of a small number of different systems, but the large amounts of single commodi-ties used in the device do mean that a change in the cost per unit of the commodity will have a large cost effect. A ten cent per gallon increase in the cost of mixed liquid scintillator translates into a $0.75 Million dollar increase in cost. Similarly, an increase of one dollar per channel of electronics would add $0.75 Million dollars. A three cent per meter increase in the cost of wavelength shifting fiber gives a similar cost increase.

As stated above, our cost estimate is in FY04 dollars since many of our vendor quotes are over a year old. We have been told by vendors that the prices of plastic and mineral oil do not follow the price of crude oil, but we have not observed enough market history to verify this fact. In addition, several of our procurements are from foreign countries and the foreign exchange rate can fluctuate during a project timescale of years. Our estimate does not include a funding profile and escalation.

We feel that at this early design stage it is important to allocate contingency in a very conservative manner. For all the reasons outlined above, we have specifically added a line to our cost estimate to reflect an exact 50% contingency of our base estimate plus overhead.

## 14.5. Operating Costs

We have used experience from the NuMI-MINOS Project to develop a model that costs those expenses incurred during the construction of a project which are not appropriate to be funded by capital equipment funds. These are items such as temporary building rental, utilities in the buildings, telephone and network expenses, etc. During the construction of the MINOS far detector these costs were about $350,000 per year. Upon completion of the construction project, a budget was developed for the annual laboratory operating expenses which is currently ~$1.3M per year to support the laboratory with a crew of 8 persons. At this time it is not obvious how the laboratory for the NOvA Far Detector would have to be staffed, but such expenses should not be in excess of those currently needed for the Soudan Laboratory. These operating costs are not included in Table 14.1.

## 14.6. Offline Computing

Data rates in the NOvA Far Detector are dominated by ~ 100 Hz of out-of-spill cosmic ray data. As explained in Chapter 7, the event record size is about 100 kB per 30 μsec long data record, so 100 Hz of out-of-spill triggers gives (100 records per sec)(3 x $10^7$ sec per year)(100 kB per record) = ~300 TB per year.

The Far Detector in-spill data consists of one event record for every 30 μsec MI spill. The event record consists of the zero-suppressed list of hits with time stamps in 500 nsec slices (see Chapter 7). There are a maximum of 1.1 x $10^7$ MI spills to NOvA per year (see Chapter 11), so there are 1.1 x $10^7$ event records per year, each 100 kB in size. This results in ~1.1 TB per year. Actual neutrino events in the in-spill data amount to only about 1 GB per year.

The NOvA Near Detector also generates a computing load. Just like the Far Detector, the thirty 500 nsec slices per 10 microsecond spill are read out as one record. Unlike the Far Detector, the record size is much smaller since the Near Detector only has 12,000 channels compared to the Far Detector's 762,000. The Near Detector data is therefore (12/762)(~1.1 TB) = ~ 17 GB per year.

Assuming ~5 GHz-sec/event reconstruction time, reconstruction of Far Detector in-spill (cosmic + beam) data can be accomplished with a few nodes of 3 GHz Linux machines. Reconstruction of the 100Hz triggered cosmic ray background should take less time per event since it is mostly straight line tracking in an unmagnetized medium. Without detailed studies of what is needed for calibration and monitoring one can only guess at the computing requirements, but a rough estimate is ~1 GHz-sec/event, thus requiring 100 GHz of dedicated CPU. This is easily achievable in a cluster of 12 nodes of four 3 GHz CPUs. Given that the NOvA read out is not multiplexed (unlike MINOS), the expansion factor for reconstruction quantities should not be larger than two to four.

We note that additional NOvA Monte Carlo samples would change the conclusions in this section. These operating costs are not included in Table 14.1.

## 14.7. Schedule

Table 14.2 shows a list of milestones for R&D and construction of the detector referenced to a time $t_0$ = construction "Project Start" date. This schedule is approximately "technically driven."

| Milestone | Date (in months relative to Project Start) | Proposed Calendar Dates | FY |
|---|---|---|---|
| Procure 32-cell test extrusions with final design | $t_0 - 12$ | October-2005 | |
| initiate R&D on APD packaging | $t_0 - 12$ | October-2005 | |
| start advanced conceptual design work on the Far Site | $t_0 - 7$ | March-2006 | 06 |
| start advanced conceptual design work on the Far Building | $t_0 - 6$ | April-2006 | |
| Site Work advanced conceptual design complete | $t_0 - 2$ | August-2006 | |
| Far Building advanced conceptual design work complete | $t_0 - 1$ | September-2006 | |
| **Project Start** | $t_0$ | **October-2006** | |
| Order extrusions | $t_0 + 1$ | November-2006 | |
| Order waveshifting fiber | $t_0 + 1$ | November-2006 | 07 |
| Notice to Proceed on Far Site Work and Building *(linked to advanced conceptual designs above)* | $t_0 + 1$ | November-2006 | |
| R&D prototype Near Detector complete | $t_0 + 3$ | March-2007 | |
| Site work complete | $t_0 + 9$ | July-2007 | |
| Begin receiving packaged APD modules *(linked to R&D start above)* | $t_0 + 12$ | October-2007 | |
| Start Extrusion Module Factories *(linked to available extrusions, manifolds, and electronics)* | $t_0 + 12$ | October-2007 | |
| Start construction of Near Detector | $t_0 + 14$ | December-2007 | 08 |
| Beneficial Occupancy of Far Building | $t_0 + 19$ | May-2008 | |
| Start Outfitting of Far Building | $t_0 + 19$ | May-2008 | |
| Start operation of Near Detector | $t_0 + 21$ | July-2008 | |
| Order scintillator oil for continuous delivery 6 months later | $t_0 + 26$ | December-2008 | |
| Far Building Outfitting complete | $t_0 + 31$ | May-2009 | 09 |
| Start construction of Far Detector | $t_0 + 31$ | May-2009 | |
| Start filling Far Detector planes with Scintillator | $t_0 + 32$ | June-2009 | |
| **First kiloton (800 modules) operational** | $t_0 + 36$ | **October-2009** | |
| 5 kilotons (4000 modules) operational | $t_0 + 40$ | February-2010 | 10 |
| 10 kilotons (8000 modules) operational | $t_0 + 43$ | May-2010 | |
| 15 kilotons (12,000 modules) operational | $t_0 + 47$ | June-2010 | |
| 20 kilotons (16,000 modules) operational | $t_0 + 50$ | October-2010 | |
| 25 kilotons (20,000 modules) operational | $t_0 + 53$ | March-2011 | 11 |
| Full 30 kilotons operational | $t_0 + 57$ | July-2011 | |

Table 14.2: Proposed NOvA schedule. Dates are shown relative to Project Start at time $t_0$. Our proposed calendar schedule and the relevant fiscal years are also shown. Most R&D tasks are shown prior to the Project Start time.

We have not started a Microsoft Project or Open Plan exercise to link our cost estimate to a schedule. The main critical path is 1) construction and outfitting of the building, followed by 2) installation of the modules at the far site. Thus the module factories have a well defined time interval in which to produce the 24,000 required modules and we have chosen a production model to match that time interval. Figure 14.1 illustrates how the two assembly tasks interleave. Liquid scintillator filling of the detector begins shortly after the far site module assembly work begins and will follow the solid red curve in Figure 14.1 with a time offset of one month.

A timely start to advanced conceptual design of the Far Site work and Far Detector building during the R&D period is a critical path to ensure that procurements can be placed soon after the project start. The other main critical path is for electronics R&D to begin in time to produce the final electronics packages needed in the module factories. Table 14.2 shows these start times during the R&D period as ($t_0$ - N month) entries.

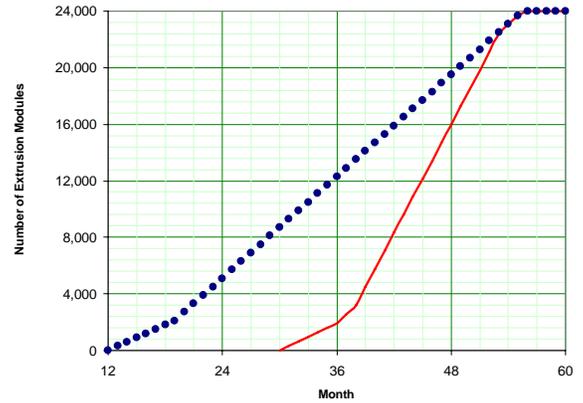

Figure 14.1: Completion schedules for modules produced in the factories (blue dots) and modules assembled into planes in the Far Detector (red line). 4,000 modules are required for each 5 kilotons of detector.

A project start in October 2006 results in NOvA data taking with the first 5 kilotons beginning in February 2010. The full detector is completed by July 2011. Figure 14.2 shows the NOvA 3 $\sigma$ sensitivity to $\sin^2(2\theta_{13})$ vs. years from the Project Start date.

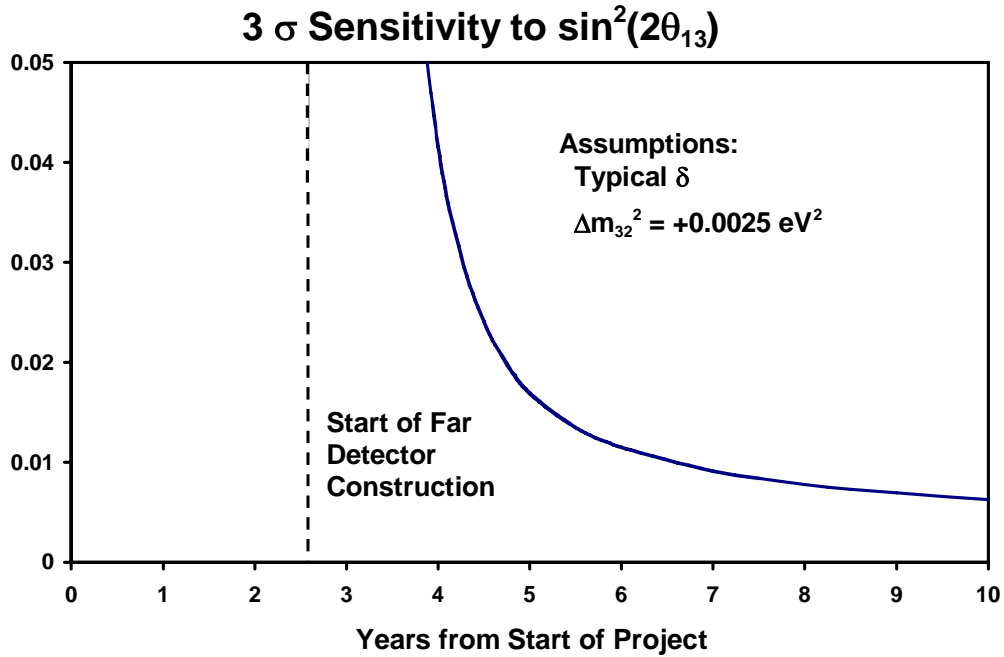

Fig. 14.2: NOvA 3 s sensitivity to $\sin^2(2\theta_{13})$ vs. years from the Project Start date.

# 15. NOνA R&D Request

## 15.1. Introduction

This chapter outlines the R&D steps we need to pursue during the next few years, FY05–FY07. Approximately $ 2 M to $ 3 M of R&D funds will be required to advance the design between now and the beginning of the construction project described in Chapter 14. Some of the R&D funds and tasks will overlap the first year of the construction project. In addition we assume contributions of effort from Fermilab and the NOνA collaborating institutions to accomplish this R&D.

The next section outlines particular subsets of the R&D, describing our progress to date and the remaining work to be done. We expect the R&D to conclude with a major effort to build a prototype NOνA Near Detector as described in Chapter 9. The final construction project Near Detector cannot be available until July 2008 (see Chapter 14), but this prototype could be complete in early 2007. The prototype Near Detector will focus our efforts to address many detailed design issues. As outlined in Chapters 9 and 10, we would run this prototype in the Fermilab test beam and in the extreme off-axis NuMI beam available in the MINOS Surface Building.

## 15.2. Specific R&D tasks

In this section we briefly discuss each of the major R&D tasks.

*15.2.1. Extrusions of Rigid PVC:* We have obtained about 3,000 feet of a 3-cell rigid PVC extrusion with a cell profile of 2.2 cm deep by 4.2 cm wide, as described in Chapter 6. These extrusions had 1.4 mm thick outer walls and 1.1 mm thick interior webs. Our next step is to repeat this 3-cell exercise with the NOνA design of cells that are 3.87 cm by 6.00 cm with 3 mm outer walls and 2 mm webs. This will give us modest lengths of extrusion with the final profile which can be used for light collection and structural studies.

The final step involves scaling up to a 32-cell extrusion with the NOνA profile. We expect to work with two vendors on these prototypes to sharpen the cost estimate and encourage competitive bidding for the final detector. The prototype 32-cell extrusions would be used in the prototype Near Detector and for structural studies of the Far Detector.

*15.2.2. Reflectivity of Rigid PVC loaded with TiO2:* The 3,000 feet of 3-cell extrusion discussed above have about 12% $TiO_2$. In pursuing these we have found that "rigid PVC" is not well defined and can contain variable amounts of acrylic impact modifiers, fillers (usually calcium carbonate), wax lubricants for the extrusion process, and organotin compounds to stabilize the extrusion process by scavenging excess HCl in the melting process.

We have seen rigid PVC products with these additives that have very poor reflectivity at 425 nm, so we need to understand the effect of each component. We want to be able to specify the composition and mechanical properties of the final NOνA rigid PVC procurement and be able to verify the product content and reflectivity.

The next step is to procure small samples from our initial vendor with additives removed one at a time so we can understand the effects, if any. The additives may change the structural properties and the reflectivity of the PVC. At the same time we will explore an increased $TiO_2$ content for better reflectivity as discussed in Chapter 6.

*15.2.3. Bottom Closures and Top Manifolds for the Extrusions:* For the design described in Chapter 5 we need to investigate injection molded top manifold parts and machined bottom closure parts. Initial prototypes of each will be custom machined during the R&D period.

*15.2.4. Wavelength Shifting Fiber:* We plan to investigate controlling the position of the fibers in the cells as discussed in Chapter 6. We will also look at products from other vendors for our large fiber order.

*15.2.5. Liquid Scintillator:* We have been investigating custom mixtures of mineral oil and fluors and will continue this work to find an optimum price scintillator. We also need to study the effect of PVC, fiber, and epoxies on the scintillator and the effect of the scintillator on the PVC, fiber and epoxies.

*15.2.6. APD Packaging:* We have been investigating a NOνA-specific APD pixel array and packaging for our thermo-electric coolers. R&D on this subject has to be concluded in time to get electronics for the final detectors as indicated in Table 14.2.

*15.2.7. ASIC Designs:* We need two ASICs for NOvA, a low noise preamp chip and a Cockroft-Walton high voltage chip. Both are being designed by the ASIC group at Fermilab. Prototype runs would likely include enough wafers to build the prototype Near Detector discussed below in 15.2.10.

*15.2.8. Site and Building Work:* We need consultant engineering help with a field inspection of the possible Ash River sites so as to narrow the choices. As noted in Chapter 14, we then need to proceed with advanced conceptual designs for the Site(s) and the Far Detector building.

*15.2.9. Structural Analysis and Prototypes of the Far Detector:* A great deal of engineering study and finite element analysis has already gone into understanding the PVC structure in Chapter 5. We still plan to hire an outside consulting firm to validate the safety factors and look for failure modes of the structure that our in-house effort may have missed. This is a unique structure and we need to verify the design.

We will continue to build small prototypes of the structure to understand its properties. We have also built a ~ half-scale prototype using a commercial garage door PVC product with a cell size about half the NOvA cell size. A picture of this structure is shown in Figure 15.1. We plan to pursue a similar structure with the "full-size" NOvA extrusions, "full" meaning a few meters wide by as tall as we can fit in an existing building at Fermilab or Argonne.

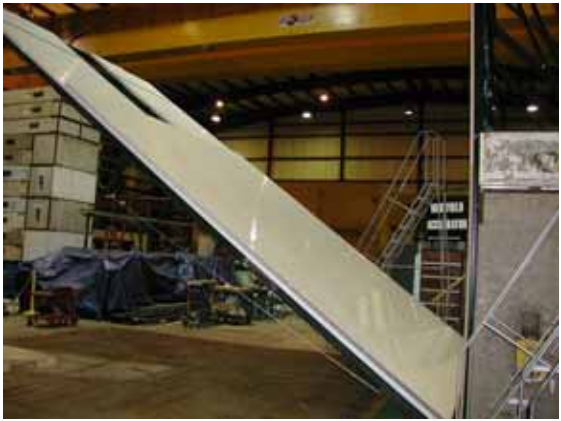

Fig. 15.1: The "half-size" prototype of NOvA at Argonne National Laboratory. The prototype is ~ 8 meters high and it is made of two layers of extrusions. The planes were assembled on a Unistrut strongback and winched up against the vertical bookend on the right while pivoting about the lower right corner.

*15.2.10. Prototype Near Detector:* Our final goal for the R&D period is to construct a prototype Near Detector like the one described in Chapter 9. This will test all our designs and our assembly procedures. The resulting device will let us get started understanding our detector response with studies in the Fermilab test beam and in the MINOS Surface Building. The device will also let us check the cosmic ray rates seen by the detector on the surface.

This prototype Near Detector might be done relatively cheaply via a loan [1] of NuTeV liquid scintillator to NOvA and by utilizing existing steel at Fermilab for the muon catcher. The electronics would be based on off-the-shelf APDs and thermo-electric coolers mated with prototype NOvA ASICs. R&D funding restrictions might limit the length of the prototype.

## 15.3. R&D Cost Estimate

Table 15.1 gives a rough cost estimate for the R&D tasks described in Section 15.2. A 33% contingency is included (mostly for uncertainties in the cost of the prototype Near Detector). The items in Table 15.1 include the costs of consultants and labor at collaborating institutions. Fermilab engineering and technical help would be in addition to this total.

| R&D Task | Approximate Materials & Services funding required (K$) |
|---|---|
| Extrusions of Rigid PVC | 325 |
| Refectivity of Rigid PVC | 50 |
| Bottom Closures and Top Manifolds | 150 |
| Liquid Scintillator studies | 50 |
| APD Packaging | 275 |
| ASIC Designs | 135 |
| Site and Building Designs | 150 |
| Structural Analysis and Prototypes | 240 |
| Prototype Near Detector | 885 |
| Contingency @ ~ 33% | 740 |
| Total | 3,000 |

Table 15.1: Estimated cost of remaining R&D.

## Chapter 15 References